\pdfoutput=1
\documentclass[aps,prd,amsmath,floats,floatfix,superscriptaddress,nofootinbib,showpacs]{revtex4}

\usepackage{aas_macros}

\usepackage[T1]{fontenc}
\usepackage[utf8]{inputenc}
\usepackage{lmodern}
\usepackage{lipsum}
\usepackage{booktabs}

\usepackage[dvipsnames, usenames]{xcolor}
\definecolor{linkcolor}{rgb}{0.0,0.3,0.5}
\usepackage[hypertexnames=false, unicode, colorlinks=true, linkcolor=linkcolor,
citecolor=linkcolor, filecolor=linkcolor,urlcolor=linkcolor,
pdfusetitle]{hyperref}

\usepackage[all]{hypcap}
\usepackage{graphicx}
\usepackage{xspace}
\usepackage{amssymb}
\usepackage[normalem]{ulem} %for \sout
\usepackage{bm} % boldmath
\usepackage{appendix}

\usepackage{microtype}

\usepackage[english]{babel}
\usepackage{blindtext}

\graphicspath{%
  {figs/}%
}

\newcommand{\hn}{\hat{n}}

\renewcommand{\vec}[1] {\bm{#1}}

\newcommand*{\df}  {\delta}

\newcommand*{\eps}  {\epsilon}

\newcommand{\hk} {\hat{k}}

\newcommand*{\non} {\nonumber}
\newcommand*{\lb} {\left(}
\newcommand*{\rb} {\right)} 
\newcommand*{\ls} {\left[}
\newcommand*{\rs} {\right]}
\newcommand*{\la} {\left\langle}
\newcommand*{\ra} {\right\rangle}

\newcommand{\eq}[1]{\begin{align}#1\end{align}}
\newcommand{\eeq}[1]{\begin{equation}#1\end{equation}}

\begin{document}

\rightline{\scriptsize RBI-ThPhys-2025-30}

\title{Intrinsic Alignments in Redshift Space I: Symmetries}

\newcommand\kazuhome{
\affiliation{Theory Center, Institute of Particle and Nuclear Studies,
High Energy Accelerator Research Organization (KEK), Tsukuba, Ibaraki 305-0801, Japan,}
}

\newcommand\stephenhome{
\affiliation{Department of Physics, Columbia University, New York, NY, USA 10027, \\
NASA Hubble Fellowship Program, Einstein Fellow ,
}
}

\newcommand\zvonehome{
\affiliation{Ru\dj er Bo\v{s}kovi\'c Institute, Bijeni\v{c}ka cesta 54, 10000 Zagreb, Croatia.}
}

\author{Kazuyuki Akitsu}
\email{kakitsu@post.kek.jp}
\kazuhome

\author{Shi-Fan Chen}
\email{sc5888@columbia.edu}
\stephenhome

\author{Zvonimir Vlah}
\email{zvlah@irb.hr}
\zvonehome

% Because hyperref only gets the *last* author, we need to be explicit.
%\hypersetup{pdfauthor={Vlah et al.}}

%\date{\today}

%Do not go over the dashed lines with the text in the .tex file. Due to formatting on the overleaf. 
%==========================================================================
\begin{abstract}

The intrinsic alignments of galaxy shapes are unique tensor tracers of large-scale structure, and their measurement is a promising avenue to both enhance current cosmological programs and detect new physics beyond the scalar sector. In this paper, we develop a general formalism to describe the full, three-dimensional structure of galaxy shapes and their statistics, including the breaking of isotropy by the observer's line of sight and redshift space distortions due to peculiar velocities. We constructively show that the nonlinear redshift-space map generates a kinematic basis whose form factors are strictly polynomial in the line-of-sight angle $\mu = \hk \cdot \hn$, and that parity selection rules restrict scalar-tensor and tensor-tensor correlators to 3 and 13 independent form factors, respectively, with the latter further reduced to 9 by exchange symmetry. We further show that polynomial form factors are preserved when transformed into a \textit{total helicity} basis of tensors with total angular momentum $M$---sourced to be nonzero by powers of the line-of-sight $\hn$---and that this is equivalent to the form factors having spin weights $(1 - \mu^2)^{|M|/2}$ and described by associated Legendre polynomials $P^{|M|}_\ell$. We further construct estimators for these form factors in the normalized total-helicity basis, showing that they provide the optimal angular weighting to extract shape information, and write down the mapping connecting the full tensor basis to the more commonly measured projected shape statistics. To validate our formalism, we study the derived tensor bases and form factors within a toy model, showing that all allowed channels are generated even within a simplified dynamical model, and apply our estimators to halo shape statistics in N-body simulations, wherein we detect all channels with up to total angular momentum $|M| \leq 2$. We anticipate that the methods developed in this work will have broad applications, ranging from optimal extraction of intrinsic-alignment information in hydrodynamical simulations to identifying new physics in tensor channels forbidden by selection rules in the standard model of cosmology.

\end{abstract}

\maketitle

%==========================================================================
\section{Introduction}
\label{sec:intro}

The advent of wide and deep galaxy surveys has established the large-scale structure as one of the main tools in precision cosmology. Galaxy positions are the most commonly used tracer of this structure, but additional information is contained in the observed shapes and orientations of galaxies. Gravitational lensing by the intervening matter distribution induces coherent distortions in galaxy images and has enabled increasingly precise constraints on the growth of structure and the expansion history of the Universe. At the same time, the physical processes that form and evolve galaxies correlate the intrinsic shapes of galaxies with the surrounding large-scale structure. These correlations, commonly referred to as intrinsic alignments, have been detected to high significance in both photometric and spectroscopic surveys \cite{Joachimi2015,Kiessling2015,Kirk2015,TroxelIshak2015,Mandelbaum2006,Singh2015}. Intrinsic alignments are therefore an important contribution to the total observed galaxy shape statistics and must be accounted for in weak-lensing analyses. Their detection is, however, on sufficiently firm grounds that they can also be considered as a cosmological observable in their own right \cite{ChisariReview2025}.

The cosmological interest in intrinsic alignments follows from the fact that galaxy shapes are tensorial tracers of the large-scale structure. At leading order, the trace-free intrinsic shape traces the local gravitational tidal field, in analogy with the relation between galaxy number counts and the matter density field \cite{Catelan2001, HirataSeljak2004, Vlah2020}. Shape correlations consequently probe a different response to the large-scale gravitational environment than do scalar clustering observables. The additional tensor structure also gives rise to a characteristic dependence on the orientation of the Fourier mode, or of the pair separation, relative to the line of sight. This angular dependence is not merely a geometrical complication, but is instead one of the main carriers of cosmological information in shape statistics, and its systematic description is our central interest in this work.

A number of late-time applications illustrate the potential of this cosmological probe. Position--shape correlations are sensitive to peculiar velocities and thus to the growth rate of structure \cite{TaruyaOkumura2020,ZwetslootChisari2022}, and the first constraints on the growth rate from redshift-space ellipticity correlations have recently been obtained from spectroscopic data \cite{OkumuraTaruya2022}. The baryon acoustic feature is retained in alignment statistics and can therefore provide geometrical information on cosmic distances and the expansion history \cite{ChisariDvorkin2013,Okumura2019,vanDompseler2023}. Evidence for this feature at the $2$--$3\sigma$ level has been reported from the density--ellipticity correlation of BOSS CMASS galaxies combined with imaging-based shape measurements \cite{Xu2023}, while density--shear correlations have been proposed, and recently applied to photometric data, as a consistency check on the distance scale inferred from conventional clustering statistics \cite{Chan2026}. Because the intrinsic shape is a spin-two object, it is also a natural observable for tests of statistical isotropy and of parity invariance, for which otherwise vanishing angular structures or correlations involving $B$-modes provide characteristic signatures \cite{Kogai2018,Vlah2021,BiagettiOrlando2020,Akitsu:2022lkl,Shiraishi2023,Kurita:2025}. Relativistic contributions to position--shape correlations generate odd multipoles, including a dipole sourced by the Doppler and gravitational-redshift effects \cite{Saga2023}; such observables may in turn provide complementary tests of gravity and the equivalence principle.

Galaxy shapes are, in addition, a sensitive probe of the physics of the early Universe, in a manner that is genuinely complementary to galaxy clustering. Because the intrinsic shape responds to the anisotropic part of the tidal field, it couples to types of primordial non-Gaussianity to which scalar number counts are not sensitive. The imprint of anisotropic primordial non-Gaussianity on shape correlations was first derived in Ref.~\cite{Schmidt2015}, where a quadrupolar, spin-two anisotropy was shown to induce in the alignment bias the analogue of the scale-dependent bias generated in number counts by isotropic non-Gaussianity. This prediction has since been confirmed in $N$-body simulations, which exhibit a characteristic  $1/k^2$ modification of the halo shape power spectra in the low-$k$ limit while the halo number-density power spectrum remains unaffected \cite{Akitsu2021}. Galaxy positions and shapes therefore carry information on different angular components of the primordial bispectrum, and a joint analysis of the density and alignment power spectra can constrain the isotropic and anisotropic contributions simultaneously, as demonstrated by the first such measurement from combined spectroscopic and imaging data \cite{KuritaTakada2023}. Since the inferred alignment strength moreover depends on the way in which galaxy shapes are measured, several shape estimators with different alignment responses could themselves be combined in a multi-tracer analysis, which is forecast to further tighten the constraints \cite{Chisari2016multi}. The cosmic microwave background bispectrum probes the same class of non-Gaussianity, but on very different physical scales, so the two probes naturally combine.

This sensitivity is emphasised further in the context of the cosmological collider programme. If massive particles with non-zero spin were present during inflation, they would imprint a characteristic scale dependence and oscillatory structure on the squeezed limit of the primordial correlators, and hence on the alignment bias \cite{Kogai2018}. Decomposing the observed galaxy shape field into its spin components can, in principle, discriminate between the spins of such particles, so that imaging surveys act as spin-sensitive detectors of the inflationary particle content \cite{Kogai2021}. Related primordial relics leave their own signatures: a primordial gravitational-wave background can leave a long-lived fossil imprint on the shapes of galaxies, observable as alignment $B$-modes \cite{SchmidtJeong2012,Schmidt2014,Akitsu:2022lkl}, although the detection prospects are tempered by current limits on the tensor-to-scalar ratio \cite{BiagettiOrlando2020}, while primordial magnetic fields source vector and tensor metric perturbations which can likewise be probed by alignments \cite{Saga2024}. A recent study of the signatures of both tensor and vector perturbations in galaxy shapes has been given in Ref.~\cite{Philcox2024}. Complementary information may also be carried by other, non-tensorial galaxy properties, such as galaxy sizes, which have been proposed as an effectively unbiased tracer of local primordial non-Gaussianity \cite{Nguyen2026}.

Spectroscopic surveys provide a particularly useful setting for this programme. Accurate redshifts allow one to identify physically associated galaxy pairs, reduce the dilution of the alignment signal from pairs separated along the line of sight, and separate intrinsic alignments more cleanly from gravitational lensing. They also allow retaining the full three-dimensional anisotropy of the correlation functions, rather than compressing the signal into projected angular statistics. Position--shape correlations have already been measured for several spectroscopic samples, and estimators for the corresponding power-spectrum and correlation-function multipoles have been developed \cite{Mandelbaum2006,Hirata2007,Singh2015,SinghMandelbaum2016,Kurita2021,KuritaTakada2022,Singh2023}. The overlap between imaging data, which provide the shape measurements, and spectroscopic surveys such as BOSS and DESI thus opens the possibility of treating galaxy shapes as genuine three-dimensional tracers of the large-scale structure, and comparable calibration efforts are under way for the forthcoming generation of surveys \cite{Hoffmann2026}. This is also of direct relevance for galaxy clustering itself: correlations between galaxy orientation and target selection generate anisotropic contributions to the observed number density and can bias measurements of redshift-space distortions if they are not accounted for \cite{Hirata2009,Martens2018,Obuljen2020,Singh2021,Lamman2023}.

A particularly important feature of spectroscopic surveys is that observed radial positions are not the real-space positions of galaxies, since peculiar velocities change observed redshifts and map galaxy positions into redshift space. For scalar tracers, the resulting redshift-space distortions are well understood and constitute one of the main probes of structure formation through the growth rate \cite{Kaiser1987,Hamilton1992}. For galaxy shapes, the situation is qualitatively different. At linear order, the intrinsic shape evaluated at the redshift-space position is unchanged by the mapping, so that the leading shape--shape correlation is not modified, while the position--shape cross-correlation acquires the familiar Kaiser contribution from the scalar density \cite{Singh2015,OkumuraTaruya2020}. This makes the joint statistics of positions and shapes sensitive to the velocity field in a way that is complementary to ordinary clustering. Beyond linear order, the displacement from real to redshift space couples the shape field to velocity moments, density weighting and nonlinear evolution, while small-scale random motions produce additional damping. A number of these contributions have been studied individually, including density weighting, the Kaiser effect beyond leading order, Fingers-of-God damping and the configuration-space multipole structure \cite{OkumuraTaruya2020,Kurita2021,Okumura2023nl,Taruya2024,Okumura2024}, and the consequences for estimator design have begun to be explored \cite{Lamman2025}. A complete treatment of perturbative alignment models in redshift space is nevertheless not yet available, and at present can only be assembled piecewise from these partial results \cite{ChenKokron2023,ChisariReview2025}.

Furthermore, a complete treatment of the redshift-space distortions for galaxy shapes cannot be obtained by some simple replacement of the density field in the standard theory of redshift-space clustering with a shape variable. The intrinsic galaxy shape transforms as a symmetric trace-free tensor, and its correlations contain several independent tensor structures even before a preferred line-of-sight direction is introduced. Assuming the plane-parallel approximation, redshift space breaks statistical isotropy down to rotations about the line of sight and thereby admits additional angular structures in scalar--tensor and tensor--tensor statistics. It is therefore useful to separate the general consequences of redshift-space mapping and residual statistical symmetries from the particular model adopted for the real-space dynamics. Systematic descriptions of the real-space shape statistics are by now well formulated, both in the Eulerian effective-field-theory and bias framework \cite{Vlah2020,Vlah2021,Bakx2023,Akitsu:2023eqa} and in its Lagrangian counterpart \cite{ChenKokron2023}, the latter having already been applied to survey data \cite{ChenDeRose2024,DeRoseChen2025,SDIP2026}; the corresponding programme has also been extended beyond two-point statistics \cite{Bakx2025a,Bakx2025b,Vedder2026}. In parallel to these, alternative perturbative models have also been developed; see, e.g., \cite{Blazek2015,Blazek2017,Schmitz:2018} and \cite{Matsubara2022,Matsubara2022II}. The geometrical and kinematic construction of the redshift-space correlators is, however, more general than any of these dynamical models, and can be applied independently of a specific perturbative truncation. Establishing this framework is a necessary step for both interpreting spectroscopic alignment measurements and developing controlled nonlinear models in future work.

The goal of this work is to develop a general framework for the statistics of galaxy shapes in redshift space. We proceed as follows:
\begin{itemize}
\item In Sec.~\ref{sec:rsd} we set up the redshift-space mapping for tensor-valued tracers and introduce the corresponding correlators.

\item In Sec.~\ref{sec:symm} we derive constraints imposed by reality, parity, exchange symmetry, homogeneity and isotropy, and obtain the resulting selection rules in an isotropic basis.

\item In Sec.~\ref{sec:IA_correlators} we specialise the construction to galaxy shapes, discussing the choice of basis for the scalar--tensor and tensor--tensor sectors.

\item In Sec.~\ref{sec:projections} we examine the projection of three-dimensional shape statistics onto the sky.

\item In Sec.~\ref{sec:measurements} we discuss shape measurements, construct the estimator, and derive the covariance matrix.

\item In Sec.~\ref{sec:toy_example} we develop a toy model, the Gaussian-field streaming model for the scalar--tensor and tensor--tensor cases, together with the corresponding loop expansion.

\item In Sec.~\ref{sec:halos}, we further measure the scalar-tensor and tensor-tensor form factors of halos in N-body simulations, and compare their statistics to our theoretical calculations in the previous sections.

\item We end by summarising our results and providing some concluding remarks in Sec.~\ref{sec:conclusion}.
\end{itemize}
Remaining technical details are delegated to the Appendices. Table~\ref{tab:notation} summarises the notation used throughout the paper for the most important physical and mathematical quantities.

%=======================================================================%
\begin{table*}
\centering
\begin{tabular}{l l}
\hline
\hline
\multicolumn{2}{l}{\textit{Conventions}} \\ [2pt]
$\df_{ij}$ &~ Kronecker symbol \\ [1pt]
$\df^{\rm D}(\vec x)$ &~ Dirac delta function \\ [1pt]
$f(\vec k) \equiv \int d^3x\, f(\vec x)\, e^{i \vec k\cdot\vec x}$ &~ Fourier transform convention \\ [1pt]
$\int_{\vec p} \equiv \int \frac{d^3 p}{(2\pi)^3}$ &~ Momentum integral \\ [1pt]
$\la A(\vec k) B(\vec k') \ra = (2\pi)^3 \df^{\rm D}(\vec k + \vec k')\, \la A(\vec k) B(\vec k') \ra'$ &~ Correlator with the momentum-conserving delta removed ($'$) \\ [3pt]
\multicolumn{2}{l}{\textit{Fields}} \\ [2pt]
$\df(\vec x)$ &~ 3D density field of matter or biased tracer \\ [1pt]
$\df_{\rm s}(\vec r)$ &~ Fractional galaxy size perturbation (trace of the shape tensor) \\ [1pt]
$I_{ij}(\vec r)$ &~ Density-weighted 3D galaxy shape tensor \\ [1pt]
$S_{ij}(\vec r)$ &~ Shape fluctuation field, $S_{ij} = g_{ij} + \tfrac13 \df_{\rm s}\df_{ij}$ \\ [1pt]
$g_{ij}(\vec r)$ &~ Trace-free part of the 3D shape tensor \\ [1pt]
$\gamma_{ij}$ &~ Shape projected on the sky (trace-free in the sky plane) \\ [1pt]
$\gamma_{E},\, \gamma_{B}$ &~ $E$- and $B$-mode components of the projected shape \\ [1pt]
$\vec u = \vec v/\mathcal H$, ~ $u_{\hn} = \hn\cdot\vec u$ &~ Scaled peculiar velocity and its line-of-sight component \\ [3pt]
\multicolumn{2}{l}{\textit{Geometry}} \\ [2pt]
$\hn$ &~ Line-of-sight direction (distant-observer limit) \\ [1pt]
$\mu \equiv \hk\cdot\hn$, ~ $s \equiv \sqrt{1-\mu^2}$ &~ Cosine of the angle between $\hk$ and $\hn$, and its complement \\ [1pt]
$m_i \equiv \hn_i - \mu \hk_i$, ~ $t_i \equiv \varepsilon_{iab}\hk_a\hn_b$ &~ Vectors transverse to $\hk$ spanning the parity-even and -odd directions \\ [1pt]
$\mathcal P_{ij} = \df_{ij} - \hn_i\hn_j$ &~ Projector onto the sky plane \\ [3pt]
\multicolumn{2}{l}{\textit{Redshift-space mapping}} \\ [2pt]
$\vec x_s = \vec r + u_{\hn}(\vec r)\, \hn$ &~ Real- to redshift-space mapping \\ [1pt]
$\hat N_{i_1\ldots i_L} = \hn_{i_1}\cdots\hn_{i_L}$ &~ Line-of-sight projection tensor of order $L$ \\ [1pt]
$T^L_{a;\, i_1\ldots i_L}$, ~ $\Xi^L_{ab;\, i_1\ldots i_L}$ &~ Velocity moments of the fields and their correlators, Eq.~\eqref{eq:Xi_L} \\ [3pt]
\multicolumn{2}{l}{\textit{Tensor bases and form factors}} \\ [2pt]
$\mathcal M_{kk}, \mathcal M_{kn}, \mathcal M_{nn}$ &~ Kinematic (polynomial) basis of rank-two STF tensors \\ [1pt]
$\mathcal Q_n$, ~ $\bar{\mathcal Q}_n$ &~ Real parity-adapted basis and its normalised form \\ [1pt]
$Y^{(m)}_{ij}(\hk)$ &~ Complex helicity basis, $m = 0,\pm1,\pm2$ \\ [1pt]
$\mathcal Y^{\parallel}_{mm'}, \, \mathcal Y^{\perp}_{mm'}$ &~ Aligned and anti-aligned total-helicity basis for rank-four correlators \\ [1pt]
$M$, ~ $|M|$ &~ Total helicity of a basis element and its magnitude \\ [1pt]
$G_\lambda(k,\mu)$ &~ Scalar--tensor form factors, $\lambda = 0,1,2$ \\ [1pt]
$H^{X}_{mm'}(k,\mu)$ &~ Tensor--tensor form factors, $X = \parallel,\perp$ \\ [1pt]
$\bar G^{\,\ell}_n, \, \bar H^{X,\ell}_{mm'}$ &~ Associated-Legendre multipoles of the normalised form factors \\ [1pt]
\hline
\hline
\end{tabular}
\caption{List of notation and most important quantities used in this paper. Fields in Fourier space are understood to be integrated over repeated momentum variables.}
\label{tab:notation}
\end{table*}

%====================================================%
\section{Redshift Space Distortions of Tensor Fields}
\label{sec:rsd}

In this section, we consider only the redshift-space remapping of the three-dimensional size and shape fields in the distant-observer approximation, in direct analogy with the usual treatment of scalar tracers in redshift space \cite{Kaiser1987, Hamilton1992, Scoccimarro04}. As in the EFT description of intrinsic alignments, we first formulate the theory for the three-dimensional tensor field in the rest frame, and only afterwards connect it to projected observables on the sky \cite{Vlah2020,Vlah2021}. The sky projection itself is a separate step; in particular, the projector onto the sky is unchanged by the redshift-space mapping in the plane-parallel limit, since the real- and redshift-space line-of-sight directions coincide \cite{Vlah2021}. We also stress that the line-of-sight dependent selection effects discussed in Ref.~\cite{Vlah2020} are conceptually distinct from the pure Doppler remapping considered here, and should be included as additional operators rather than being absorbed into the mapping itself. Full projection effects for spectroscopic samples, together with subleading light-cone and GR contributions, are beyond the scope of this section and are deferred to future work, in the same spirit as in Refs.~\cite{Vlah2020,Vlah2021,Bakx2023,ChenKokron2023}.

\subsection{Tensor Fields in Redshift Space}

We define a tensor field of intrinsic, i.e. unlensed, galaxy shapes as a function of comoving location,
\eeq{
  I_{ij}(\vec r) = \sum_\alpha I^\alpha_{ij}\, \delta^{\rm D}(\vec r-\vec r_\alpha)\, ,
}
where the sum runs over all galaxies in the survey and $I^\alpha_{ij}$ is the symmetric shape tensor of each galaxy. The trace of the field returns a scalar field, which describes the number-weighted galaxy size, i.e.
\eeq{
{\rm tr}[I_{ij}(\vec r)] = \overline{s^2} (1+\delta_{\rm s}(\vec r)),
}
where $\overline{s^2}$ is defined through $\langle I_{ij} \rangle = \delta_{ij}\, \overline{s^2}/3$ and $\delta_{\rm s}(\vec x)$ is the corresponding size fluctuation field.  We can then define the shape fluctuation field as the trace-free fluctuations of $I_{ij}$
\eq{
S_{ij}(\vec r) =  \frac{ I_{ij}(\vec r) - \langle I_{ij}(\vec r) \rangle}{{\rm tr} \langle I_{ij} \rangle }
= g_{ij}(\vec r) + \frac13 \delta_{\rm s}(\vec r) \delta_{ij}\,,
}
where we introduced the trace-free tensor $g_{ij}$, which describes galaxy shape perturbations.

The mapping from the real to the redshift space is given by the Doppler shift
\eeq{
\vec x_s = \vec r + u_{\hat n}(\vec r) \hat n\, ,
}
where $u_{\hat n} = \hat n \cdot \vec u$, $\vec u = \vec v /\mathcal H$, and we have discarded the explicit time variable. The density-weighted redshift-space shape field is therefore 
\eq{
  I^s_{ij}(\vec x_s)
   = \sum_\alpha I_{ij}^\alpha\,  \delta^{\rm D}(\vec x_s-\vec x_\alpha - \hat n u_{\hat n, \alpha} )  = \int d^3 r ~ I_{ij}(\vec r)\, \delta^{\rm D}(\vec x_s - \vec r - \hat n u_{\hat n} (\vec r) ) \, .
}
Since the redshift-space map only translates the support of the density-weighted tensor field the mean is unchanged $\langle I^s_{ij} \rangle = \langle I_{ij} \rangle $. We can therefore analogously define the shape fluctuation field in redshift space as
\eq{
S^s_{ij}(\vec x_s) =  \frac{ I^s_{ij}(\vec x_s) - \langle I^s_{ij}(\vec x_s) \rangle}{{\rm tr} \langle I_{ij} \rangle }
= g^s_{ij}(\vec x_s) + \frac13 \delta^s_{\rm s}(\vec x_s) \delta_{ij}\, .
\label{eq:def_shape_field}
}
such that
\eq{
  \frac{1}{3} \delta_{ij} + S^s_{ij}(\vec x_s)
  & = \int d^3 r ~ \lb \frac{1}{3} \delta_{ij} + S_{ij}(\vec r)  \rb \delta^{\rm D}(\vec x_s-\vec r - \hat n u_{\hat n} (\vec r) ) \, .
}
The trace part then gives us the size fluctuation in redshift space
\eeq{
  1 + \delta^s_{\rm s}(\vec x_s)
  = \int d^3 r ~ \lb 1 + \delta_{\rm s} (\vec r)  \rb \delta^{\rm D}(\vec x_s-\vec r - \hat n u_{\hat n} (\vec r) ) \, ,
}
while the trace-free part gives the redshift space shape fluctuations
\eeq{
  g^s_{ij}(\vec x_s)
  = \int d^3 r ~ g_{ij}(\vec r) \delta^{\rm D}(\vec x_s-\vec r - \hat n u_{\hat n} (\vec r) ) \, .
}

It is especially convenient to express the above redshift-space fields in Fourier space. For the trace-free part we have
\eq{
  g^s_{ij}(\vec k) =  \int d^3 r ~ g_{ij}(\vec r) e^{i \vec k \cdot  \lb \vec r + \hat n u_{\hat n} (\vec r) \rb} =  \sum_{L=0}^\infty \frac{( i k_{\hat n} )^L}{L!} \int d^3 r ~ g_{ij}(\vec r) u^L_{\hat n} (\vec r) e^{i \vec k \cdot \vec r} \, ,
}
and the trace part can be similarly obtained by substituting $(1 + \delta_s)$ for $g_{ij}$. An important structural point is that, unlike the scalar trace part, the trace-free field does not contain a homogeneous (``$1$'') contribution. Therefore the first RSD correction to $g^s_{ij}$ starts only at nonlinear order. At linear order one therefore obtains
\eq{
\delta^{{s},(1)}_{\rm s}(\vec k)
&=
\delta^{(1)}_{\rm s}(\vec k) + i k_{\hat n} u^{(1)}_{\hat n}(\vec k) = \lb b_1^{(s)} + f\mu^2 \rb \delta_m^{(1)}(\vec k)\, ,
\non\\
g^{{s},(1)}_{ij}(\vec k)
&=
g^{(1)}_{ij}(\vec k) = c_s \lb \frac{k_i k_j}{k^2} - \frac13 \delta_{ij} \rb \delta_m^{(1)}(\vec k)\, .
}
where we have defined the scalar and tensor linear biases $b_1^{(s)}$ and $c_s$, and we recover the usual Kaiser-type structure for the trace part since it behaves identically to the galaxy density in redshift space.

\subsection{Redshift-Space 2-point Functions and The Moment Expansion}

In analogy to the velocity moment expansion for the scalar field \cite{Seljak:2011,Vlah:2012,Vlah:2013,Vlah19}, the structure of the redshift-space mapping thus makes it convenient to define the size and shape velocity moments
\eq{
T^L_{i_1\ldots i_L} (\vec r) &=
\begin{cases}
\delta_{\rm s}(\vec r) \, , & L=0 \, , \\
( 1 + \delta_{\rm s}(\vec r) )  u_{i_1}(\vec r) \cdots u_{i_L}(\vec r) \, , & L\geq 1 \, ,
\end{cases}
\non\\
T^L_{ij; i_1\ldots i_L} (\vec r) &=
\begin{cases}
g_{ij}(\vec r) \, , & L=0 \, , \\
g_{ij}(\vec r)  u_{i_1}(\vec r) \cdots u_{i_L}(\vec r) \, , & L\geq 1 \, .
\end{cases}
}
such that
\eq{
 \delta^s_{\rm s}(\vec k)  &= \sum_{L=0}^\infty \frac{(i k_{\hat n} )^L}{L!} \hat N_{i_1\ldots i_L}T^L_{i_1\ldots i_L} (\vec k) \, , \non\\
 g^s_{ij}(\vec k)  &=  \sum_{L=0}^\infty \frac{(i k_{\hat n} )^L}{L!}  \hat N_{i_1\ldots i_L} T^L_{ij;\, i_1\ldots i_L} (\vec k) \, ,
}
where we introduced the projection tensor $\hat N_{i_1 \ldots i_L} = \hat n_{i_1} \ldots \hat n_{i_L}$. The scalar and tensor correlator we are interested in computing for redshift-space galaxy shapes can therefore be re-expressed in terms of the cross-spectra of these velocity moments
\eq{
\la \delta^s_{\rm s}(\vec k)  \delta^s_{\rm s}(\vec k') \ra  &=  
 \sum_{L L'} (-1)^{L'} \frac{( i k_{\hat n} )^{L+L'}}{L! L'!} \hat N_{i_1\ldots i_L} \hat N_{j_1 \ldots j_{L'}}
 \la T^L_{i_1\ldots i_L} (\vec k) T^{L'}_{j_1\ldots j_{L'}} (\vec k') \ra \, , \non\\
\la \delta^s_{\rm s}(\vec k)   g^s_{ij}(\vec k') \ra  &=  
 \sum_{L L'} (-1)^{L'} \frac{( i k_{\hat n} )^{L+L'}}{L! L'!} \hat N_{i_1\ldots i_L} \hat N_{j_1 \ldots j_{L'}}
 \la T^L_{i_1\ldots i_L} (\vec k) T^{L'}_{ij;\, j_1\ldots j_{L'}} (\vec k') \ra \, ,  \non\\
 \la g^s_{ij} (\vec k)   g^s_{lm}(\vec k') \ra  &=  
 \sum_{L L'} (-1)^{L'} \frac{( i k_{\hat n} )^{L+L'}}{L! L'!} \hat N_{i_1\ldots i_L} \hat N_{j_1 \ldots j_{L'}}
 \la T^L_{ij;\, i_1\ldots i_L} (\vec k) T^{L'}_{lm ;\, j_1\ldots j_{L'}} (\vec k') \ra \, . 
}
More generally, denoting either the scalar field $\delta_{\rm s}$ or the tensor field $g_{ij}$ by the
abstract label $a$, and similarly for $b$, we can write
\eq{
\la \delta^s_a(\vec k)  \delta^s_b(\vec k') \ra'
&= \sum_{L L'} (-1)^{L'} \frac{( i k_{\hat n} )^{L+L'}}{L! L'!} \hat N_{i_1\ldots i_L} \hat N_{j_1 \ldots j_{L'}}
 \la T^L_{a;\, i_1\ldots i_L} (\vec k) T^{L'}_{b;\, j_1\ldots j_{L'}} (\vec k') \ra'  \non\\
&= \sum_{L} \frac{(- i k_{\hat n} )^L}{L!} \hat N_{i_1\ldots i_{L}} \, \Xi^L_{ab;\, i_1\ldots i_L}(\vec k) \, ,
}
where, in analogy with the scalar case \cite{Vlah19}, we introduced 
\eeq{
\Xi^L_{ab;\, i_1\ldots i_L}(\vec k)
=
\sum_{\ell=0}^L (-1)^\ell \binom{L}{\ell}
 \la T^{\ell}_{a;\, i_1\ldots i_{\ell}} (\vec k) T^{L-\ell}_{b;\, i_{\ell+1}\ldots i_L} (\vec k') \ra' \, .
 \label{eq:Xi_L}
}
For brevity, we have used an apostrophe ($'$) to denote expectation values with the momentum-conserving Dirac delta removed. 
The combinatorial factors in the sum reflect that we can equivalently express Eq.~\eqref{eq:Xi_L} in terms of pairwise velocities $\Delta \vec u = \vec u(\vec r_2) - \vec u(\vec r_1)$,
\begin{equation}
    \Xi^L_{ab;\, i_1\ldots i_L}(\vec k) =
 \la F_a(\vec r_1) F_b(\vec r_2) \Delta \vec u_{i_1} \ldots \Delta \vec u_{i_L} \ra(\vec k)\, .
\end{equation}
and the redshift-space power spectrum of the two fields as
\begin{equation}
    \la \delta^s_a(\vec k)  \delta^s_b(\vec k') \ra' = \int d^3 \vec r \ e^{i\vec k \cdot \vec r} \la e^{i k_{\hat n} \Delta u_{\hat n}}\ F_a(\vec r_1) F_b(\vec r_2) \ra\, ,
\end{equation}
where 
\eeq{
F_a(\vec r) \equiv
\begin{cases}
1 + \delta_{\rm s}(\vec r) \, , & {\rm for~scalars} \, , \\
g_{ij}(\vec r) \, , & {\rm for~tensors} \, . 
\end{cases}
}
These expressions make manifest that redshift-space correlators preserve the Galilean invariance of their real-space counterparts.

The above overview is the direct generalization of the familiar scalar velocity-moment expansion \cite{Seljak:2011,Vlah:2012,Vlah19} to the case of the density-weighted trace-free tensor field. In practice, for two-point statistics, the sums are truncated at a finite $L$ depending on the perturbative order of interest, exactly as in the scalar redshift space EFT treatment \cite{Chen2020}. The corresponding short-scale contributions are
then absorbed into counterterms and stochastic contributions for the redshift-space moments, in complete analogy with the real-space EFT expansions for the size and shape fields
\cite{Vlah2020,Bakx2023}.

Finally, in order to connect to observations, one has to project the redshift-space tensor field onto the sky. Since the sky projector is invariant under the redshift-space map in the distant-observer limit \cite{Vlah2021}, the projected helicity fields can be obtained by applying the same projection operators as in real space (now to the redshift-space field $g^s_{ij}$). This provides the natural starting point for computing redshift-space corrections to the projected $E/B$-mode statistics and to spectroscopic shape-density observables, which we do in Sec.~\ref{sec:projections}.

%======================================================%
\section{Symmetry constraints and the selection rules}
\label{sec:symm}

In the previous section we showed that redshift-space two-point functions of the scalar size field and the trace-free tensor shape field can be written in terms of real-space velocity-moment correlators given in Eq.~\eqref{eq:Xi_L}.
There the labels $a,b$ denote either a scalar or a tensor field, and the remaining free spatial indices are inherited from the tensor nature of the fields and from the velocity moments.  Before choosing explicit redshift-space bases, it is useful to record the general symmetry constraints obeyed by such two-point correlators.  These constraints are purely kinematic: they follow from homogeneity, isotropy, exchange symmetry, and parity, while the dynamics enter only through scalar coefficient functions.

The discussion below is written in configuration space, where the role of exchange and parity is most transparent.  The same statements apply in Fourier space after the replacement $\hat r_i\to \hat k_i$, and they are related to the redshift-space correlators we are interested in this work after the contraction of the velocity-moment indices with the line-of-sight tensor $\hat N_{i_1\ldots i_L}=\hat n_{i_1}\cdots \hat n_{i_L}$.

%------------------------------------------------------%
\subsection{Correlators and isotropic tensor bases}

Let $A_\alpha$ and $B_\beta$ be two classical random fields, where the multi-indices $\alpha$ and $\beta$ denote any set of spatial indices.  We define the two-point correlator
\eeq{
C^{AB}_{\alpha\beta}(\vec x,\vec y)
\equiv
\la A_\alpha(\vec x) B_\beta(\vec y)\ra \, .
}
The fields are classical commuting variables, and hence
\eeq{
\la A_\alpha(\vec x)B_\beta(\vec y)\ra
=
\la B_\beta(\vec y)A_\alpha(\vec x)\ra \, .
}
This statement should be distinguished from the reality condition.  For real fields the correlator is real, while for complex fields complex conjugation relates the correlator to one involving the conjugated fields.  Reality alone does not impose an exchange relation between the free indices or the spatial arguments.

Statistical homogeneity implies that the correlator depends only on the separation
\eeq{
\vec r\equiv \vec y-\vec x\, ,
\qquad
C^{AB}_{\alpha\beta}(\vec x,\vec y)=C^{AB}_{\alpha\beta}(\vec r)\, .
}
Combining homogeneity with commutativity gives the basic exchange relation
\eeq{
C^{AB}_{\alpha\beta}(\vec r)
=
C^{BA}_{\beta\alpha}(-\vec r)\, .
\label{eq:commu}
}
For genuine cross-correlations this relates the $AB$ and $BA$ correlators.  For autocorrelations it becomes an additional constraint on a single correlator.

Statistical isotropy restricts the angular dependence of the correlator.  A generic isotropic tensor correlator can be expanded as
\eeq{
C^{AB}_{\alpha\beta}(\vec r)
=
\sum_n c^{AB}_n(r)\,
\mathcal S^{(n)}_{\alpha\beta}(\hat r)\, ,
\label{eq:expansion}
}
given rotational invariance about $\hat{r}$, where the $\mathcal S^{(n)}_{\alpha\beta}$ are isotropic tensor structures built from $\delta_{ij}$, $\hat r_i$, and, if parity-odd structures are allowed, one Levi--Civita tensor $\varepsilon_{ijk}$.  The scalar functions $c^{AB}_n(r)$ are unconstrained by isotropy alone.  The finite set of allowed structures is fixed entirely by the free indices and by their symmetry or trace constraints.

Under parity the fields transform as
\eeq{
A_\alpha(\vec x)\to (\Pi^A)_\alpha{}^{\alpha'} A_{\alpha'}(-\vec x)\, ,
\qquad
B_\beta(\vec y)\to (\Pi^B)_\beta{}^{\beta'} B_{\beta'}(-\vec y)\, ,
}
where $\Pi^A$ and $\Pi^B$ encode the intrinsic parity properties of the fields.  In the simple case in which each field carries a definite intrinsic parity, we write this as $\Pi^A=\eta_A$ and $\Pi^B=\eta_B$, with $\eta_A,\eta_B=\pm1$.  Parity invariance of the ensemble then implies
\eeq{
C^{AB}_{\alpha\beta}(\vec r)
=
(\Pi^A)_\alpha{}^{\alpha'}(\Pi^B)_\beta{}^{\beta'}
C^{AB}_{\alpha'\beta'}(-\vec r)\, ,
\label{eq:parity_invariance_II}
}
and this relation should be imposed in addition to the exchange relation in Eq.~\eqref{eq:commu}.

%------------------------------------------------------%
\subsection{Exchange and parity selection rules}
\label{sec:exchange_symm}

The constraints become especially transparent if the isotropic basis is chosen to diagonalize the exchange map
\eeq{
X:\ (\alpha,\beta,\hat r)\mapsto (\beta,\alpha,-\hat r)\, .
}
Since $X^2=1$, each basis tensor can be chosen to have a definite exchange eigenvalue,
\eeq{
\lb X\mathcal S\rb^{(n)}_{\alpha\beta}(\hat r)
\equiv
\mathcal S^{(n)}_{\beta\alpha}(-\hat r)
=
\sigma_n\,\mathcal S^{(n)}_{\alpha\beta}(\hat r)\, ,
\qquad
\sigma_n=\pm1\, .
\label{eq:exchange_eigenbasis}
}
Substituting this expansion into Eq.~\eqref{eq:commu} gives
\eeq{
c^{AB}_n(r)=\sigma_n c^{BA}_n(r)\, .
\label{eq:exchange_coeff_rule}
}
Thus exchange-even structures have coefficients symmetric under $A\leftrightarrow B$, while exchange-odd structures have coefficients antisymmetric under $A\leftrightarrow B$.  In particular, for an autocorrelation $A=B$,
\eeq{
\sigma_n=-1 \quad \Rightarrow \quad c^{AA}_n(r)=0\, .
\label{eq:exchange_auto_rule}
}
Exchange-odd structures are therefore specific to genuine cross-correlations.

Similarly, we can assign a geometric parity eigenvalue to each basis tensor,
\eeq{
\mathcal P\mathcal S^{(n)}_{\alpha\beta}(\hat r)
=
\rho_n\,\mathcal S^{(n)}_{\alpha\beta}(\hat r)\, ,
\qquad
\rho_n=\pm1\, .
}
For fields with definite intrinsic parities $\eta_A$ and $\eta_B$, parity invariance imposes the selection rule
\eeq{
\eta_A\eta_B\rho_n=-1
\quad \Rightarrow \quad
c^{AB}_n(r)=0\, .
\label{eq:parity_selection_rule}
}
Equivalently, only structures with total parity $\pi_n\equiv\eta_A\eta_B\rho_n=+1$ survive in a parity-invariant ensemble, and the two labels $(\sigma_n,\pi_n)$ are independent.  Exchange invariance determines how $AB$ and $BA$ are related, while parity invariance determines whether a given angular structure is allowed at all.

A simple example is the correlator of two polar vector fields.  Homogeneity and isotropy give
\eeq{
C^{AB}_{ij}(\vec r)
=
c^{AB}_\delta(r)\,\delta_{ij}
+
c^{AB}_r(r)\,\hat r_i\hat r_j
+
c^{AB}_\varepsilon(r)\,\varepsilon_{ijk}\hat r_k\, .
\label{eq:vector-vector_corr}
}
All three basis tensors are exchange-even: under $i\leftrightarrow j$ together with $\hat r\to-\hat r$, the first two structures are manifestly invariant, and the two minus signs in $\varepsilon_{jik}(-\hat r_k)$ cancel.  Hence all three are allowed by exchange symmetry even in an autocorrelation.  However, the last term is geometrically parity-odd, while the first two are parity-even.  Therefore, for two polar vectors in a parity-invariant ensemble, $c^{AB}_\varepsilon(r)$ vanishes.  If instead one of the two fields is a pseudovector, the same bookkeeping would allow the $\varepsilon_{ijk}\hat r_k$ structure and forbid the two parity-even structures.

For correlations of two symmetric rank-two tensor fields, $\la A_{ij}(\vec x)B_{lm}(\vec y)\ra$, the same logic applies, but the basis is larger.  The parity-even sector is spanned by rank-four tensors constructed from $\delta_{ij}$ and $\hat r_i$, while the parity-odd sector contains one Levi--Civita tensor.  These structures can be further projected onto trace and trace-free pieces, and then organized according to their exchange eigenvalue $\sigma$. In Section \ref{sec:IA_correlators} we will also re-derive the explicit rank-four basis, which is equivalent to the tensor decomposition used in the EFTofIA treatment of intrinsic-shape correlators \cite{Vlah2020}. The direct constructive algorithm relevant for the redshift-space moments is given in Appendix~\ref{app:direct_tensor}.  The only fact we will use below is that parity-invariant tensor autocorrelations keep only the $(\sigma,\pi)=(+,+)$ sector, while tensor cross-correlations can also contain exchange-odd structures with $\sigma=-1$.

%------------------------------------------------------%
\subsection{Irreducible tensors and Cartesian tensor bases}
\label{sec:irreducible_cartesian_basis}

The selection rules above determine which tensor structures are allowed once a basis has been chosen.  It remains to specify how such a basis can be constructed.  One systematic way of doing this is to organise all free indices into irreducible representations of $SO(3)$, equivalently into symmetric trace-free (STF) tensors or spherical tensors.  Irreducible tensor products can then be constructed using Clebsch--Gordan coefficients,
\eeq{
\lb A \otimes B \rb_{L M}=\sum_{m_1,m_2} C^{LM}_{\ell_1 m_1\,\ell_2 m_2}\, A^{(\ell_1)}_{m_1}B^{(\ell_2)}_{m_2} \, ,
}
where
\eeq{
L\in\{|\ell_1-\ell_2|,\ldots,\ell_1+\ell_2\}\, .
}
This is the representation-theory origin of the tensor bases used in isotropic correlators.
In this representation the angular dependence is absorbed into the spherical tensor basis (as introduced in \cite{Vlah2020}), and the correlator is then described by scalar coefficient functions multiplying fixed tensor structures.  This is essentially the viewpoint we adopt throughout the paper, even though our tensorial basis is constructed directly from $\hat r_i$, together with $\delta_{ij}$ (and $\varepsilon_{ijk}$ for parity-odd structures), which we call a direct Cartesian form. The latter is often simpler for practical construction of the same basis, and we present this explicitly in Appendix~\ref{app:direct_tensor}. For example, consider a parity-even isotropic tensor function $\xi_{i_1\cdots i_n}(\vec r)$ of rank $n$. A spanning set is then obtained by choosing $p=0,\ldots,\lfloor n/2\rfloor$, inserting $p$ Kronecker deltas, and filling the remaining indices with factors of $\hat r_i$.  Symbolically,
\eeq{
\xi_{i_1\cdots i_n}(\vec r)
=
\sum_{p=0}^{\lfloor n/2\rfloor}
\sum_{\alpha}
f_{p,\alpha}(r)\,
\Big(\delta\cdots\delta\Big)
\Big(\hat r\cdots\hat r\Big)_{i_1\cdots i_n,\alpha}\, ,
\label{eq:cartesian_rank_n_basis}
}
where $\alpha$ labels inequivalent contractions modulo the imposed index symmetries.  Trace constraints, such as those appropriate for shape tensors, are then imposed by projecting onto the STF part.  Parity-odd structures can be obtained analogously by adding a sector that contains a single Levi--Civita tensor $\varepsilon_{ijk}$, together with the required factors of $\delta_{ij}$ and $\hat r_i$.

As we stated above, this Cartesian recipe is equivalent to the irreducible-tensor construction, but we find it more convenient for our redshift-space application.  The explicit rank-four decompositions used for intrinsic-shape correlators have been discussed in the EFTofIA literature, where the spherical-tensor basis gives a particularly transparent helicity organisation \cite{Vlah2020,Vlah2021}.  In Appendix~\ref{app:direct_tensor} we give the direct Cartesian implementation needed for the redshift-space moment correlators used in Sec.~\ref{sec:IA_correlators}.

%------------------------------------------------------%
\subsection{Consequences for redshift-space tensor correlators}

We can now return to the redshift-space moments of Sec.~\ref{sec:rsd}. Before the line-of-sight contractions, the correlators $\Xi^L_{ab;\,i_1\ldots i_L}(\vec k)$ have the free indices associated with the two fields, together with $L$ symmetric velocity-moment indices.  Homogeneity fixes the momentum-conserving delta function, while isotropy fixes the allowed tensor structures in terms of $\hat k_i$, $\delta_{ij}$, and, if parity-odd terms are retained, $\varepsilon_{ijk}$.  After contraction with $\hat N_{i_1\ldots i_L}$, the correlators become axisymmetric functions of the two directions $\hat k_i$ and $\hat n_i$.  The only scalar angular dependence is $\mu\equiv \hat k\cdot \hat n$. At fixed velocity-moment order $L$, the contraction with $\hat N_{i_1\ldots i_L}$ produces a finite set of tensors built from $\hat k_i$, $\hat n_i$, $\delta_{ij}$, and possibly $\varepsilon_{ijk}$, with form factors that are finite polynomials in $\mu$ multiplying scalar functions of $k$.

This observation is the bridge to the explicit construction in the next section.  In the scalar--scalar sector, the result is simply a polynomial expansion in $\mu$.  In the scalar--tensor sector, the remaining free indices form a symmetric trace-free rank-two tensor, so the answer can be expanded in a five-dimensional STF tensor basis.  In the tensor--tensor sector, the remaining object is symmetric and trace-free in each index pair, and the corresponding rank-four basis must also respect exchange symmetry when the two fields are identical.  In Sec.~\ref{sec:IA_correlators} we implement this classification using bases adapted to redshift space: the kinematic basis $\mathcal M$, the parity-adapted real basis $\mathcal Q$, and the helicity basis $\mathcal Y$.

%======================================================%
\section{Galaxy Shapes Correlators in Redshift Space}
\label{sec:IA_correlators}

%------------------------------------------------------%
\subsection{Geometry and the choice of the basis}
\label{subsec:IA_rsd_geometry}

The redshift-space shape problem is governed by two preferred directions in Fourier space: the wavevector direction $\hat k_i$ and the line-of-sight direction $\hat n_i$. All angular dependence is therefore encoded in the single cosine $\mu \equiv \hat k \cdot \hat n$.  Once the velocity moments are contracted with $\hat N^{(L)}_{i_1\cdots i_L}$ the resulting correlators depend only on $k$ and $\mu$. Our goal in this section is to organise the redshift-space correlators of size and shape fields into a basis of STF tensors built from $\hat k_i$, $\hat n_i$, and $\delta_{ij}$. We will consider three cases: scalar--scalar, scalar--tensor, and tensor--tensor. The scalar--scalar case is completely described by polynomials in $\mu$, while the scalar--tensor and tensor--tensor cases require a basis of STF rank-two and rank-four tensors, respectively.

A convenient form of our result should have several properties. First, the basis should be \emph{complete}, so that any axisymmetric correlator can be expanded in it. Second, it should make the \emph{parity properties} manifest. Since in this section we restrict attention to the parity-even sector, only parity-even combinations of the basis tensors will contribute. Third, for tensor--tensor auto-correlations the basis should allow for the \emph{exchange symmetry} between the two tensor fields. Fourth, it is desirable to work with an \emph{orthogonal} basis, since this separates independent angular structures.
Finally, for practical applications, it is useful that the corresponding form factors remain \emph{polynomial in $\mu$}, so that the multipole expansion is finite and manifest.

These requirements are not all optimised on the same basis. We therefore proceed in stages. We begin with the natural \emph{kinematic} basis $\mathcal M$, constructed directly from $\hat k_i$ and $\hat n_i$. This basis makes completeness and the general tensor structure transparent, but it is not orthogonal. We then pass to a \emph{parity-adapted} orthogonal basis $\mathcal Q$, which describes the scalar--tensor sector with the desired properties, and  makes manifest the even--even/odd--odd decomposition of the tensor--tensor sector. Finally, we introduce the \emph{helicity} basis $\mathcal Y$ for the tensor---tensor sector, which satisfies all the above properties. Finally, we show that both $\mathcal Q$ and $\mathcal Y$ in the corresponding sectors are examples of \emph{total-helicity bases}.

\subsection{Rank-2 Symmetric Trace-Free Shape Bases}
\label{ssec:field_basis}

We begin in this section by constructing general rank-2, symmetric and trace-free (STF) tensor bases to describe galaxy shape fields. Our first basis, $\mathcal M$, is, by construction, kinematic and polynomial. This basis is the most direct one can write down from $\hat k_i$, $\hat n_i$, $\delta_{ij}^{\rm K}$ and $\varepsilon_{ijk}$. In the scalar--tensor sector, it is given by
\eq{
[\mathcal M_{kk}]_{ij} &= \hat k_i\hat k_j - \frac13\delta_{ij}\, , \non\\
[\mathcal M_{kn}]_{ij} &= \frac12\lb \hat k_i\hat n_j + \hat n_i\hat k_j \rb - \frac{\mu}{3}\delta_{ij}\, , \non\\
[\mathcal M_{nn}]_{ij} &= \hat n_i\hat n_j - \frac13\delta_{ij}\, , \non\\
[\mathcal M_{\varepsilon k}]_{ij} &= \frac{1}{2}\lb \varepsilon_{iab} \hat k_j + \varepsilon_{jab} \hat k_i  \rb  \hat k_a \hat n_b \, , \non\\
[\mathcal M_{\varepsilon n}]_{ij} &= \frac{1}{2} \lb \varepsilon_{iab} \hat n_j + \varepsilon_{jab} \hat n_i  \rb  \hat k_a \hat n_b  \, ,
\label{eq:M_basis}
}
where $\mu = \hk \cdot \hn$. These span the 5-dimensional space of symmetric trace-free rank-2 tensors built from $\hat k$ and $\hat n$. The main advantage of this basis is its direct dependence on the two distinguished directions of the problem, and completeness is explicit from construction. However, the $\mathcal M$ basis is not orthogonal, and its elements do not carry definite helicity. With respect to the geometric parity $\mathcal {P} $, the basis elements are eigenstates, and it is clear that the first three tensors are even, while the last two are odd. Thus, it is the natural basis for writing the most general redshift-space tensor, but not optimal for organising the independent sectors.

Our second basis is real, orthogonal, and parity-adapted, and we label it $\mathcal Q$. Using the two preferred directions in redshift space, the wavevector $\hat k_i$ and the line-of-sight direction $\hat n_i$, or equivalently from $\hat k_i$ and the two vectors transverse to it, we can build the 3D basis
\eeq{
m_i \equiv \hat n_i - \mu \hat k_i \, , \qquad
t_i \equiv \varepsilon_{iab}\hat k_a \hat n_b \, ,
}
where  $\mu$ is defined as before and $s \equiv \sqrt{1-\mu^2}$. These vectors satisfy
\eeq{
\hat k\cdot \vec m = 0\, , \qquad
\hat k\cdot \vec t = 0\, , \qquad
\vec m\cdot \vec t = 0\, ,
}
and their norms are $\vec m\cdot \vec m = \vec t\cdot \vec t = s^2 = 1- \mu^2$. The basis is obtained by separating the directions longitudinal and transverse to $\hat k$. This leads to the real STF tensors
\eq{
[\mathcal Q_0]_{ij}
&\equiv
N_0 \left(\hat k_i\hat k_j-\frac13\delta_{ij} \right)\, , \non\\
[\mathcal Q_1]_{ij}
&\equiv
\frac{N_1}{2}\lb \hat k_i m_j + m_i \hat k_j \rb \, , \non\\
[\mathcal Q_2]_{ij}
&\equiv
\frac{N_2}{2}\lb m_i m_j - t_i t_j \rb \, , \non\\
[\mathcal Q_3]_{ij}
&\equiv
\frac{N_3}{2}\lb \hat k_i t_j + t_i \hat k_j \rb \, , \non\\
[\mathcal Q_4]_{ij}
&\equiv
\frac{N_4}{2}\lb m_i t_j + t_i m_j \rb \, ,
\label{eq:Q_basis}
}
where we have inserted normalizing factors $N_0 = \sqrt{3/2}$ and $N_{1,2,3,4} = \sqrt{2}$.
Each basis tensor carries a definite number of directions transverse to $\hat k$, i.e. $\mathcal Q_0$ is purely longitudinal, $\mathcal Q_1$ and $\mathcal Q_3$ carry one transverse direction, while $\mathcal Q_2$ and $\mathcal Q_4$ carry two. The basis elements are mutually orthogonal, and related to the original $\mathcal M$ basis by polynomial transformations in $\mu$. Since $t_i=\varepsilon_{iab}\hat k_a\hat n_b$ is an axial vector we have
\eeq{
\mathcal Q_0,\mathcal Q_1,\mathcal Q_2
\quad \text{are parity-even,}
\qquad
\mathcal Q_3,\mathcal Q_4
\quad \text{are parity-odd.}
}
Note that $\mathcal Q_3$ and $\mathcal Q_4$ span the same parity-odd subspace as $\mathcal M_{\varepsilon k}$ and $\mathcal M_{\varepsilon n}$. Since it is also orthogonal, the $\mathcal Q$ basis is therefore a useful \emph{real} basis for organizing physics. We have defined the $\mathcal{Q}$ basis such that each element has norm
\eeq{
\mathcal Q_0\cdot \mathcal Q_0
=
1\, , \qquad
\mathcal Q_1\cdot \mathcal Q_1
=
(1-\mu^2)\, , \qquad
\mathcal Q_2\cdot \mathcal Q_2
=
(1-\mu^2)^2\, .
}
equal to unity when $\mu = 0$. We can further define the angularly normalized basis
\eq{
\bar{\mathcal Q}_0 \equiv \mathcal Q_0 \, ,~~~
\bar{\mathcal Q}_{1,3} \equiv \frac{1}{s}\,\mathcal Q_{1,3} \, , ~~~
\bar{\mathcal Q}_{2,4} \equiv \frac{1}{s^2}\,\mathcal Q_{2,4} \, ,
\label{eq:normal_Q}
}
such that $\bar{\mathcal Q}_a\cdot \bar{\mathcal Q}_b = \delta_{ab}$. This, of course, also preserves all the nice properties of the non-normalised basis, except that the transformation to the $\mathcal{M}$ basis is no longer polynomial.

The third basis is the complex helicity basis $Y^{(m)}$, constructed from vectors $\hat k$ and $e_\pm = \mp \frac{1}{\sqrt2}(e_1 \mp i e_2)$, where
\eeq{
e_1 = \frac{\hat k\times \hat n}{|\hat k\times \hat n|} = \frac{t}{s}\, , \qquad
e_2 = \hat k\times e_1 = -\,\frac{m}{s}\, .
\label{eq:def_e1e2}
}
The corresponding spherical-tensor basis is then
\eq{
Y^{(0)}_{ij}
&=
\sqrt{\frac32}\lb \hat k_i\hat k_j-\frac13\delta_{ij}\rb\, , \non\\
Y^{(\pm1)}_{ij}
&=
\frac{1}{\sqrt2}\lb \hat k_i e^\pm_j + \hat k_j e^\pm_i \rb\, , \non\\
Y^{(\pm2)}_{ij}
&=
e^\pm_i e^\pm_j \, .
\label{eq:def_helicity}
}
This basis has definite helicity with respect to rotations around $\hat k$, and $Y^{(m)}_{ij}$ thus has helicity $m=0,\pm1,\pm2$. In other words, the $\mathcal {Y} $ basis diagonalises the residual ${\rm SO}(2)$ of rotations about the wavevector. It is therefore the natural basis for matching to the spherical-tensor and helicity language used in the intrinsic-alignment literature, where statistical isotropy constrains correlators most directly in helicity space \cite{Vlah2020}.

The relation to the $\mathcal Q$ basis is simple, $\mathcal Q$ is just the real, non-normalized version of the helicity basis. More precisely, $\mathcal Q_0$ corresponds to helicity zero, while
\eq{
\mathcal Q_1,\mathcal Q_3 &\propto Y^{(+1)} \pm Y^{(-1)} \, , \non\\
\mathcal Q_2,\mathcal Q_4 &\propto Y^{(+2)} \pm Y^{(-2)} \, .
}
Thus $\mathcal Q_1$ and $\mathcal Q_2$ are the parity-even real combinations of the $\pm1$ and $\pm2$ helicity states, and $\mathcal Q_3$ and $\mathcal Q_4$ are the parity-odd real combinations of the $\pm1$ and $\pm2$ helicity states. So the $\mathcal Q$ basis is the parity-adapted real form of the complex helicity basis, while conversely the helicity basis is not the eigenstates of the parity operator.

%------------------------------------------------------%
\subsection{Scalar-tensor sector}
\label{subsec:scalar-tensor}

We now consider the redshift-space correlator of one scalar and one trace-free tensor field. After contracting the velocity moments with $\hat N^{(L)}_{i_1\cdots i_L}$, the remaining correlator is a rank-two STF tensor depending only on $k$ and $\mu$. Since we restrict our attention here to the parity-even sector, only the three parity-even rank-two tensors contribute. The parity-odd structures $\mathcal M_{\varepsilon k}$ and $\mathcal M_{\varepsilon n}$ introduced above are therefore absent.

In the original kinematic basis $\mathcal M$, the most general scalar--tensor correlator is
\eeq{
\lb \hat N^{(L)} \cdot \Xi^L_{ab}(\vec k) \rb_{ij}
=
F_{kk}(k,\mu)\,[\mathcal M_{kk}]_{ij}
+
F_{kn}(k,\mu)\,[\mathcal M_{kn}]_{ij}
+
F_{nn}(k,\mu)\,[\mathcal M_{nn}]_{ij}\, ,
\label{eq:Mbasis_scalar_tenros}
}
with $\mathcal M$ terms given in Eq.~\eqref{eq:M_basis}. Moreover, the three associated form factors can be expanded as polynomials in $\mu$,
\eq{
F_{kk}(k,\mu)
&=
\sum_{m=0}^{\lfloor L/2\rfloor}
P^{(L,m)}_{kk}(k)\,\mu^{L-2m}\, , \non\\
F_{kn}(k,\mu)
&=
\sum_{m=0}^{\lfloor (L-1)/2\rfloor}
P^{(L,m)}_{kn}(k)\,\mu^{L-2m-1}\, , \non\\
F_{nn}(k,\mu)
&=
\sum_{m=0}^{\lfloor (L-2)/2\rfloor}
P^{(L,m)}_{nn}(k)\,\mu^{L-2m-2}\, .
\label{eq:FF_M_basis}
}
The exact angular dependence of these form factors follows from the direct tensorial correlator construction and subsequent contraction with the line of sight, described in App.~\ref{app:direct_tensor}. The pattern of powers is fixed by counting explicit factors of $\hat n$ already carried by the basis tensors: $\mathcal M_{kk}$ contains none, $\mathcal M_{kn}$ contains one, and $\mathcal M_{nn}$ contains two. Thus the corresponding form factors must supply $L$, $L-1$, and $L-2$ powers of $\hat n\cdot \hat k=\mu$, respectively. As we have noted earlier, the basis $\{\mathcal M_{kk},\mathcal M_{kn},\mathcal M_{nn}\}$ is complete, but not orthogonal. Thus the $\mathcal M$ basis is useful for writing the most general tensor structure and for making polynomiality manifest, but not for separating independent angular sectors.  This basis plays a similar role for tensor correlators that the monomial basis $\{1,\mu,\mu^2,\ldots\}$ plays for scalar correlators. 

A more transparent organization is obtained by passing to the orthogonal $\mathcal Q$ parity-even basis given in Eq.~\eqref{eq:Q_basis}. The mapping between the two basis is simple and reads
\eq{
N_0^{-1} [\mathcal Q_0]_{ij}
&\equiv
[\mathcal M_{kk}]_{ij}\, , \non\\
N_1^{-1} [\mathcal Q_1]_{ij}
&\equiv
[\mathcal M_{kn}]_{ij}-\mu[\mathcal M_{kk}]_{ij}\, , \non\\
N_2^{-1} [\mathcal Q_2]_{ij}
&\equiv
[\mathcal M_{nn}]_{ij}
-2\mu[\mathcal M_{kn}]_{ij}
+\frac{1+\mu^2}{2}[\mathcal M_{kk}]_{ij}\, .
}
This also makes the physical interpretation clearer: $\mathcal Q_0$ is purely longitudinal, $\mathcal Q_1$ carries one direction transverse to $\hat k$, and $\mathcal Q_2$ carries two. Thus the $\mathcal Q$ basis separates the scalar--tensor sector according to transverse content.

In the $\mathcal{Q}$ basis the correlator becomes
\eeq{
\lb \hat N^{(L)} \cdot \Xi^L_{ab}(\vec k) \rb_{ij}
=
G_0(k,\mu)\,[\mathcal Q_0]_{ij}
+
G_1(k,\mu)\,[\mathcal Q_1]_{ij}
+
G_2(k,\mu)\,[\mathcal Q_2]_{ij}\, ,
}
with
\eq{
N_0 G_0(k,\mu)
&=
F_{kk}(k,\mu)
+\mu F_{kn}(k,\mu)
+\frac{3\mu^2-1}{2}F_{nn}(k,\mu)\, , \non\\
N_1 G_1(k,\mu)
&=
F_{kn}(k,\mu)+2\mu F_{nn}(k,\mu)\, , \non\\
N_2 G_2(k,\mu)
&=
F_{nn}(k,\mu)\, .
}
Since the mapping between the two bases is polynomial in $\mu$, the new form factors remain polynomial:
\eq{
G_0(k,\mu)
&=
\sum_{m=0}^{\lfloor L/2\rfloor}
P^{(L,m)}_{0}(k)\,\mu^{L-2m}\, , \non\\
G_1(k,\mu)
&=
\sum_{m=0}^{\lfloor (L-1)/2\rfloor}
P^{(L,m)}_{1}(k)\,\mu^{L-2m-1}\, , \non\\
G_2(k,\mu)
&=
\sum_{m=0}^{\lfloor (L-2)/2\rfloor}
P^{(L,m)}_{2}(k)\,\mu^{L-2m-2}\, .
}
Thus, in the scalar--tensor sector, the $\mathcal Q$ basis preserves the polynomial structure of the $\mathcal M$ basis while rendering the angular sectors orthogonal. We also see that in case when $\hat k$ and $\hat n$ are collinear ($\mu=1$) only the $Q_0$ contributions survives, which agrees with the real space case \cite{Vlah2020}. In the $\mathcal M$ basis, on the other hand, the same limit manifests as a degeneracy of the three basis elements in the collinear limit.

Finally, it is useful to rewrite the scalar--tensor sector in the complex spherical-tensor basis $Y^{(m)}$, introduced above. Since we are in the parity-even sector, only the parity-even helicity combinations appear. Using
\eeq{
    \mathcal Q_0 = Y^{(0)}\,, \quad \mathcal Q_1 = \frac{i s}{\sqrt{2}} (Y^{(1)} + Y^{(-1)})\,, \quad \mathcal Q_2 = \frac{-s^2}{\sqrt{2}} (Y^{(2)} + Y^{(-2)})\,,
    \label{eq:Q_Y_relation}
}
the same correlator can then be written using the helicity basis. 
This form makes the helicity content explicit:
\begin{itemize}
\item $Y^{(0)}$ is the helicity-$0$ component,
\item $Y^{(+1)}+Y^{(-1)}$ is the parity-even combination of the helicity $\pm1$ states,
\item $Y^{(+2)}+Y^{(-2)}$ is the parity-even combination of the helicity $\pm2$ states,
\end{itemize}
and parity-odd combinations $Y^{(+1)}-Y^{(-1)}$ and $Y^{(+2)}-Y^{(-2)}$ do not appear. The advantage of the helicity basis is that it diagonalizes the residual ${\rm SO}(2)$ of rotations about $\hat k$ and matches directly to the spherical-tensor language used in the intrinsic-alignment literature \cite{Vlah2020}. For the scalar--tensor sector, the $\mathcal Q$ basis therefore provides the most convenient practical representation, while the $Y^{(m)}$ basis gives the cleanest interpretation in terms of helicity. Indeed, we can observe that, by construction, each $\mathcal{Q}_n$ has total helicity $|M| = n$, is constructed from $n$ powers of the line of sight $\hn$, and involves a spin-weight $s^n$. We will see in Section~\ref{sec:spin_weights} that the parity-adapted basis in the scalar-tensor sector is also an example of the \textit{total-helicity} basis we will construct independently in the tensor-tensor sector in Section~\ref{subsec:tensor-tensor}.

Yet another pragmatic choice is to use the normalized basis introduced in Eq.~\eqref{eq:normal_Q}, which we can also call the \emph{normalized} total-helicity basis. As the $\mathcal{Q}_n$ are normalized up to spin weights $s^n$, the normalized total-helicity basis corresponds simply to parity-even combinations of spherical harmonic tensors at fixed $|M|$. In the normalized basis the corresponding form factors cease to be polynomial in $\mu$ because of the explicit factors of $s^{-1}$ and $s^{-2}$. Instead, these coefficients satisfy the property that
\begin{equation}
    \bar{G}_n(k,\mu) = s^n G_n(k,\mu) = \sum_\ell \bar{G}_n^\ell (k) P^n_\ell(\mu)\, ,
    \label{eq:def_bar_G}
\end{equation}
where, in the last line, we have used the fact that the associated Legendre polynomials are polynomials multiplied by $s^n$. We therefore see that having $\bar{G}_n$ spanned by associated Legendre polynomials up to $L$ is equivalent to $G_n$ being a polynomial up to $L-n$. Adopting the associated Legendre expansion for form factors in the normalized basis thus replaces the vanishing basis elements in the collinear limit with vanishing angular coefficients. We will see in Section~\ref{sec:measurements} that this basis has practical advantages at the level of estimators.

%------------------------------------------------------%
\subsection{Tensor-tensor sector}
\label{subsec:tensor-tensor}

We now turn to the auto-correlation of two STF rank-two tensor fields. After contraction with $\hat N^{(L)}=\hat n_{i_1}\cdots \hat n_{i_L}$, the result is an axisymmetric rank-four tensor, symmetric and trace-free in each pair of indices separately, $ \lb \hat N^{(L)}\cdot \Xi^L_{ab}(\vec k)\rb_{ij,lm}$. Since we restrict attention to the parity-even sector, the final decomposition contains both even-even and odd-odd products of rank-two basis tensors, while mixed even-odd products are absent.

Following the construction presented in Appendix~\ref{app:direct_tensor}, a direct kinematic basis decomposition gives
\eq{
\hat N^{(L)} \cdot \Xi^L_{ab} (\vec k)
=&
F_{kk,kk}\, \mathcal M_{kk} \mathcal M_{kk}
+ F_{\delta kk}\, \mathcal M_{\delta kk}
+ F_{\delta\delta}\, \mathcal M_{\delta\delta} \non\\
&
+ F_{kk,kn}\, \mathcal M_{kk} \mathcal M_{kn}
+ F_{kn,kk}\, \mathcal M_{kn} \mathcal M_{kk} \non\\
&
+ F_{\delta kn}\, \mathcal M_{\delta kn}
+ F_{\delta nk}\, \mathcal M_{\delta nk} \non\\
&
+ F_{kk,nn}\, \mathcal M_{kk} \mathcal M_{nn}
+ F_{nn,kk}\, \mathcal M_{nn} \mathcal M_{kk}
+ F_{kn,kn}\, \mathcal M_{kn} \mathcal M_{kn} \non\\
&
+ F_{nn,kn}\, \mathcal M_{nn} \mathcal M_{kn}
+ F_{kn,nn}\, \mathcal M_{kn} \mathcal M_{nn} \non\\
&
+ F_{nn,nn}\, \mathcal M_{nn} \mathcal M_{nn}
+ F_{\delta nn}\, \mathcal M_{\delta nn} \, ,
}
where the additional basis tensors, relative to those used in the scalar-tensor case in Eq.~\eqref{eq:Mbasis_scalar_tenros}, are 
\eq{
[\mathcal M_{\delta kk}]_{ij,lm}
&=
\frac{1}{4} \Big( \hat k_m \hat k_j \delta_{il} + \hat k_m \hat k_i \delta_{jl}
+ \hat k_j \hat k_l \delta_{im} + \hat k_i \hat k_l \delta_{jm} \Big)
- \frac{1}{3}\lb \hat k_l \hat k_m \delta_{ij}  + \hat k_i \hat k_j \delta_{lm} \rb
+ \frac{1}{9} \delta_{ij} \delta_{lm} \, , \\
[\mathcal M_{\delta\delta}]_{ij,lm}
&=
\frac{1}{2}\lb \delta_{im} \delta_{jl} + \delta_{il} \delta_{jm} \rb
- \frac{1}{3} \delta_{ij}\delta_{lm} \, , \non\\
[\mathcal M_{\delta kn}]_{ij,lm}
&=
\frac{1}{4} \lb \hat n_m \hat k_j \delta_{il} + \hat n_l \hat k_j\delta_{im}
+ \hat n_m \hat k_i \delta_{jl} + \hat n_l \hat k_i \delta_{jm}\rb
- \frac{1}{6}\lb \hat n_m \hat k_l \delta_{ij} + \hat n_l \hat k_m \delta_{ij}
+ \hat n_j \hat k_i \delta_{lm} + \hat n_i \hat k_j \delta_{lm}\rb
+ \frac{1}{9} \mu\delta_{ij} \delta_{lm} \, . \non
}
The tensors $\mathcal M_{\delta nk}$ and $\mathcal M_{\delta nn}$ are obtained from $\mathcal M_{\delta kn}$ and $\mathcal M_{\delta kk}$ by the replacements $k\leftrightarrow n$ and $k\rightarrow n$, respectively. This basis is direct and polynomial, but it is not minimal. Indeed, the 14 structures above satisfy the exact identity
\eeq{
\mathcal M_{\delta kk}
+\mathcal M_{\delta nn}
-\mu \lb \mathcal M_{\delta kn}+\mathcal M_{\delta nk}\rb
+\frac{1}{2}\lb \mathcal M_{kk}\mathcal M_{nn}+\mathcal M_{nn}\mathcal M_{kk}\rb
-\mathcal M_{kn}\mathcal M_{kn}
-\frac{1}{2}(1-\mu^2)\mathcal M_{\delta\delta} = 0\, ,
}
so the original 14-term representation is overcomplete, giving us in total thirteen independent terms.

The true advantage of this basis lies in deriving the angular form of associated form factors, which are polynomial in $\mu$, as is clear from the construction shown in Appendix \ref{app:direct_tensor}. The same counting logic as in the scalar--tensor sector gives:
\eq{
\{ F_{kk,kk},\, F_{\delta kk}, \, F_{\delta \delta} \}
&=
\sum_{m=0}^{\lfloor L/2 \rfloor} P^{(L,m)}_{\alpha}(k) \mu^{L-2m} \, , \non\\
\{F_{kk,kn},\, F_{\delta kn},\,F_{\delta nk}\}
&=
\sum_{m=0}^{\lfloor (L-1)/2 \rfloor } P^{(L,m)}_{\alpha}(k) \mu^{L-2m-1} \, , \non\\
\{F_{kk,nn},\, F_{kn,kn}, \, F_{\delta nn},\,F_{nn,kk} \}
&=
\sum_{m=0}^{\lfloor (L-2)/2 \rfloor } P^{(L,m)}_{\alpha}(k) \mu^{L-2m-2} \, , \non\\
\{F_{nn,kn},\,F_{kn,nn}\}
&=
\sum_{m=0}^{\lfloor (L-3)/2 \rfloor  } P^{(L,m)}_{\alpha}(k) \mu^{L-2m-3} \, , \non\\
F_{nn,nn}
&=
\sum_{m=0}^{\lfloor (L-4)/2 \rfloor  } P^{(L,m)}_{\alpha}(k) \mu^{L-2m-4} \, .
}
For identical tracers, one further has exchange symmetry between the two tensor slots,
\eeq{
F_{kk,kn} = F_{kn,kk}\, , ~~
F_{\delta kn} = F_{\delta nk}\, , ~~
F_{kk,nn} = F_{nn,kk}\, , ~~
F_{nn,kn} = F_{kn,nn}\, ,
}
which reduces the number of independent form factors from thirteen to nine.

It is useful to interpret this basis in terms of the direct product of the rank two basis elements given in Eq.~\eqref{eq:M_basis}. It is clear that nine of these terms are given as the direct cross and auto products of the first three basis elements $\mathcal M_{kk}$, $\mathcal M_{kn}$ and $\mathcal M_{nn}$.  Since we restrict our analysis to the parity-even sector of the rank-four tensor correlator, these nine can be considered as the even-even contributions. However, the parity-even sector also receives contributions from the odd-odd correlations, i.e. cross and auto products of the $\mathcal M_{\varepsilon k}$ and $\mathcal M_{\varepsilon n}$ elements. The latter gives rise to the remaining four terms in the above basis. In other words, the parity decomposition is already present in the original $[\mathcal M]_{ij}$ language if one works instead with the rank-two basis
\eeq{
\{\mathcal M_{kk},\,\mathcal M_{kn},\,\mathcal M_{nn}\}
\qquad\text{and}\qquad
\{\mathcal M_{\varepsilon k},\,\mathcal M_{\varepsilon n}\}\, .
}
The first three tensors are parity-even, and the last two are parity-odd. Hence, the parity-even tensor--tensor sector consists precisely of even--even and odd--odd products. The 13-term decomposition above is simply a convenient repackaging of this statement in terms of explicit rank-four STF tensors. The only thing that remains to be done is express $\mathcal M_{\delta kk}$, $\mathcal M_{\delta\delta}$ and $\mathcal M_{\delta kn}$ in terms of products of $[\mathcal M]_{ij}$ basis elements in Eq.~\eqref{eq:M_basis}.  We obtain
\eq{
s^2\mathcal M_{\delta kk} &= \mathcal M_{kk} \mathcal M_{kk} - \mu \lb \mathcal M_{kk} \mathcal M_{kn} + \mathcal M_{kn} \mathcal M_{kk} \rb +  \mathcal M_{kn} \mathcal M_{kn} +  \mathcal M_{\varepsilon k} \mathcal M_{\varepsilon k}    \, , \\
s^2 \mathcal M_{\delta kn} &= \mathcal M_{kk} \mathcal M_{kn} - \mu \lb \mathcal M_{kk} \mathcal M_{nn} + \mathcal M_{kn} \mathcal M_{kn} \rb +  \mathcal M_{kn} \mathcal M_{nn} +  \mathcal M_{\varepsilon k} \mathcal M_{\varepsilon n}    \, , \non\\
s^4 \mathcal M_{\delta\delta} &=  2\, \mathcal M_{kk} \mathcal M_{kk} - 4\mu \lb \mathcal M_{kk} \mathcal M_{kn} + \mathcal M_{kn} \mathcal M_{kk} \rb + \lb 1 + \mu^2 \rb \lb \mathcal M_{kk} \mathcal M_{nn} + \mathcal M_{nn} \mathcal M_{kk} \rb + \lb 2 + 6\mu^2 \rb \mathcal M_{kn} \mathcal M_{kn} \non\\
&\quad - 4\mu \lb \mathcal M_{nn} \mathcal M_{kn} + \mathcal M_{kn} \mathcal M_{nn} \rb + 2\, \mathcal M_{nn} \mathcal M_{nn} + 2\, \mathcal M_{\varepsilon k} \mathcal M_{\varepsilon k} + 2\, \mathcal M_{\varepsilon n} \mathcal M_{\varepsilon n} - 2\mu \lb \mathcal M_{\varepsilon k} \mathcal M_{\varepsilon n} + \mathcal M_{\varepsilon n} \mathcal M_{\varepsilon k} \rb \, . \non
}
From these relations, it is clear that the associated form factors in the new basis would no longer be polynomial, since the factors of $s$ on the left-hand side would enter into the denominator of the new-basis form factors.

The orthogonal basis $\mathcal Q_a$ introduced above also makes the parity structure explicit. The first three tensors are parity-even, while $\mathcal Q_3$ and $\mathcal Q_4$ are parity-odd:
\eeq{
\mathcal Q_0,\mathcal Q_1,\mathcal Q_2
\quad \text{even,}
\qquad
\mathcal Q_3,\mathcal Q_4
\quad \text{odd.}
}
Thus, the parity-even rank-four sector splits into two sectors: even-even and odd-odd. We can thus introduce a convenient ordered basis 
\eq{
[\vec{\mathcal Q}^{\rm e}]_{nn'}
&=
\big\{
\mathcal Q_0\mathcal Q_0,\,
\mathcal Q_0\mathcal Q_1,\,
\mathcal Q_1\mathcal Q_0,\,
\mathcal Q_0\mathcal Q_2,\,
\mathcal Q_2\mathcal Q_0,\,
\mathcal Q_1\mathcal Q_1,\,
\mathcal Q_1\mathcal Q_2,\,
\mathcal Q_2\mathcal Q_1,\,
\mathcal Q_2\mathcal Q_2
\big\}_{\alpha},
\qquad n, n' = 0, 1, 2
\non\\
[\vec{\mathcal Q}^{\rm o}]_{nn'}
&=
\big\{
\mathcal Q_3\mathcal Q_3,\,
\mathcal Q_3\mathcal Q_4,\,
\mathcal Q_4\mathcal Q_3,\,
\mathcal Q_4\mathcal Q_4
\big\}_{\beta},
\qquad n, n' = 3, 4 \, .
}
In this basis, the correlator reads
\eeq{
\hat N^{(L)} \cdot \Xi^L_{ab} (\vec k)
=
\sum_{\sigma=\rm e, o} \sum_{(n,n')} G^\sigma_{nn'}(k,\mu) [\mathcal \mathcal Q^\sigma_{nn'}]_{ab}\, .
}
Since the basis tensors are mutually orthogonal, the parity-even and parity-odd rank-two subspaces remain separated. This is the main practical advantage of the $\mathcal Q$ basis. The same can, of course, be written for the normalised basis $\bar{\mathcal Q}$.  However, the basis suffers from the same issue as described above for the products of $[\mathcal M]_{ij}$, i.e. the corresponding form factors are not polynomial.  We would end up with the same issue if we had chosen the helicity basis $Y^{m}$ instead. 

Finally, we present the \emph{total-helicity basis}, which has all the desired properties: it is orthogonal, and the associated form factors remain polynomial. Moreover, it can be constructed out of products of helicity or normalised parity-adapted basis elements. We can again identify the two sectors, \emph{aligned}, $\mathcal Y^{\parallel}_{mm'}$ spanned by nine elements 
\eq{
{\mathcal Y}^{\parallel}_{00}
&=
Y^{(0)}Y^{(0)}
=
\bar{\mathcal Q}_0 \bar{\mathcal Q}_0\, , \non\\
{\mathcal Y}^{\parallel}_{01}
&=
 i N \, s\,Y^{(0)}\lb Y^{(+1)}+Y^{(-1)}\rb
=
s\,\bar{\mathcal Q}_0 \bar{\mathcal Q}_1\, , \non\\
{\mathcal Y}^{\parallel}_{10}
&=
 i N \,s\,\lb Y^{(+1)}+Y^{(-1)}\rb Y^{(0)}
=
s\,\bar{\mathcal Q}_1 \bar{\mathcal Q}_0\, , \non\\
{\mathcal Y}^{\parallel}_{02}
&=
- N \,s^2\,Y^{(0)}\lb Y^{(+2)}+Y^{(-2)}\rb
=
s^2\,\bar{\mathcal Q}_0 \bar{\mathcal Q}_2\, , \non\\
{\mathcal Y}^{\parallel}_{20}
&=
- N \,s^2\,\lb Y^{(+2)}+Y^{(-2)}\rb Y^{(0)}
=
s^2\,\bar{\mathcal Q}_2 \bar{\mathcal Q}_0\, , \non\\
{\mathcal Y}^{\parallel}_{11}
&=
- N \,s^2\lb Y^{(+1)}Y^{(+1)}+Y^{(-1)}Y^{(-1)}\rb
=
N s^2\lb \bar{\mathcal Q}_1 \bar{\mathcal Q}_1 - \bar{\mathcal Q}_3 \bar{\mathcal Q}_3 \rb\, , \non\\
{\mathcal Y}^{\parallel}_{12}
&=
- i N \,s^3\lb Y^{(+1)}Y^{(+2)}+Y^{(-1)}Y^{(-2)}\rb
=
N \,s^3\lb \bar{\mathcal Q}_1 \bar{\mathcal Q}_2 - \bar{\mathcal Q}_3 \bar{\mathcal Q}_4 \rb\, , \non\\
{\mathcal Y}^{\parallel}_{21}
&=
- i N \,s^3\lb Y^{(+2)}Y^{(+1)}+Y^{(-2)}Y^{(-1)}\rb
=
N s^3\lb \bar{\mathcal Q}_2 \bar{\mathcal Q}_1 - \bar{\mathcal Q}_4 \bar{\mathcal Q}_3 \rb\, , \non\\
{\mathcal Y}^{\parallel}_{22}
&=
N s^4\lb Y^{(+2)}Y^{(+2)}+Y^{(-2)}Y^{(-2)}\rb
=
N s^4\lb \bar{\mathcal Q}_2 \bar{\mathcal Q}_2 - \bar{\mathcal Q}_4 \bar{\mathcal Q}_4 \rb\, ,
\label{eq:y_parallel}
}
and \emph{anti-aligned} $\mathcal Y^{\perp}_{mm'}$, with additional four elements
\eq{
{\mathcal Y}^{\perp}_{11}
&=
- N \lb Y^{(+1)}Y^{(-1)}+Y^{(-1)}Y^{(+1)}\rb
=
N \lb \bar{\mathcal Q}_1 \bar{\mathcal Q}_1 + \bar{\mathcal Q}_3 \bar{\mathcal Q}_3 \rb \, , \non\\
{\mathcal Y}^{\perp}_{12}
&=
- i N \,s\,\lb Y^{(+1)}Y^{(-2)}+Y^{(-1)}Y^{(+2)}\rb
=
N s\lb \bar{\mathcal Q}_1 \bar{\mathcal Q}_2 + \bar{\mathcal Q}_3 \bar{\mathcal Q}_4 \rb\, , \non\\
{\mathcal Y}^{\perp}_{21}
&=
- i N \,s\,\lb Y^{(+2)}Y^{(-1)}+Y^{(-2)}Y^{(+1)}\rb
=
N s\lb \bar{\mathcal Q}_2 \bar{\mathcal Q}_1 + \bar{\mathcal Q}_4 \bar{\mathcal Q}_3 \rb\, , \non\\
{\mathcal Y}^{\perp}_{22}
&=
N \lb Y^{(+2)}Y^{(-2)}+Y^{(-2)}Y^{(+2)}\rb
=
N \lb \bar{\mathcal Q}_2 \bar{\mathcal Q}_2 + \bar{\mathcal Q}_4 \bar{\mathcal Q}_4 \rb \, ,
\label{eq:y_perp}
}
where we also introduced the norm $N = \sqrt 2/2$. The  $\mathcal Y$ basis is most naturally understood as corresponding to total helicity, where each element is a real, parity-even combination of the helicity products $Y^{(m)}Y^{(m')}$ that is diagonal in total helicity $M=m+m'$ about $\hat k$, and is labelled by $|M|$. As shown before, we can link helicity states to the normalised parity-adapted basis through $Y^{(0)}\sim\bar{\mathcal Q}_0$, $Y^{(\pm1)}\sim\bar{\mathcal Q}_1\mp i\,\bar{\mathcal Q}_3$ and $Y^{(\pm2)}\sim\bar{\mathcal Q}_2\mp i\,\bar{\mathcal Q}_4$; the parity-even $\bar{\mathcal Q}_{0,1,2}$ and parity-odd $\bar{\mathcal Q}_{3,4}$ tensors therefore appear as even and odd combinations of the helicity states under $m \rightarrow -m$. The two sectors are then distinguished not by parity (both are parity-even) but by the relative orientation of the two helicities: the aligned combinations $Y^{(\pm m)}Y^{(\pm m')}$ carry $|M|=m+m'$, while the anti-aligned ones $Y^{(\pm m)}Y^{(\mp m')}$ carry $|M|=|m-m'|$. This can also be seen in the pre-factors, $\mathcal Y^{\parallel}_{mm'}\propto s^{m+m'}$ and $\mathcal Y^{\perp}_{mm'}\propto s^{|m-m'|}$, where the power of $s$ counts $|M|$ and reflects the spin-weight suppression along the line of sight. 

By construction, the full 13-element basis
\eq{
[\vec{\mathcal Y}^\parallel]_{mm'}
&=
\big\{
\mathcal Y^\parallel_{00},\,
\mathcal Y^\parallel_{01},\,
\mathcal Y^\parallel_{10},\,
\mathcal Y^\parallel_{02},\,
\mathcal Y^\parallel_{20},\,
\mathcal Y^\parallel_{11},\,
\mathcal Y^\parallel_{12},\,
\mathcal Y^\parallel_{21},\,
\mathcal Y^\parallel_{22}
\big\}_{\alpha}\, ,
\qquad m,m' = 0, 1, 2
\non\\
[\vec{\mathcal Y}^{\perp}]_{mm'}
&=
\big\{
\mathcal Y^\perp_{11},\,
\mathcal Y^\perp_{12},\,
\mathcal Y^\perp_{21},\,
\mathcal Y^\perp_{22}
\big\}_{\beta}\, ,
\qquad m,m' = 1,2
}
is orthogonal with the norms $\|{\mathcal Y}^\parallel_{mm'}\|^2= s^{2(m+m')}$ and $\|{\mathcal Y}^\perp_{mm'}\|^2= s^{2|m-m'|}$. 
Correlators can thus be written as 
\eeq{
\hat N^{(L)} \cdot \Xi^L_{ab} (\vec k)
=
\sum_{\alpha=\parallel, \perp} \sum_{(m,m')} H^\alpha_{mm'}(k,\mu) [\mathcal Y^\alpha_{mm'}]_{ab}\, .
}
The final basis is orthogonal and still polynomial, i.e. the explicit factors of $s$ have been
chosen precisely so that the associated form factors remain polynomials in $\mu$.
Indeed, if $m,m'\in\{0,1,2\}$ label the absolute helicities in the even sector and
$m,m'\in\{1,2\}$ in the odd sector, then the corresponding form factors have the generic
structure
\eeq{
H^{\alpha}_{m m'}(k,\mu)
=
\sum_{\ell=0}^{\lfloor (L-|M|)/2 \rfloor} P^{(L,\ell)}_{mm'}(k)\,\mu^{L-|M|-2\ell}\, .
}
For identical tracers the exchange symmetry of the two tensor slots implies
\eeq{
H^{\parallel}_{mm'}(k,\mu)= H^{\parallel}_{m'm}(k,\mu)\, ,
\quad
H^{\perp}_{mm'}(k,\mu)= H^{\perp}_{m'm}(k,\mu)\, ,
}
and the basis reduces to the $6+3=9$ elements, as expected.

Analogously to the scalar-tensor sector, we can define the normalized total-helicity basis
\begin{equation}
    \bar{\mathcal{Y}}^\alpha_{mm'} = s^{-|M|} \mathcal{Y}^\alpha_{mm'}\, .
    \label{eq:def_bar_y}
\end{equation}
In this basis, the coefficients are
\eeq{
    \bar{H}^{\alpha}_{mm'}(k,\mu) = s^{|M|} H^\alpha_{mm'}(k,\mu) = (1 - \mu^2)^{|M| / 2} \sum_{\ell=0}^{\lfloor (L-|M|)/2 \rfloor} P^{(L,\ell)}_{mm'}(k)\,\mu^{L-|M|-2\ell}\, .
\label{eq:assoc_legendre_TTa}
}
The above form again suggests that a natural basis is given by the associated Legendre polynomials
\eeq{
    \bar{H}^{\alpha}_{mm'}(k,\mu) = \sum_\ell \bar{H}^{\alpha, \ell}_{mm'}(k) P^{|M|}_\ell(\mu)\, .
\label{eq:assoc_legendre_TTb}
}
Since the $P^M_\ell$ take the form of $s^M$ times a polynomial, this is equivalent to the statement that the un-normalized $H^\alpha_{mm'}$ are spanned by polynomials in $\mu$. It is also clear that these coefficients correspond directly to the coefficients in the helicity basis.

Lastly, let us provide a transformation matrix linking the original $\mathcal M$ basis with the $\mathcal Y$ basis. 
The kinematic basis used above, we can label as
\eq{
\vec{\mathcal M}_{\rm e}
&=  \big\{
\mathcal M_{kk}\mathcal M_{kk}, 
\mathcal M_{kk}\mathcal M_{kn}, 
\mathcal M_{kn}\mathcal M_{kk}, 
\mathcal M_{kk}\mathcal M_{nn}, 
\mathcal M_{nn}\mathcal M_{kk}, \non\\
&\hspace{5.7cm} \mathcal M_{kn}\mathcal M_{kn}, 
\mathcal M_{nn}\mathcal M_{kn}, 
\mathcal M_{kn}\mathcal M_{nn}, 
\mathcal M_{nn}\mathcal M_{nn}
\big\} \, , ~ \alpha = 1,\dots,9\, ,\non\\ 
\vec{\mathcal M}_{\rm o}
&=
 \big\{
\mathcal M_{\delta kk}, 
\mathcal M_{\delta kn},
\mathcal M_{\delta nk},
\mathcal M_{\delta\delta}
\big\} \, , \qquad \beta = 1,\dots,4 \, .
}
The transformation matrix is then
\eeq{
\begin{pmatrix}
\vec{\mathcal M}_{\rm e}\\
\vec{\mathcal M}_{\rm o}
\end{pmatrix}
=
M
\begin{pmatrix}
\vec{\mathcal Y}_{\parallel}\\
\vec{\mathcal Y}_{\perp}
\end{pmatrix}\, ,
}
with $q = 3 \mu^2 -1$ and 
\eq{
M \;=\; \hat M \,\mathcal N\,,
\qquad
\mathcal N \;=\;
\mathrm{diag}\!\Big(
\tfrac13,\;
\underbrace{\tfrac1{\sqrt3},\tfrac1{\sqrt3},\tfrac1{\sqrt3},\tfrac1{\sqrt3}}_{4},\;
\underbrace{\tfrac1{\sqrt2},\dots,\tfrac1{\sqrt2}}_{8}
\Big)\, ,
}
and
{\setcounter{MaxMatrixCols}{13}
\eq{
\hat M \;=\;
\begin{pmatrix}
2 & 0 & 0 & 0 & 0 & 0 & 0 & 0 & 0 & 0 & 0 & 0 & 0 \\
2\mu & 1 & 0 & 0 & 0 & 0 & 0 & 0 & 0 & 0 & 0 & 0 & 0 \\
q & 2\mu & 0 & 1 & 0 & 0 & 0 & 0 & 0 & 0 & 0 & 0 & 0 \\
2\mu & 0 & 1 & 0 & 0 & 0 & 0 & 0 & 0 & 0 & 0 & 0 & 0 \\
2\mu^2 & \mu & \mu & 0 & 0 & \tfrac12 & 0 & 0 & 0 & \tfrac{s^2}{2} & 0 & 0 & 0 \\
\mu q & 2\mu^2 & \tfrac{q}{2} & \mu & 0 & \mu & \tfrac12 & 0 & 0 & \mu s^2 & \tfrac{s^2}{2} & 0 & 0 \\
q & 0 & 2\mu & 0 & 1 & 0 & 0 & 0 & 0 & 0 & 0 & 0 & 0 \\
\mu q & \tfrac{q}{2} & 2\mu^2 & 0 & \mu & \mu & 0 & \tfrac12 & 0 & \mu s^2 & 0 & \tfrac{s^2}{2} & 0 \\
\tfrac{q^2}{2} & \mu q & \mu q & \tfrac{q}{2} & \tfrac{q}{2} & 2\mu^2 & \mu & \mu & \tfrac12 & 2\mu^2 s^2 & \mu s^2 & \mu s^2 & \tfrac{s^4}{2} \\
2 & 0 & 0 & 0 & 0 & 0 & 0 & 0 & 0 & 1 & 0 & 0 & 0 \\
2\mu & 1 & -\tfrac12 & 0 & 0 & 0 & 0 & 0 & 0 & \mu & 1 & 0 & 0 \\
2\mu & -\tfrac12 & 1 & 0 & 0 & 0 & 0 & 0 & 0 & \mu & 0 & 1 & 0 \\
3 & 0 & 0 & 0 & 0 & 0 & 0 & 0 & 0 & 2 & 0 & 0 & 2
\end{pmatrix} \, .
}
}

Summarizing, the practical logic of the tensor--tensor sector is as follows:
the kinematic $\mathcal M$ basis is the most direct and polynomial one, but is redundant and does not
organize the parity structure transparently. The parity-adapted $\mathcal Q$ basis separates the parity-even
sector into even--even and odd--odd pieces. Finally, the total-helicity basis
$\vec{ \mathcal Y}$ provides an orthogonal and polynomial implementation of this same
decomposition, together with a direct connection to helicity states.

%------------------------------------------------------%
\subsection{Kinematic Origin of the Spin Weights and Associated Legendre Expansion}
\label{sec:spin_weights}

We close this section by re-deriving the basis spin weights directly from the redshift-space mapping, which makes their origin transparent. Expanding the rank-four correlator in the products $Y^{(m)}_{ij}(\hk)\, Y^{(m')}_{kl}(\hk)$, each product acquires a phase under a rotation about $\hk$, with total helicity $|M|=|m + m'|$ for the aligned combinations and $|M|=|m-m'|$ for the anti-aligned. In the scalar--tensor sector $m=0$, and thus $|M|=0,1,2$ label $G_0$, $G_1$ and $G_2$ form factors. By statistical isotropy, $\Xi^L$ is built solely from $\hk_i$ and $\delta_{ij}$, and is therefore invariant under such rotations, so that its projection onto a helicity-$M$ element vanishes unless $M=0$. Every structure with $M\neq0$ is thus generated by the $L$ contractions with $\hn$, and survives even when the underlying fields are exactly Gaussian, as explicitly shown in Sec.~\ref{sec:toy_example}.

At order $L$ the angular dependence, besides the prefactor $(-ik_{\hat n})^L\propto\mu^L$ carrying no helicity, arises via the contraction $\hat N_{i_1\ldots i_L}\Xi^L_{ijkl;\,i_1\ldots i_L}$ where we can rewrite each line-of-sight factor into components of definite helicity about $\hk$
\eeq{
\hn = \mu \hk + \frac{i s}{\sqrt 2}\lb e_+ + e_- \rb \, ,
}
so that a single $\hn$ supplies no helicity through the longitudinal piece $\mu\hk$, or one unit through the transverse pieces $e_\pm$. Since $\Xi^L$ carries none of its own, the net helicity is fixed entirely by the transverse selections, and because each $\hn$ contributes at most one unit we obtain the selection rule $|M|\leq L$.

The powers of $s$ and $\mu$ follow from the same counting. Of the $L$ factors of $\hn$, let 
\begin{itemize}
    \item $|M|+2p$ be selected transverse and deposited on the free indices, the extra $p$ pairs carrying cancelling helicity
    \item $2N_{\rm pairs}$ be contracted against one another through a Kronecker delta
    \item the remainder be selected longitudinal.
\end{itemize}
Each transverse selection carries a power of $s$ and each longitudinal one a power of $\mu$, while pairs contracted with a Kronecker delta return unity, so
\eeq{
\hat N_{i_1\ldots i_L} \Xi^L_{ijkl;\, i_1\ldots i_L}(\vec k)
\supset
s^{|M|+2p}\, \mu^{L-|M|-2p-2N_{\rm pairs}}\,
\lb \Xi^L_{M,p,N_{\rm pairs}}(\vec k) \rb_{ijkl} \, .
}
Transverse selections must land on the free indices, so that $p$ is bounded by $|M|+2p\leq4$ in the tensor--tensor sector, and by $|M|+2p\leq2$ in the scalar--tensor sector.

The excess power $s^{2p}=(1-\mu^2)^p$ is polynomial and can be reabsorbed, leaving $s^{|M|}$ multiplying a polynomial in $\mu$ of degree $L-|M|$. The form factor of a given harmonic is therefore a finite sum of associated Legendre polynomials of order $|M|$,\footnote{No dependence on the azimuthal angle can appear, since the basis is built from the single plane shared by $\hn$ and $\hk$.}
\eeq{
\bar H^{L,\alpha}_{mm'}(k,\mu)
= s^{|M|} \sum_{\ell=0}^{\lfloor(L-|M|)/2\rfloor} \tilde H^{L,\alpha}_{mm',\ell}(k)\, \mu^{L-|M|-2\ell}
= \sum_{\ell=0}^{\lfloor(L-|M|)/2\rfloor} H^{L,\alpha}_{mm',\ell}(k)\, P^{|M|}_{L-2\ell}(\mu) \, ,
}
the two sets of coefficients being related by the change of basis from monomials to associated Legendre polynomials, in line with results obtained in Eqs.~\eqref{eq:assoc_legendre_TTa} and \eqref{eq:assoc_legendre_TTb}. This recovers the spin weights $s^{|M|}$ of the $\mathcal Y$ basis and $s^n$ of the $\mathcal Q_n$ basis, and shows the residual angular dependence to be polynomial in $\mu$. The weights are thus not imposed by hand, but forced by the mapping: the real-space correlator is invariant under rotations about $\hk$, and it is the powers of $\hn$ that break this invariance, each able to shift the helicity by at most one unit.

Lastly, it is worth emphasizing again that the only property of $\Xi^L_{ijkl;\,i_1\ldots i_L}$ that has been used is that every tensor structure available to it is built from $\hk_i$ and $\delta_{ij}$, so that each carries zero helicity about $\hk$. That this exhausts all the possibilities precisely follows from the completeness of the isotropic decomposition constructed in Appendix~\ref{app:direct_tensor}. The entire angular structure - the spin weight $s^{|M|}$, the selection rule $|M|\leq L$, and the degree of the accompanying polynomial, then follows from counting powers of $\hn$ alone.

%==========================================================================
\section{Projection of 3D shapes onto the sky} 
\label{sec:projections}

In order to make contact with most observations we will have to convert from the three-dimensional shape of galaxies to projected shapes
\begin{equation}
    \gamma_{ij} = \Big( \mathcal P_{ik} \mathcal P_{jl} - \frac12 \mathcal P_{ij} \mathcal P_{kl} \Big) g_{kl}\, , 
\end{equation}
where the projection tensor is $\mathcal P_{ij} = \delta_{ij} - \hn_i \hn_j$. Analogously to the 3D case we can define the projected basis
\begin{equation}
    \vec m^{(1)} = \frac{\hk - \mu \hn}{\sqrt{1 - \mu^2}}\, , \quad \vec m^{(2)} = \hn \times \vec m^{(1)}\, , \quad \vec m^{\pm} = \mp \frac{1}{\sqrt{2}} (\vec m^{(1)} \mp i \vec m^{(2)}) \, ,
\end{equation}
and shape components
\begin{equation}
    \gamma_\pm = \vec M_{ij}^{(\pm 2) \ast} \gamma_{ij}\, , \quad \vec M_{ij}^{(\pm 2)} = \vec m^{\pm}_{i} \vec m^\pm_j\, .
\end{equation}
Note that $\vec M^{\pm 2}$ are spin-2 and therefore even under $\vec k \rightarrow - \vec k$. The commonly used $E$- and $B$- modes are defined to be the real and imaginary parts of $\gamma_\pm$:
\begin{equation}
    \gamma_E = \frac12 \left( \gamma_+ + \gamma_-\right)\, , \quad \gamma_B = -\frac{i}{2} \left( \gamma_+ - \gamma_- \right)\, ,
\end{equation}
such that the 2-point function of these projected modes with themselves or scalars can be obtained by contracting with appropriate powers of $\vec M$ \cite{Vlah2021}.

It is convenient to write the $E$ and $B$ modes of the tensor shape field in terms of its decomposition in the parity-adapted $\mathcal{Q}_n$ basis. Defining $\bar{\mathcal{Q}}_n^{E/B}$ to be these projections for a given $\hk, \hn$ we have
\begin{equation}
    \bar{\mathcal Q}_n^E = \left(\frac{N_0}{2} s^2, -\frac{N_1}{2} \mu s, \frac{N_2}{4} (1 + \mu^2), 0, 0 \right)\, , \quad \bar{\mathcal Q}_n^B = \left(0,0,0,-\frac{N_3}{2} s, \frac{N_4}{2} \mu \right)\, ,
\end{equation}
i.e. the parity even and odd channels project precisely into the E and B modes, respectively. Since $\bar{\mathcal{Q}}_n$ are orthonormal, these projections satisfy $\sum_n | \bar{Q}^{E/B}_n |^2 = \frac12$. Finally, we note that in this work we always work at the 2-point function level and refer to the $\mathcal{Q}_n$ basis solely in reference to one wave-vector $\vec k$---however, unlike $\vec M^{\pm 2}$ described above $\mathcal{Q}_{1,4}$ flip sign for opposite modes and therefore require some caution when operating at the level of the field instead.

For the scalar-tensor sector we then have in terms of the $\mathcal Q_n$ basis
\begin{equation}
    P_{\delta E}(k,\mu) = \frac{N_0}{2} (1 - \mu^2) G_0(k,\mu) - \frac{N_1}2 \mu (1 - \mu^2) G_1(k,\mu) + \frac{N_2}4 (1 - \mu^4) G_2(k,\mu)\, .
    \label{eq:deltaE_proj}
\end{equation}
We can immediately note two important features: firstly, projecting from 3D shape to $E$ modes mixes the three helicity channels $G_{0,1,2}$, such that it is in general not possible to disentangle them, other than at $\mu = 0$ where the $|M| > 0$ vanish due to the inaction of the redshift-space mapping. Secondly, we recover the observation from ref.~\cite{KuritaTakada2022} that the scalar-projected tensor power spectrum is best described by an associated Legendre expansion with $P^{m=2}_\ell(\mu)$, since each polynomial form factor is accompanied by a factor of $s^2 = (1-\mu^2)$, whose coefficients indeed inherit the covariance properties we will derive below in 3D in Section~\ref{sec:measurements}.

We can similarly write down contributions to projected shape autospectra in terms of form factor coefficients. Adopting the ${\mathcal Y}$ basis for example we can write
\begin{equation}
    P_{XY}(k,\mu) = \sum_{ p = \parallel, \perp} \sum_{(a,b)} H^{p}_{ab}(k,\mu)\ \pi^{p}_{ab, XY}(\mu), \quad X, Y \in \{E, B\}\, ,
\end{equation}
where $\pi^{\parallel/\perp}_{ab,XY}(\mu)$ are the form factors projected onto the sky, given by
\begin{align*}
\pi^\parallel_{00,EE}(\mu) &= \frac38 (1 - \mu^2)^2,
&
\pi^\parallel_{00,BB}(\mu) &= 0,
\\
\pi^\parallel_{01,EE}(\mu) &= -\sqrt{\frac38} N \mu (1 - \mu^2)^2,
&
\pi^\parallel_{01,BB}(\mu) &= 0,
\\
\pi^\parallel_{02,EE}(\mu) &= \sqrt{\frac{3}{32}} N (1 + \mu^2) (1 - \mu^2)^2,
&
\pi^\parallel_{02,BB}(\mu) &= 0,
\\
\pi^\parallel_{11,EE}(\mu) &= \frac12 N \mu^2 (1 - \mu^2)^2,
&
\pi^\parallel_{11,BB}(\mu) &= -\frac12 N (1 - \mu^2)^2,
\\
\pi^\parallel_{12,EE}(\mu) &= -\frac14 N\mu (1 + \mu^2)(1 - \mu^2)^2,
&
\pi^\parallel_{12,BB}(\mu) &= \frac12 N \mu (1 - \mu^2)^2,
\\
\pi^\parallel_{22,EE}(\mu) &= \frac18 N (1 - \mu^4)^2,
&
\pi^\parallel_{22,BB}(\mu) &= -\frac12 N \mu^2 (1 - \mu^2)^2,
\\
\pi^\perp_{11,EE}(\mu) &= \frac12 N \mu^2 (1 - \mu^2),
&
\pi^\perp_{11,BB}(\mu) &= \frac12 N (1 - \mu^2),
\\
\pi^\perp_{22,EE}(\mu) &= \frac18 N (1 + \mu^2)^2,
&
\pi^\perp_{22,BB}(\mu) &= \frac12 N \mu^2,
\\
\pi^\perp_{12,EE}(\mu) &= -\frac14 N \mu (1 - \mu^4),
&
\pi^\perp_{12,BB}(\mu) &= -\frac12 N \mu (1 - \mu^2)
\end{align*}
and $\pi_{EB}=0$ since we have only written down parity-even form factors in the $\mathcal Y$ basis. The projected form factors $\pi^\parallel_{00}, \pi^\perp_{11}, \pi^\perp_{22}$ correspond to form factors that do not vanish when $\hk = \hn$ and match the real-space contributions derived in ref.~\cite{Singh2015,KuritaTakada2022}. Indeed, our results above generalize their observation for projected tensor-tensor spectra to all channels $\mathcal Y^\alpha_{mm'}$ beyond real-space or linear theory: in the basis $P_{\pm} = P_{EE} \pm P_{BB}$, the above projection kernels imply that $P_+$ follows the $P^{m=0}_\ell(\mu)$ basis while $P_-$ follows the $P^{m=4}_\ell(\mu)$ basis, again with covariance properties that inherit the optimal weighting of the 3D form factors that we will derive in the next section.\footnote{
An alternative derivation the angular momentum $m$ of the associated Legendre expansion of projected shapes follows by noting that the harmonic bases $\vec M ^{(\pm 2)}$ and $Y^{(m)}_{ij}$ are related by a rotation of $\theta$ in the $\hk, \hn$ plane, where $\cos\theta = \mu$. In particular, they are related by the Wigner matrices
\begin{equation}
    \vec M^{(\pm 2)*}_{ij} Y^{(m)}_{ij} = i^{\pm m} d^2_{\pm 2, m}(\theta) \sim i^{\pm m} \sin\left(\frac\theta 2\right)^{2\mp m} \cos\left(\frac\theta 2\right)^{2\pm m}
\end{equation}
where the powers of $i$ come from an additional $\pi/2$ rotation stemming from our definition of the basis vectors $e_{1,2}$ and $\vec m^{(1,2)}$. Using the double-angle sine identity $s = \sin(\theta) = \sin(\theta/2) \cos(\theta/2)$, this means that the above product is proportional to $s^{2-|m|}$ times a polynomial of $\mu$. For the scalar-tensor sector, this implies that in the normalized basis 
\begin{equation}
    \langle \delta g^{(m)} \rangle \sim \underbrace{s^{2 - |m|}}_{\text{ Wigner-$d$}} \times \underbrace{s^{|m|}}_{\text{spin weight}} \times \text{``polynomial''} \sim s^2 \times \text{``polynomial''}
\end{equation}
i.e. that the projected scalar-tensor spectrum is described by associated Legendre polynomials $P^{m=2}_\ell$. A corollary of the canceling angular momentum factors is that the three form factors in the total-helicity basis collapse into a single $P^{m=2}_\ell$ expansion, and are only distinguishable by the lowest power of $\mu$ allowed by the redshift-space mapping for each channel. The tensor-tensor case ($P_\pm$) is more involved, requiring the product $d^2_{\pm2,m} d^2_{\pm2,m'}$ of Wigner matrices, but follows the same logic, yielding that $P_{\pm}$ are described by $P^{m=0,4}_\ell$, respectively. Note that the projection removes any radicals in odd powers of $s$ in favor of purely even powers that project onto polynomials.
}

%==========================================================================
\section{Measurements}
\label{sec:measurements}

%------------------------------------------------------%
\subsection{Estimator}

In this section we construct estimators for the form factors, $G_n(k, \mu)$ and $H^{\parallel,\perp}_{mm'}(k, \mu)$  defined in Sec.~\ref{sec:IA_correlators}, in terms of the associated Legendre multipole moments $\bar{G}_n^\ell (k)$ and $\bar{H}^{\alpha}_{mm',\ell}(k)$.
Thanks to the orthogonality of the $\mathcal Q$ and $\mathcal Y$ bases---in contrast with the kinematic ${\mathcal M}$ basis---our strategy will be to project the shape field $g_{ab}$ onto individual field-level basis elements (Section~\ref{ssec:field_basis}) and construct form factors out of their cross correlations.
While they may seem like the most natural candidates, the normalized parity-adapted basis $\bar{{\mathcal Q}}$ and helicity basis $Y^{(m)}_{ab}$, as we have defined them in the $\hk, \hn$ plane, become singular in the collinear limit ($s=0$), and are hence unreliable for the required projections. Instead, in this section we will use the \textit{unnormalized} helicity basis, wherein we substitute in $m$ and $t$ for the transverse basis vectors $e_1$ and $e_2$ in Equation~\ref{eq:def_helicity}, as the projection basis. The former have the advantage that their construction does not rely on dividing by vanishing norms in the collinear limit. For most of the estimated form factors we therefore first project the shape field $g_{ab}$ onto the un-normalized helicity basis $\bar{Y}^{(m)}_{ab} = s^{|m|}Y_{ab}^{(m)}$, and define
\begin{equation}
    \bar{g}^{(m)}({\vec k}) = \sum_{ab} g_{ab}(\vec k) \bar{Y}^{(m)}_{ab}\,,
\end{equation}
As we will see, the diverging factors of $s^{-|M|}$ relating the correlators in this basis to the desired form factors are canceled out by the spin weights $s^{|M|}$ we derived in previous sections. For the \textit{anti-aligned} channels in the tensor-tensor sector, we will also need the corresponding projection with the usual, \textit{normalized}, helicity basis
\begin{equation}
    g^{(m)}(\vec k) = \sum_{ab} g_{ab}(\vec k)Y_{ab}^{(m)}\,, \quad \bar{g}^{(m)}(\vec k) = s^{|m|} g^{(m)}(\vec k)\, .
\end{equation}
We will discuss how to deal with the pole at $s=0$ for $g^{(m)}$ at the end of this subsection. Note that yet another, equivalent possibility would be to project onto the un-normalized parity-adapted basis $\mathcal Q$.
%Note that $\bar{Y}^{m}\cdot\bar{Y}^{m'} = s^{2|m|}\delta_{mm'}$.

For the scalar-tensor sector, the $s^n$ factors relating the parity-adapted $\mathcal Q$ basis to the helicity basis (Eq.~\eqref{eq:Q_Y_relation}) are naturally absorbed into the \textit{unnormalized} basis
\eeq{
    \mathcal Q_0 =  Y^{(0)}\, , 
    \quad \mathcal Q_1 = \frac{i}{\sqrt 2} (\bar{Y}^{(1)} + \bar{Y}^{(-1)})\, , 
    \quad \mathcal Q_2 = -\frac{1}{\sqrt 2} (\bar{Y}^{(2)} + \bar{Y}^{(-2)})\, 
}
with $\|Q_n\|^2 = s^{2n}$. Introducing 
\begin{equation}
    C_0(\vec k) = \mathrm{Re}[\bar{g}^{(0)}(\vec k) \delta^*(\vec k)]\, , 
    \; C_1(\vec k) = -\frac{1}{\sqrt{2}} \mathrm{Im}[(\bar{g}^{(1)}(\vec k) + \bar{g}^{(-1)}(\vec k)) \delta^*(\vec k)]\, ,
    \; C_2(\vec k) = -\frac{1}{\sqrt2}\mathrm{Re}[(\bar{g}^{(2)}(\vec k) + \bar{g}^{(-2)}(\vec k)) \delta^*(\vec k)]\, , \non
\end{equation}
such that 
\begin{equation}
    \langle C_n(\vec k)\rangle = s^{2n} G_n(k,\mu) = s^n \bar{G}_n(k,\mu)\, 
\end{equation}
leads to the estimator for $\bar{G}_n^\ell(k)$ (Eq.~\eqref{eq:def_bar_G})
\begin{equation}
    \hat{\bar{G}}_n^\ell(k_i) = \frac{V}{N_{k_i}} \sum_{\vec k\in k_i} C_n(\vec k) D_\ell^n(\mu_{\vec k})\, ,
    \label{eq:estimator_G}
\end{equation}
where the angular weight $D_L^M(\mu)$ is the \textit{reduced} associated Legendre polynomial
\begin{align}
    D_L^M (\mu) = (2L+1)\frac{(L-M)!}{(L+M)!} \frac{P^M_L(\mu)}{s^M}
    =  (2L+1)\frac{(L-M)!}{(L+M)!} (-1)^M\frac{\mathrm d^M}{\mathrm d \mu^M} \mathcal L_L(\mu) \, .
\end{align}
Here $\mathcal L_L(\mu)$ is the usual Legendre polynomial of order $L$. Importantly, $D^m_L$ is a polynomial in $\mu$ and does not have any poles.

For the \textit{aligned} tensor-tensor sector, we additionally use
\eeq{
    \mathcal Q_3 = -\frac{1}{\sqrt 2} (\bar{Y}^{(1)} - \bar{Y}^{(-1)}), \quad \mathcal Q_4 = -\frac{i}{\sqrt 2} (\bar{Y}^{(2)} - \bar{Y}^{(-2)})\, ,
}
to express the projected shape-shape fields onto $\mathcal Y^\parallel_{mm'}$ (Eq.~\eqref{eq:y_parallel}) as
\begin{align}
    C^\parallel_{00}(\vec k) =&  |\bar{g}^{(0)}(\vec k)|^2\, ,
    \quad C^\parallel_{01}(\vec k) = \frac{1}{\sqrt 2} \mathrm{Im}[\bar{g}^{(0)}(\vec k)(\bar{g}^{(+1)*}(\vec k) + \bar{g}^{(-1)*}(\vec k))]\, , \non \\
    \quad C^\parallel_{02}(\vec k) =& -\frac{1}{\sqrt 2} \mathrm{Re}[\bar{g}^{(0)}(\vec k)(\bar{g}^{(+2)*}(\vec k) + \bar{g}^{(-2)*}(\vec k))]\, ,\quad C^\parallel_{11}(\vec k) = \sqrt{2}\mathrm{Re}[\bar{g}^{(1)}(\vec k) \bar{g}^{(-1)*}(\vec k)]\, , \non \\
    C^\parallel_{12}(\vec k) =& \frac{1}{\sqrt 2} \mathrm{Im}[\bar{g}^{(+1)}(\vec k)\bar{g}^{(-2)*}(\vec k) + \bar{g}^{(-1)}(\vec k)\bar{g}^{(+2)*}(\vec k)]\, , \quad 
    C^\parallel_{22}(\vec k) = \sqrt{2}\mathrm{Re}[\bar{g}^{(2)}(\vec k) \bar{g}^{(-2)*}(\vec k)]\, .
\end{align}
Similarly to the scalar-tensor case, given $\|\mathcal Y^\parallel_{mm'} \|^2 = s^{2M_\parallel}$ with $M_\parallel = m+m'$ we have 
\begin{equation}
    \langle C^\parallel_{mm'} (\vec k)\rangle = s^{2M_\parallel} H^\parallel_{mm'}(k, \mu) = s^{M_\parallel} \bar{H}^\parallel_{mm'}(k, \mu)\, ,
\end{equation}
leading naturally to the estimator
\begin{equation}
    \hat{\bar{H}}_{mm',\ell}^\parallel(k_i) = \frac{V}{N_{k_i}} \sum_{\vec k\in k_i} C^\parallel_{mm'}(\vec k) D_\ell^{M_\parallel}(\mu_{\vec k})\, ,
    \label{eq:estimator_H_para}
\end{equation}
again avoiding any poles in the collinear limit.

On the other hand, for the \textit{anti-aligned} tensor-tensor sector, the total spin weight $s^{M_\perp} = s^{|m-m'|}$ does not, in general, factorize into the product of the two individual fields. In other words, tensor-tensor correlators projected onto $\mathcal Y^\perp_{mm'}$ cannot be rewritten in terms of projected fields $\bar{g}^{m}$ alone. Instead, they can be measured as power spectra of both $g^{m}$ and $\bar{g}^{m}$
\begin{align}
    C^\perp_{11}(\vec k) =&  \frac{1}{\sqrt 2} (|g^{(1)}(\vec k)|^2 + |g^{(-1)}(\vec k)|^2)\, ,
    \nonumber \\
    C^\perp_{12}(\vec k) = & \frac{1}{\sqrt 2} \mathrm{Im}[\bar{g}^{(+1)}(\vec k)g^{(+2)*}(\vec k) + \bar{g}^{(-1)}(\vec k)g^{(-2)*}(\vec k)]\, ,
    \nonumber \\
    C^\perp_{22}(\vec k) =&  \frac{1}{\sqrt 2} (|g^{(2)}(\vec k)|^2 + |g^{(-2)}(\vec k)|^2)\, .
    \label{eq:C_perp}
\end{align}
Using $\|\mathcal Y^\perp_{mm'} \|^2 = s^{2M_\perp}$ leads to 
\begin{equation}
    \langle C^\perp_{mm'} (\vec k)\rangle = s^{2M_\perp} H^\perp_{mm'}(k, \mu) = s^{M_\perp} \bar{H}^\perp_{mm'}(k, \mu)\, ,
\end{equation}
and therefore the similar form
\begin{equation}
    \hat{\bar{H}}_{mm',\ell}^\perp(k_i) = \frac{V}{N_{k_i}} \sum_{\vec k\in k_i} C^\perp_{mm'}(\vec k) D_\ell^{M_\perp}(\mu_{\vec k})\, .
    \label{eq:estimator_H_perp}
\end{equation}

However, in order to make use of the anti-aligned estimators above we must specify the treatment of $g^{m}(\vec k)$ in the collinear limit.
When $\hat{k}$ and $\hat{n}$ are parallel, the choice of orthonormal transverse frame becomes arbitrary, and $g^{(m)}$ become basis-dependent.\footnote{Specifically, we employ $e_1 = (\hat{k}\times\hat{x})/|\hat{k}\times\hat{x}|$ and $\vec e_2 = \hat{k}\times e_1$ at the pole, taking $\hat{n} = \hat{z}$. Otherwise we follow Eq.~\eqref{eq:def_e1e2}.}
Fortunately, under a rotation of the transverse frame by $\psi$ about $\hat{k} = \hat{n}$, the projection changes only by the helicity phase
\begin{equation}
    g^{(m)} \to e^{-im\psi}g^{(m)}\, .
\end{equation}
As a result, the combinations $|g^{(+1)}|^2 + |g^{(-1)}|^2$ and $|g^{(+2)}|^2 + |g^{(-2)}|^2$ in Eq.~\eqref{eq:C_perp} are independent of the choice of basis and the 11 and 22 channels in the \textit{anti-aligned} tensor-tensor sector are well-behaved.
In contrast, the phase of the 12 channel with $M_\perp=1$ does not cancel, but since $\bar{g}^{(\pm 1)} = s g^{(\pm 1)}$ vanishes at the pole, the estimator is independent of the frame choice as well.

%------------------------------------------------------%
\subsection{Covariances}

Let us now extract the Gaussian covariances of the redshift-space shape 2-point functions above. For a wedge $(k,\mu)$ the density-shape and shape-shape power spectra have covariances
\begin{align}
    \text{Cov}\left[ \hat{P}^{\delta g}_{ab}(k,
    \mu), \hat{P}^{\delta g}_{cd}(k,
    \mu)  \right] &= \left( \frac{2}{N_k \Delta \mu} \right) \left( P^{\delta }(k,\mu) P^{g}_{abcd}(k,\mu) + P^{\delta g}_{ab}(k,
    \mu) P^{\delta g}_{cd}(k,\mu) \right)\,, \nonumber \\
    \text{Cov}\left[ \hat{P}^{g}_{abcd}(k,\mu), \hat{P}^{g}_{efgh}(k,\mu)  \right] &= \left( \frac{2}{N_k \Delta \mu} \right) \left( P^{g}_{abef}(k,\mu) P^{g}_{cdgh}(k,\mu) + P^{g}_{abgh}(k,\mu) P^{g}_{cdef}(k,\mu) \right)\, , \nonumber \\
    \text{Cov}\left[ \hat{P}^{\delta g}_{ab}(k, \mu), \hat{P}^{g g}_{cdef}(k, \mu)  \right] &= \left( \frac{2}{N_k \Delta \mu} \right) \left( P^{\delta g}_{cd}(k,\mu) P^{g}_{abef}(k,\mu) + P^{\delta g}_{ef}(k,\mu) P^{g}_{abcd}(k,\mu) \right)\, , \nonumber
\end{align}
where $N_k = V_{\rm obs} k^2 \Delta k / (2\pi^2)$ is the number of modes per momentum bin $\Delta k$. The covariance between form factors is then given by contracting the above with the appropriate basis tensors, e.g. $\bar{Q}, \bar{\mathcal{Y}}$ for normalized form factors, and the covariances of the associated Legendre multipoles given by angular integration.\footnote{Specifically, in terms of the multipole moments of the covariance $\text{Cov}\left[ A(\vec k), B(\vec k) \right] = \sum_L C_L(k) \mathcal L_L(\mu)$ we have the associated Legendre covariance
\begin{equation}
     \text{Cov}\left[ A_{\ell m}, B_{\ell' m} \right] =  \frac{(-1)^m}{N_k} (2\ell+1)(2\ell'+1) \sqrt{\frac{(\ell-m)!(\ell'-m)!}{(\ell+m)!(\ell'+m)!}}  \Bigg( \sum_{L} C_L(k) \, 
     \begin{pmatrix}
     \ell & \ell' & L\\
     m & -m & 0
     \end{pmatrix}
     \begin{pmatrix}
     \ell & \ell' & L\\
     0 & 0 & 0
     \end{pmatrix} \Bigg)\, .
\end{equation}
}

Rather than work out the general structure, let us consider the limit where the shape noise dominates the shape-shape autospectrum \cite{Vlah2020}, i.e.
\begin{equation}
    P^{\rm stoch}_{abcd} \approx P_\epsilon (\delta_{ac} \delta_{bd} + \delta_{ad} \delta_{bc} - \frac23 \delta_{ab} \delta_{cd})\, ,
\end{equation}
or $H^\parallel_{00} = 2 P_\epsilon$, $H^\perp_{11} = H^\perp_{22} = 2\sqrt2 P_\epsilon$. Since the shape noise is diagonal in helicity, the only nonzero covariances involve pairs of indices with matching angular momenta.  For example, the density-shape spectrum has diagonal covariance in the form factors
\begin{equation}
    \text{Cov}\left[ \hat{G}_n, \hat{G}_m \right] = 
\left( \frac{2}{N_k \Delta \mu} \right) 2 ||Q_n ||^{-2} P_\epsilon P^\delta(k,\mu) \delta_{nm}\, ,
\end{equation}
similar to the real-space covariance between helicity spectra. For auto spectra, the diagonality of the shape noise similarly implies that the covariance is diagonal in the $\mathcal{Y}$ basis
\begin{equation}
    \text{Cov}\left[ \hat{H}^{\alpha}_{m_1 m_2}, \hat{H}^{\beta}_{m_3 m_4} \right] = \left( \frac{2}{N_k \Delta \mu} \right) 4 P_\epsilon^2  ||\mathcal{Y}^\alpha_{m_1 m_2} ||^{-2} \delta_{\alpha \beta} (\delta_{m_1 m_3} \delta_{m_2 m_4} + \delta_{m_1 m_4} \delta_{m_2 m_3})\, .
\end{equation}
The $\hat{G}, \hat{H}$ cross spectrum is similar, and we give a more complete expression below.

The covariances above have the apparently undesirable property that they blow up as $\mu \rightarrow 1$ at finite shape noise even while $\hat{G}, \hat{H}$ are finite and polynomial in $\mu$. Intuitively, we can understand this as follows: the form factors $Q_n$ and $\mathcal{Y}^\alpha_{ab}$ simplify to the real-space and diagonal helicity basis for 2-point functions when $\mu \rightarrow 1$, rendering all but $1$ and $3$ basis elements zero; the \textit{finite} coefficients of the vanishing basis elements therefore become unmeasurable at finite noise levels, since their contribution to the full correlators vanish in the collinear limit. This divergence is cancelled by adopting the normalized basis and coefficients, which precisely cancel the inverse-basis norm in the above expressions. In this case, we can also derive the covariances of the associated Legendre moments e.g.
\begin{align}
     \text{Cov}\left[ \hat{\bar{G}}_n^\ell, \hat{\bar{G}}_m^{\ell'} \right] &= \left( \frac{2 P_\epsilon}{N_k} \right) \delta_{nm} \left( \frac{(2\ell+1)(\ell-n)!}{(\ell+n)!} \right) \left(  \frac{(2\ell'+1)(\ell'-n)!}{(\ell'+n)!} \right) \frac12 \int  P^n_\ell(\mu) P^m_{\ell'}(\mu) P^\delta(k,\mu) d\mu \non \\
     &= \left( \frac{2 P_\epsilon}{N_k} \right)  (-1)^n \Bigg( \sum_{L} P_L(k) \, (2\ell+1)(2\ell'+1) \sqrt{\frac{(\ell-n)!(\ell'-m)!}{(\ell+n)!(\ell'+m)!}}
     \begin{pmatrix}
     \ell & \ell' & L\\
     n & -n & 0
     \end{pmatrix}
     \begin{pmatrix}
     \ell & \ell' & L\\
     0 & 0 & 0
     \end{pmatrix} \Bigg)\ \delta_{nm}\, ,
\end{align}
where $P_L(k)$ are the multipoles of the scalar-scalar power spectrum $P^\delta(k,\mu)$. Note that while the form factors are polynomial in the un-normalized basis, measuring the usual Legendre multipoles in that basis would produce non-convergent noise levels; in the normalized basis the \textit{signal} vanishes while the covariance remains finite, preserving the relative signal-to-noise in both cases. In the limit that the scalar-scalar autospectrum is dominated by the monopole this reduces to
\begin{equation}
    \text{Cov}\left[ \hat{\bar{G}}_n^\ell, \hat{\bar{G}}_m^{\ell'} \right] = \frac{ 2 P_\epsilon P_0}{N_k} \frac{(2\ell+1)(\ell-n)!}{(\ell+n)!} \delta_{\ell \ell'} \delta_{n m}\, ,
\end{equation}
where the normalization reflects that $P^m_\ell$ do not have unit norm. In this limit, each $\hat{G}^\ell_m$ is statistically independent. The case for the auto-spectrum is essentially identical, noting that $P_\epsilon$ is monopole only: 
\begin{equation}
    \text{Cov}\left[ \hat{\bar{H}}_{m_1 m_2}^{\alpha,\ell}, \hat{\bar{H}}_{m_3 m_4}^{\beta,\ell'} \right] =  \left( \frac{4 P_\epsilon^2}{N_k} \right) \frac{(2\ell+1)(\ell-|M_{12}|)!}{(\ell+|M_{12}|)!}   (\delta_{m_1 m_3} \delta_{m_2 m_4} + \delta_{m_1 m_4} \delta_{m_2 m_3}) \delta_{\alpha \beta} \delta_{\ell \ell'}\, ,
\end{equation}
where $M_{12}$ is the total angular momentum of $\mathcal Y^\alpha_{m_1 m_2}$. In this limit, the tensor-tensor form factors are diagonal and constant in covariance. 

It is also straightforward to add the linear theory contribution to the covariance, since it contributes to only one channel each in the cross and auto spectra
\begin{equation}
    P^{g \delta}_{ab} = \sqrt{\frac23} (b_1 + f \mu^2) c_s P_{\rm lin}(k) [\mathcal{Q}_{0}]_{ab}\, , \quad P^{g g}_{abcd} = \frac23 c_s^2 P_{\rm lin}(k) [\mathcal{Y}^\parallel_{00}]_{abcd}\, .
\end{equation}
meaning that only form factors involving $Y^{(0)}$ are affected. Specifically, this modifies
\eq{
    &\text{Cov}\left[ \hat{G}_n, \hat{G}_m \right] =  \left( \frac{2}{N_k \Delta \mu} \right) \left( 2 P_\epsilon P^\delta + \left( \frac23 (c_s^2 P_{\rm lin}) P^\delta + \frac23 ( (b_1 + f\mu^2) c_s P_{\rm lin})^2\right) \delta_{n0} \right) ||Q_n||^{-2} \delta_{nm}\, , \non \\
    &\text{Cov}\left[ \hat{H}^{\alpha}_{m_1 m_2}, \hat{H}^{\beta}_{m_3 m_4} \right] = \left( \frac{2}{N_k \Delta \mu} \right) 4 \lb P_\epsilon + \frac13 c_s^2 P_{\rm lin} \delta_{m_1 0}\rb \lb P_\epsilon + \frac13 c_s^2 P_{\rm lin} \delta_{m_2 0}\rb  \nonumber \\
    & \qquad \quad \qquad \qquad \qquad \quad \qquad \qquad \qquad \quad \qquad \qquad ||\mathcal{Y}^\alpha_{m_1 m_2} ||^{-2} \delta_{\alpha \beta} (\delta_{m_1 m_3} \delta_{m_2 m_4} + \delta_{m_1 m_4} \delta_{m_2 m_3}) \, ,
    \label{eqn:cov_theory}
}
while keeping the diagonal structure. For autospectra, the different form factors are independent to leading order, with correlations between multipoles due to the anisotropy of the galaxies in linear theory (cross) and the anisotropic normalization of the basis (auto). In the absence of shape noise, in linear theory only the $n = 0$ and $a = 0$ or $b=0$ spectra have nonzero covariance; this is expected since in this regime $g_{ab} = \sqrt{2/3}\ c_s Y^{(0)}_{ab} (\hk) \delta(\vec k)$, i.e. the $|m| > 0$ helicities are not dynamical degrees of freedom and therefore do not have any variance. In general, form factors $(\alpha,a,b)$ and $(\beta,c,d)$ will have nonzero covariance only if $H^{\sigma}_{ab}$, $H^{\sigma'}_{ad}$ etc are nonzero, e.g. $\hat{H}^\parallel_{12}$ has a variance sourced by $H^{\sigma}_{11} H^{\sigma}_{22}$ and $H^\sigma_{12} H^{\sigma}_{12}$ where we have used $\sigma$ to denote the relative orientations $\parallel, \perp$.

For completeness, we can also write down the cross-covariance
\begin{equation}
    \text{Cov}\left[ \hat{G}_n, \hat{H}^{\alpha}_{m_1 m_2} \right] = \left( \frac{2}{N_k \Delta \mu} \right) \sqrt{\frac{8}{3}} \lb P_\epsilon + \frac13 c_s^2 P_{\rm lin} \delta_{m_1 0} \delta_{m_2 0} \rb (b_1 + f\mu^2) c_s P_{\rm lin} ||Q_{n} ||^{-1} ||\mathcal{Y}^\alpha_{m_1 m_2} ||^{-1} (\delta_{0m_1} \delta_{n m_2} + \delta_{0 m_2} \delta_{n m_1})\, ,
\end{equation}
which can be seen as a special case of the auto covariance with one of the angular momenta equal to zero.

%==========================================================================
\section{Toy Example: Gaussian-field Streaming Model}
\label{sec:toy_example}

In order to explore the interplay between the nonlinear mapping between real and redshift space and the angular form factor basis, let us consider a toy example for galaxy shapes in redshift-space where the real-space shape, density and velocity fields are all given by linear theory, i.e. $\delta_g = b_1 \delta_0, g_{ij} = c_s s_{0,ij}, \vec u_i = - f \nabla^{-1}_i \delta_0.$ In this limit, the only relevant nonzero connected correlators are the 2-point functions of the pairwise velocity coupled to shape and density fields
\eq{
    \langle s_{ab}(\vec r) \delta(\vec 0) \rangle &= S_2(r) \mathcal{L}_{2,ab}(\hat{r})\, , \non \\
    \langle \Delta \vec u_i \delta(\vec 0) \rangle &= V_1(r) \mathcal{L}_{1,i}(\hat{r})\, , \non\\
    \langle \Delta \vec u_i s_{ab}(\vec r) \rangle &= B_1(r) \mathcal{L}_{1,iab}(\hat{r}) + B_3(r) \mathcal{L}_{3,iab}(\hat{r})\, , \non \\
    \langle \Delta \vec u_i \Delta \vec u_j \rangle &= A_0(r) \mathcal{L}_{0,ij}(\hat{r}) + A_2(r) \mathcal{L}_{2,ij}(\hat{r})\, ,
\label{eqn:linear_correlators}
}
where $S_n,~V_n,~B_n,~A_n$ are scalar functions of $r = |\vec r|$ (Appendix~\ref{app:simple_model}) and $\mathcal{L}_n$ are tensors corresponding to Legendre polynomials
\eq{
\mathcal{L}_{0,ij}(\hat{r}) &= \delta_{ij}\, , \quad
\mathcal{L}_{1,i}(\hat{r}) = \hat{r}_i\, , \quad
\mathcal{L}_{1,iab}(\hat{r}) = \hat{r}_i \delta_{ab} - 3 \delta_{i(a} \hat{r}_{b)}\, , \non\\ 
\mathcal{L}_{2,ij}(\hat{r}) &= \frac32 \left( \hat{r}_i \hat{r}_j - \frac13 \delta_{ij} \right)\, , \quad
\mathcal{L}_{3,iab}(\hat{r}) = \frac12 \left( 5 \hat{r}_i \hat{r}_a \hat{r}_ b - 3 \delta_{(ab} \hat{r}_{i)} \right) \, ,
}
which Fourier transform into $i^\ell \mathcal{L}_\ell(\hk)$. Within this toy model, any mode coupling arises strictly from the nonlinearity of the redshift-space mapping itself. In what follows, we will refer to this toy model as the Gaussian-field Streaming Model (GfSM)---it is closely related, but not exactly equivalent, to the Gaussian Streaming Model discussed in the context of galaxy clustering \cite{Fisher95,Bharadwaj01,Scoccimarro04,Vlah16,Vlah19}.

%------------------------------------------------------%
\subsection{One-loop result}

Let us begin with the scalar-tensor cross spectrum. Since all underlying real-space quantities in the GfSM are Gaussian, the redshift-space cross spectrum can be obtained exactly via the integral
\eq{
    P_{ab}(\vec k) &= \int d^3 \vec r \ e^{i \vec k \cdot \vec r} \langle  e^{i k_{\hat n} \Delta u_{\hat n}} \ c_s \ s_{ab}(\vec r) (1 + b_1 \delta(\vec 0)) \rangle \non \\
    &= c_s \int d^3 \vec r  \ e^{i \vec k \cdot \vec r - \frac12 (k \mu)^2 \hn_i \hn_j \langle \Delta \vec u_i \Delta \vec u_j \rangle } \Big( b_1 \langle s_{ab}(\vec r) \delta(\vec 0) \rangle + i (k\mu) \hn_i \langle \Delta \vec u_i s_{ab}(\vec r) \rangle  \non\\
    &\hspace{8cm} - b_1 (k \mu)^2 \hn_i \hn_j \langle \Delta \vec u_i s_{ab}(\vec r) \rangle \langle \Delta \vec u_j \delta(\vec 0) \rangle \Big)\, . 
}
A notable feature of the above expression is that even Gaussian velocities can generate arbitrarily high powers of $\hat{n}$, and therefore all spherical harmonics of angular momentum $|M| \leq 2$.

It is instructive to first look at the cross spectrum at 1-loop order 
\eq{
 P^{\rm 1-loop}_{ab}(k) =  c_s \int  d^3 \vec r \ e^{i \vec k \cdot \vec r} \ \Big( & b_1 \langle s_{ab}(\vec r) \delta(\vec 0) \rangle + i (k\mu) \hn_i \langle \Delta \vec u_i s_{ab}(\vec r) \rangle \non \\
    & - \frac12 (k \mu)^2 \hn_i \hn_j \langle \Delta \vec u_i \Delta \vec u_j \rangle \left( b_1 \langle s_{ab}(\vec r) \delta(\vec 0) \rangle + i (k\mu) \hn_i \langle \Delta \vec u_i s_{ab}(\vec r) \rangle  \right) \non\\
    &- b_1 (k \mu)^2 \hn_i \hn_j \langle \Delta \vec u_i s_{ab}(\vec r) \rangle \langle \Delta \vec u_j \delta(\vec 0) \rangle \Big)\, ,
\label{eqn:toy_model_cross}
}
as an example of how the redshift-space mapping generates polynomial form factors in the $\mathcal M$ basis. For example, the first 1-loop term proportional to $b_1$ can be written as
\begin{equation}
    P_{ab}(k) \supset - c_s \frac12 (k\mu)^2 b_1 \hn_i \hn_j \int dr\ r^2 \int d\Omega_{\hat{r}} \ e^{i \vec k \cdot \vec r}  \left( A_0(r) \mathcal{L}_{0,ij}(\hat{r}) + A_2(r) \mathcal{L}_{2,ij}(\hat{r}) \right) S_2(r) \mathcal{L}_{2,ab}(\hat{r})\, .
\end{equation}
The angular integral involves products of Legendre tensors composed purely of $\hat{r}$ and therefore evaluates to products of $\hat{k}$ and $\delta_{ij}$ by symmetry, with coefficients given by Hankel transforms of $A_{0,2}(r) S_2(r)$ depending on $k = |\vec k|$. Contracting with $\hn_i$ and $\hn_j$ therefore yields traceless combinations of products of $\hk, \hn, \delta_{ij}$ multiplied by $(k \mu)^2$ and additional powers $\mu = \hk \cdot \hn$. This is precisely the form of the $\mathcal M$ basis. For example, the $A_2(r) S_2(r)$ piece contains
\begin{equation*}
    -\frac12 (k\mu)^2 b_1 \hn_i \hn_j \int dr\ r^2 \int d\Omega_{\hat{r}} \ e^{i\vec k \cdot \vec r} \ A_2(r) S_2(r) \mathcal{L}_{4,ijab}(\hat{r}) = -\frac12 (k\mu)^2 b_1   \hn_i \hn_j \mathcal{L}_{4,ijab}(\hk) (4\pi) \int dr\ r^2 A_2(r) S_2(r) j_4(kr) \, ,
\end{equation*}
where the contracted tensors yield
\begin{equation}
    \hn_i \hn_j \mathcal{L}_{4,ijab}(\hk) = \frac18 \left( (35 \mu^2 - 5) [\mathcal{M}_{kk}]_{ab} - 20 \mu [\mathcal{M}_{kn}]_{ab} + 2 [\mathcal{M}_{nn}]_{ab} \right)\, ,
\end{equation}
giving rise to all three parity-even basis elements in the scalar-tensor sector with polynomial coefficients in $\mu$, as expected, with similar results for the other contributions.

We can also study the GfSM tensor-tensor autospectrum, given by
\eeq{
    P_{abcd}(k) = c_s^{2} \int  d^3 \vec r \ e^{i \vec k \cdot \vec r - \frac12 (k\mu)^2 \hn_i \hn_j \langle \Delta \vec u_i \Delta \vec u_j \rangle}  \Big( \langle s_{ab}(\vec r) s_{cd}(\vec 0) \rangle  - (k \mu)^2 \hn_i \hn_j \langle \Delta \vec u_i s_{ab}(\vec r) \rangle \langle \Delta \vec u_j s_{cd}(\vec 0) \rangle \Big)\, .
}
The velocity two-point function in the exponent leads at order $m$ in the exponential to contributions
\eq{
    P_{abcd}(\vec k) 
    &\supset c_s^{2} \frac{(-1)^m (k\mu)^{2m} }{2^m m!}   \hn_{i_1} \hn_{j_1} \ldots \hn_{i_m} \hn_{j_m} \int d^3 \vec r \ e^{i\vec k \cdot \vec r} \ (\hat{r}_{i_1} \hat{r}_{j_1}) \ldots (\hat{r}_{i_m} \hat{r}_{j_m}) \left( f(r) \hat{r}_a \hat{r}_b \hat{r}_c \hat{r}_d + \ldots \right) \non \\
    &=  \frac{(-1)^m (k\mu)^{2m} }{2^m m!} \Big( 4\pi \int dr \ r^2\ \tilde{f}(r) j_{2m+4}(kr) \Big) \  \hn_{i_1} \hn_{j_1} \ldots \hn_{i_m} \hn_{j_m} \mathcal{L}_{2m+4, i_1 j_1 \ldots abcd}(\hk) + \ldots \, .
}
Similarly to the cross, the leading contraction shown above is sufficient to generate the tensor-tensor $\mathcal{M}$ basis, though terms with more than two $\hn$ are generated in the toy model only at 2-loop order.

For practical computations, it is more straightforward to compute form factors in the orthogonal bases $\mathcal{Q}$ and $\mathcal{Y}$. Unsurprisingly, in these bases we find again that the coefficients are polynomials of $\mu$. We give details for the 1-loop and exact calculations within the GfSM in Appendix~\ref{app:simple_model}. As we showed in Section~\ref{sec:IA_correlators}, this is equivalent to the statement that the form factors in the normalized $\bar{\mathcal Q}, \, \bar{\mathcal{Y}}$ bases can be expressed via generalized multipole moments multiplying generalized Legendre polynomials.

%------------------------------------------------------%
\subsection{Full solution}
\label{subsec:gfsm_full}

The one-loop expressions above can be extended to all orders in the streaming exponential with little extra work. Since the velocity-dispersion term depends on $\hat r$ only through $\hn\cdot\hat r$, it is convenient to split it into an isotropic piece and an angular one,
\eeq{
\exp\ls -\tfrac12 (k\mu)^2\, \hn_i \hn_j \la \Delta \vec u_i \Delta \vec u_j \ra \rs
= D_0(r,q)\; e^{-a(r,q)\, \mathcal L_2(\hn \cdot \hat r)} \, ,
\qquad q \equiv k\mu \, ,
}
where, using Eq.~\eqref{eqn:linear_correlators}, we have
\eeq{
D_0(r,q) = e^{-q^2 A_0(r)/2} \, , \qquad\qquad a(r,q) = q^2 A_2(r)/2 \, .
\label{eqn:gfsm_damping_split}
}
The isotropic factor $D_0$ passes through the angular integrals, while the remaining angular dependence is carried by the coefficients
\eeq{
\mathcal E^{(n)}_\ell(a) \equiv \frac{2\ell+1}{2} \int_{-1}^{1} dy\; y^n\, e^{-a \mathcal L_2(y)}\, \mathcal L_\ell(y) \, ,
\qquad n = 0,1,2 \, ,
\label{eqn:gfsm_E_coefficients}
}
in terms of which $(\hn\cdot\hat r)^n e^{-a \mathcal L_2(\hn\cdot\hat r)} = \sum_\ell \mathcal E^{(n)}_\ell(a)\, \mathcal L_\ell(\hn\cdot\hat r)$. Because $e^{-a\mathcal L_2}$ is even, these obey the selection rule $\mathcal E^{(n)}_\ell(a) = 0$ whenever $\ell+n$ is odd, which is what separates the even and odd Bessel channels below. Closed forms are collected in Appendix~\ref{app:simple_model}.

Performing the angular integrals and reindexing the resulting spherical Bessel functions by their final order, as detailed in Appendix~\ref{app:simple_model}, the scalar--tensor form factors reduce to a single radial integral per Bessel order,
\eeq{
G_\lambda(k,\mu) = -4\pi c_s \sum_{m=0}^{\infty} (-1)^m \int_0^\infty dr\; r^2\, D_0(r,q)
\ls \mathcal R^{(\lambda)}_{{\rm E},m}\, j_{2m}(kr) + \mathcal R^{(\lambda)}_{{\rm O},m}\, j_{2m+1}(kr) \rs \, ,
\label{eqn:gfsm_full_cross}
}
with the two manifestly real radial kernels
\eq{
\mathcal R^{(\lambda)}_{{\rm E},m}(r;\mu,q) &= b_1 \left\{ -\frac32 S_2(r)\, \widetilde{C}^{(0)}_{\lambda,2m}
- \frac{5}{2} q^2 V_1(r) \ls \big(3B_1(r)+B_3(r)\big) \widetilde{D}^{(1)}_{\lambda,2m} - B_3(r)\, \widetilde{C}^{(2)}_{\lambda,2m} \rs \right\} \, , \non \\
\mathcal R^{(\lambda)}_{{\rm O},m}(r;\mu,q) &= \frac52 q \left\{ B_3(r)\, \widetilde{C}^{(1)}_{\lambda,2m+1}
- \big(3B_1(r)+B_3(r)\big) \widetilde{D}^{(0)}_{\lambda,2m+1} \right\} \, .
\label{eqn:gfsm_full_kernels}
}
Here $\widetilde{C}^{(n)}_{\lambda L}$ and $\widetilde{D}^{(n)}_{\lambda L}$ are the shifted angular factors
\eq{
\widetilde{C}^{(n)}_{\lambda L} &= C^{(\lambda,0)}_{L}\, \mathcal E^{(n)}_{L} - C^{(\lambda,2)}_{L-2}\, \mathcal E^{(n)}_{L-2} - C^{(\lambda,-2)}_{L+2}\, \mathcal E^{(n)}_{L+2} \, , \non \\
\widetilde{D}^{(n)}_{\lambda L} &= D^{(\lambda,1)}_{L-1}\, \mathcal E^{(n)}_{L-1} - D^{(\lambda,-1)}_{L+1}\, \mathcal E^{(n)}_{L+1} \, ,
\label{eqn:gfsm_shifted_angular}
}
where all quantities carrying a negative angular index are understood to vanish, and the purely angular coefficients $C^{(\lambda,\Delta)}_{L}(\mu)$, $D^{(\lambda,\Delta)}_{L}(\mu)$ are collected in Appendix~\ref{app:simple_model}. Note that,  in the form given in  Eq.~\eqref{eqn:gfsm_full_cross}, our result contains no inverse powers of $kr$, and is suited to evaluation using FFTLog algorithm \cite{Hamilton1999, Vlah2014}.

As a check, at linear order only the $\ell=2$ term of $\widetilde{C}^{(0)}$ and the $\ell=1,3$ terms of $\widetilde{C}^{(1)},\widetilde{D}^{(0)}$ survive, and the Bessel closure relation reduces the radial integrals onto $P_{\rm lin}(k)$, giving
\eeq{
G_0^{\rm lin}(k,\mu) = N_0^{-1} c_s \lb b_1 + f\mu^2 \rb P_{\rm lin}(k) \, , \qquad G_1^{\rm lin} = G_2^{\rm lin} = 0 \, ,
}
i.e. the Kaiser result with the tensor structure collapsing onto $\mathcal Q_0$ alone, as anticipated in Sec.~\ref{sec:IA_correlators}. It is worth noting that $G_1$ vanishes at linear order not term by term, but through an exact cancellation between the $B_1$ contribution entering at $j_1$ and the $B_3$ contribution entering at $j_3$.

The tensor--tensor autospectrum is derived analogously, the only structural difference is that the selection rule now requires every contributing Bessel order to be even, resulting with a single sum,
\eeq{
H^{X}_{\gamma}(k,\mu) = 4\pi c_s^2 \sum_{m=0}^{\infty} (-1)^m \int_0^\infty dr\; r^2\, D_0(r,q)\, \mathcal R^{X}_{\gamma m}(r;\mu,q)\, j_{2m}(kr) \, ,
\label{eqn:gfsm_full_auto}
}
where $X = \parallel, \perp$ labels the aligned and anti-aligned sectors and $\gamma$ the total-helicity pair. The radial kernels $\mathcal R^{X}_{\gamma m}$ are built from $S_2$, $B_1$, $B_3$ and the same coefficients $\mathcal E^{(n)}_\ell(a)$, and are given in Appendix~\ref{app:simple_model}, and their derivation is a straightforward, albeit lengthier, repetition of the scalar--tensor case and we do not reproduce it here. At linear order a single form factor survives
\eeq{
H^{\parallel,\rm lin}_{00}(k,\mu) = \frac23 c_s^2\, P_{\rm lin}(k) \, ,
}
all other aligned and anti-aligned form factors being generated by the nonlinear streaming terms beyond linear order, consistent with the $|M| \leq L$ selection rule of Sec.~\ref{sec:spin_weights}.

%------------------------------------------------------%
\subsection{Validation against Simulations}

\begin{figure}
    \centering
    \includegraphics[width=\linewidth]{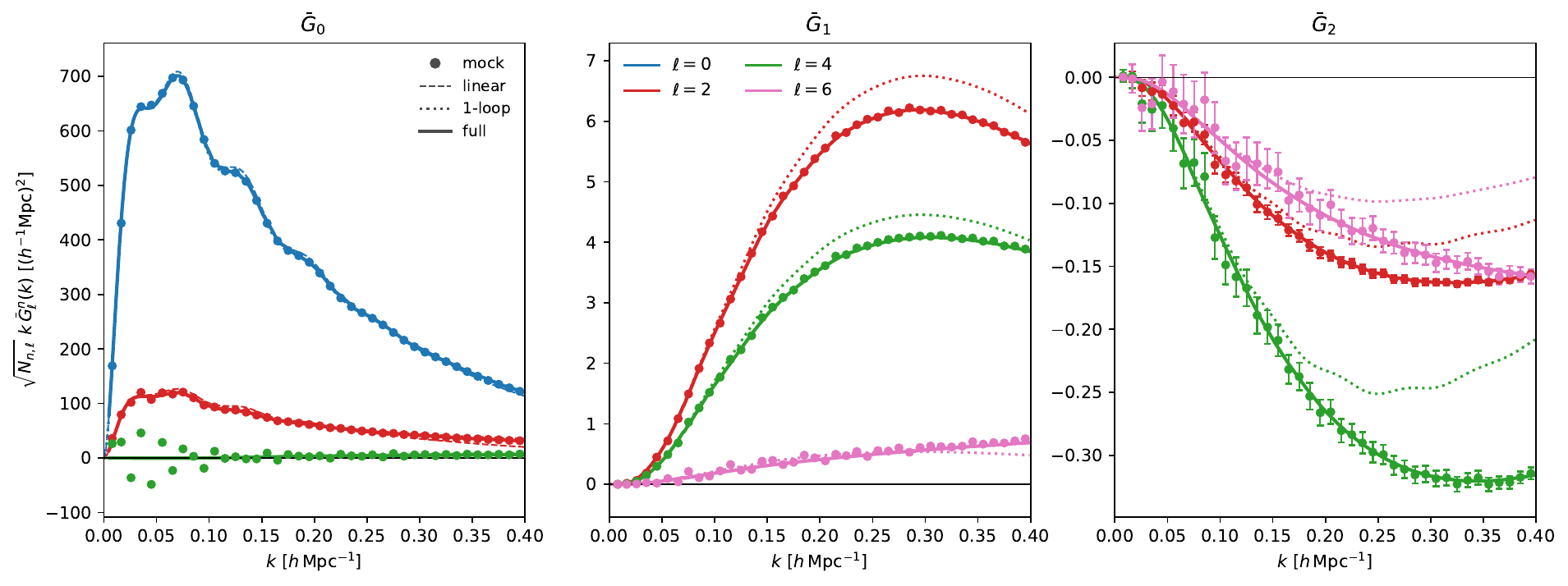}
    \caption{Multipole moments of the form factors of the scalar-tensor cross correlation in the Gaussian streaming model, as measured from simulated mocks (points) and computed to linear (dashed), 1-loop (dotted), and all (solid) orders. All three form factors are generated in this simple model, with $n =1, 2$, corresponding to higher total-angular momentum form factors, generated by nonlinearities at 1-loop order and above.}
    \label{fig:GfSM_cross}
\end{figure}

\begin{figure}
    \centering
    \includegraphics[width=1.0\textwidth]{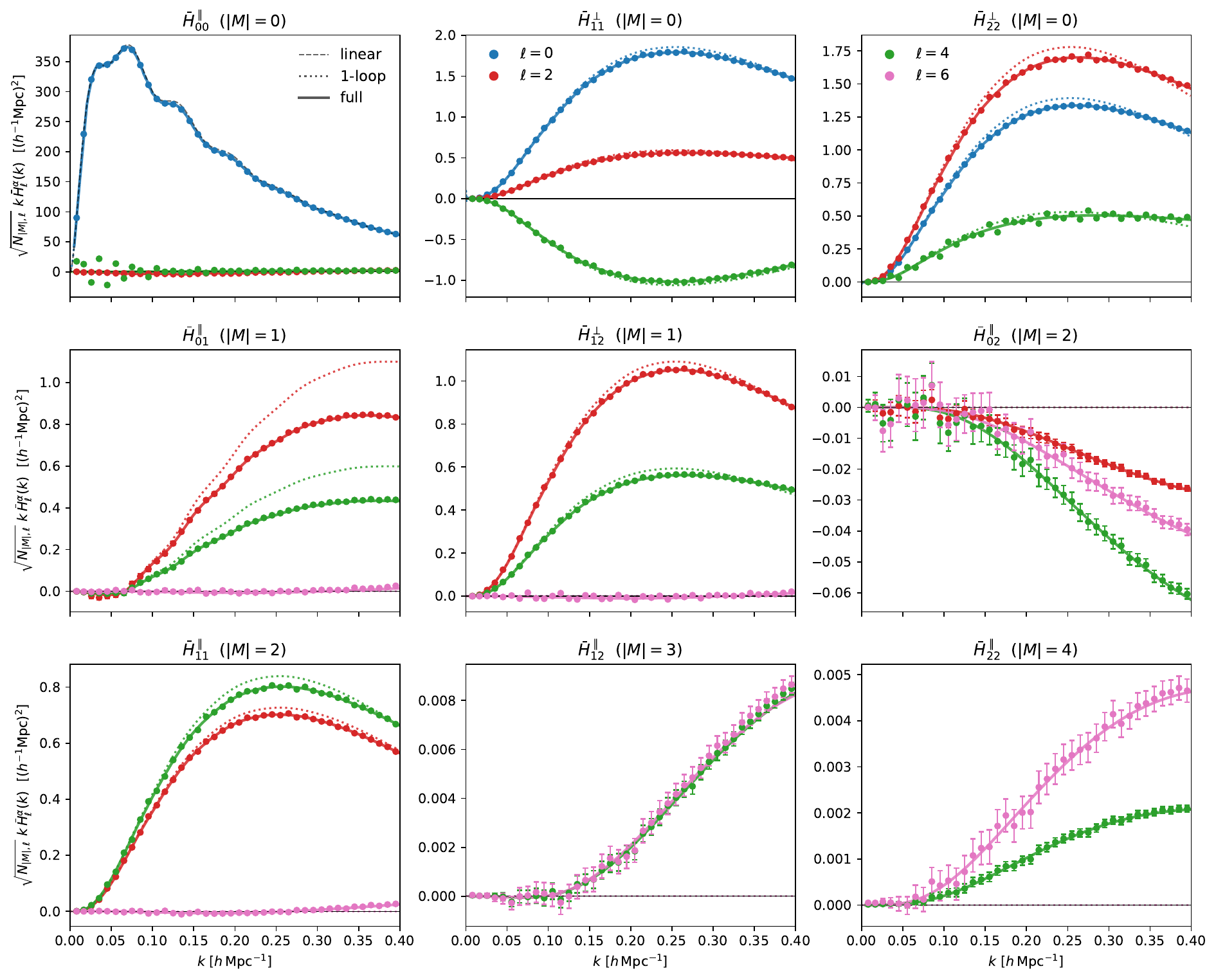}
    \caption{Same as Figure~\ref{fig:GfSM_cross} but for tensor-tensor power spectra. All form factors are similarly generated to be nonzero by the GfSM, but $|M| > 2$ arise only at 2-loop order.}
    \label{fig:GfSM_auto}
\end{figure}

As a test of both the above calculations and the estimators developed in Section~\ref{sec:measurements}, we generate 5000 mocks at $z = 0.5$ with $L=1~\mathrm{Gpc}/h$ and the number of grid being $N_g = 512^3$ within our fiducial cosmology, displacing particles weighted by the linear density and shear fields by their corresponding line-of-sight velocities according to GfSM.
Specifically, we use the field-level forward modeling code developed in Ref.~\cite{Akitsu:2025boy}.
One small caveat of the GfSM mock generation is that we must not apply the CIC interpolation along $x$ and $y$ axis since particles are displaced only along the line-of-sight (i.e., $z$-axis) and stay exactly on the lattice in $x$ and $y$ axis.

Figures~\ref{fig:GfSM_cross} and \ref{fig:GfSM_auto} compare the cross and auto multipole moments measured in these simulations to the analytic results above. The first notable feature is that the GfSM mocks generate clustering in all channels $\mathcal{Q}_n$ and $\mathcal Y^\alpha_{nm}$, including those that cross-correlate unequal helicities required to vanish in real space, as predicted from our calculations above. On the other hand, while nonzero, higher helicity $|M|$ moments are highly suppressed compared to the real-space channels where $M = 0$, even when we plot the multipole moments accounting for the normalization 
\begin{equation}
    N_{m,\ell} = \int_{-1}^1 d\mu \ P^m_\ell(\mu)^2 = \frac{2(\ell + m)!}{(2\ell+1)(\ell-m)!} \, ,
\end{equation}
of the associated Legendre polynomials, as can be readily seen from both plots.

The dashed, solid, and dotted curve in Figures~\ref{fig:GfSM_cross} and \ref{fig:GfSM_auto} show the linear, 1-loop and fully nonlinear theory predictions for the form factors outlined in the previous subsections. In all cases linear theory describes only the $M=0$ form factors. The 1-loop predictions show good qualitative agreement with all form factors $|M| < 3$, with asymptotically better behavior towards lower $k$, while the fully nonlinear prediction agrees with the mocks on all scales and at all angular momenta. This hierarchical behavior is expected since higher helicity modes are necessarily generated by powers of $\vec u_{\hn}$, and therefore must appear at higher loop orders than the channels that already appear in real space.

\begin{figure}
    \centering
    \includegraphics[width=\textwidth]{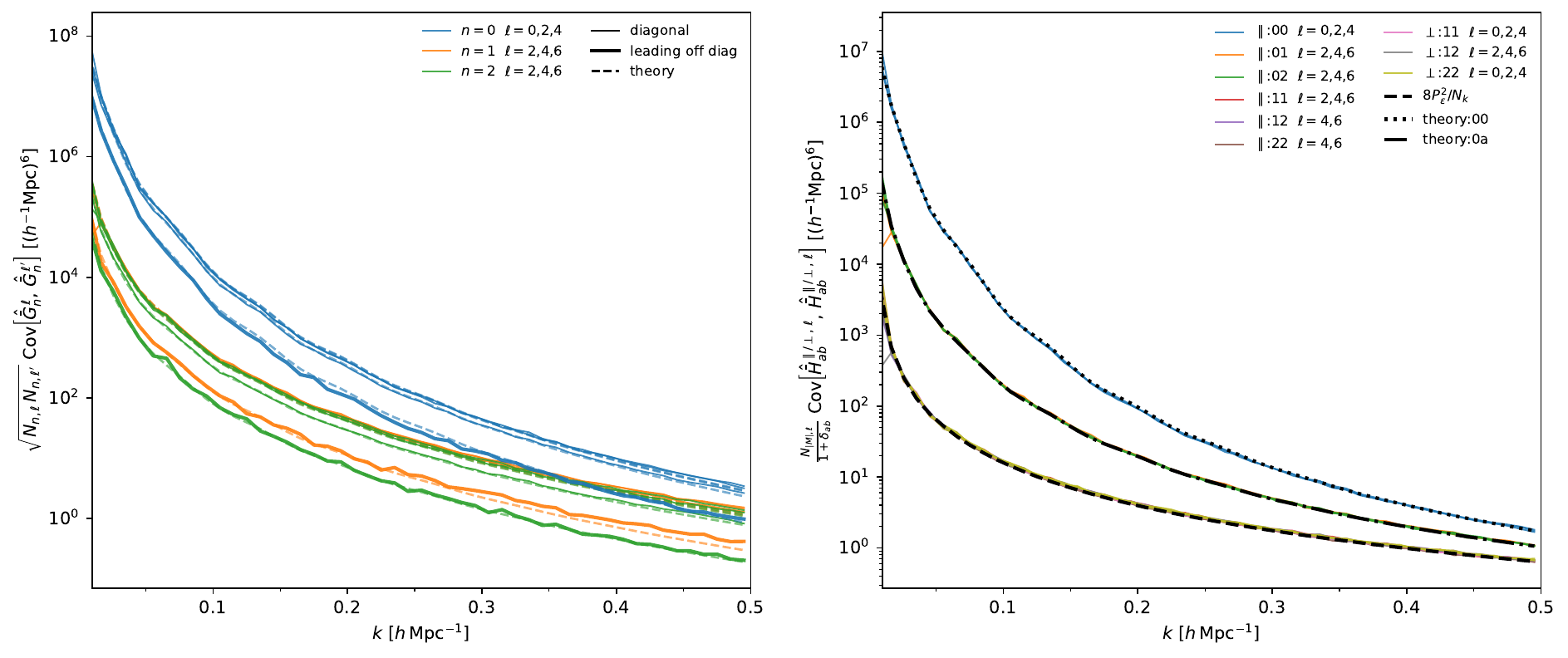}
    \caption{Diagonal and leading off-diagonal (cross between two lowest multipoles) covariances of the scalar-tensor (left) and tensor-tensor (right) power spectra in the Gaussian streaming model, as measured from mock simulations (solid) vs. predictions in the disconnected limit (dashed), which are shown to be in excellent agreement with the former.}
    \label{fig:cov-mocks}
\end{figure}

Finally, Figure~\ref{fig:cov-mocks} shows the covariance matrices for the scalar-tensor and tensor-tensor power spectra as measured from the 5000 mocks. In order to test Eq.~\eqref{eqn:cov_theory}, including the effect of stochasticity, we generated 5000 realizations of shape noise on top of the deterministic mocks above. The covariance in these mocks is in excellent agreement with the theory predictions in the disconnected Gaussian limit, and off-diagonal covariances between different angular momenta $M$ are small or negligible compared to those shown here as expected given that the leading (linear-theory) signal in the 2-point functions only correlates equal angular momenta. The well-behaved-ness of these covariances across multipole moment and $\mu$ is a result of the implicit angular weighting in the normalized bases $\bar{G}, \bar{H}$, which are inverse-variance weightings of the form factors in the Gaussian limit. Had we chosen to measure these spectra in, for example, the un-normalized basis, the higher multipoles and $\mu$ measurements would have diverging covariances due to vanishing factors of $s = \sqrt{1-\mu^2}$. Finally, it is worth noting that the covariances shown in Figure~\ref{fig:cov-mocks} for $|M| > 0$ are significantly larger than the error bars shown in  Figures~\ref{fig:GfSM_cross} and \ref{fig:GfSM_auto} due to the presence of shape noise; in the absence of shape noise these higher angular momenta spectra covariances arise only from 2-point correlations at higher loop order, such that the relative error on the corresponding form factors is comparable to those with lower $M = 0$.

%==========================================================================

%==========================================================================
\section{Measurements from Halos}
\label{sec:halos}

\begin{table}[t]
\centering
\begin{tabular}{lcccc}
\hline
$\log M$ $\left[h^{-1} M_\odot \right]$ & $b_1$ & $c_s$ & $1/\bar{n}$ $\left[ (\text{Mpc}/h)^3 \right]$ & $\sigma_\gamma^2/\bar{n}$ $\left[ (\text{Mpc}/h)^3 \right]$ \\
\hline
$(12.0,13.0)$ & 1.23 & $-0.097$ & 270 & 3.6\\
$(13.0, 14.0)$ & 1.93 & $-0.17$ & 2800 & 34.5\\
\hline
\end{tabular}
\caption{Properties of the low and high mass halo bins.}
\label{tab:halo_properties}
\end{table}

As a final validation of the formalism developed in this paper, we measure the redshift-space statistics of the shapes of halos in fully nonlinear $N$-body simulations. 
We run 15 realizations of $N$-body simulation of volume $1\ (\text{Gpc}/h)^3$ with $1536^3$ particles using the \texttt{2LPTIC}~\cite{Crocce:2006ve} and \texttt{Gadget-4} code~\cite{Springel:2020plp}.
We use the shapes of halos at $z = 0.5$ in two mass bins $\log(M/M_\odot) \in (12.0, 13.0)$ and $(13.0,14.0)$---which we will refer to as the low and high mass bins below---identified via \texttt{Rockstar} halo finder~\cite{Behroozi:2011ju}. 
We adopt the \textit{reduced} inertia tensor as the halo shape and construct the shape field normalized by their mean trace (see Refs.~\cite{Akitsu:2020fpg, Akitsu:2023eqa} for details). 
We measure the statistics of these halos in 15 boxes and list halo properties including the linear density and shape biases and shot and shape noise in Table~\ref{tab:halo_properties}.

\begin{figure}
    \centering
    \includegraphics[width=\linewidth]{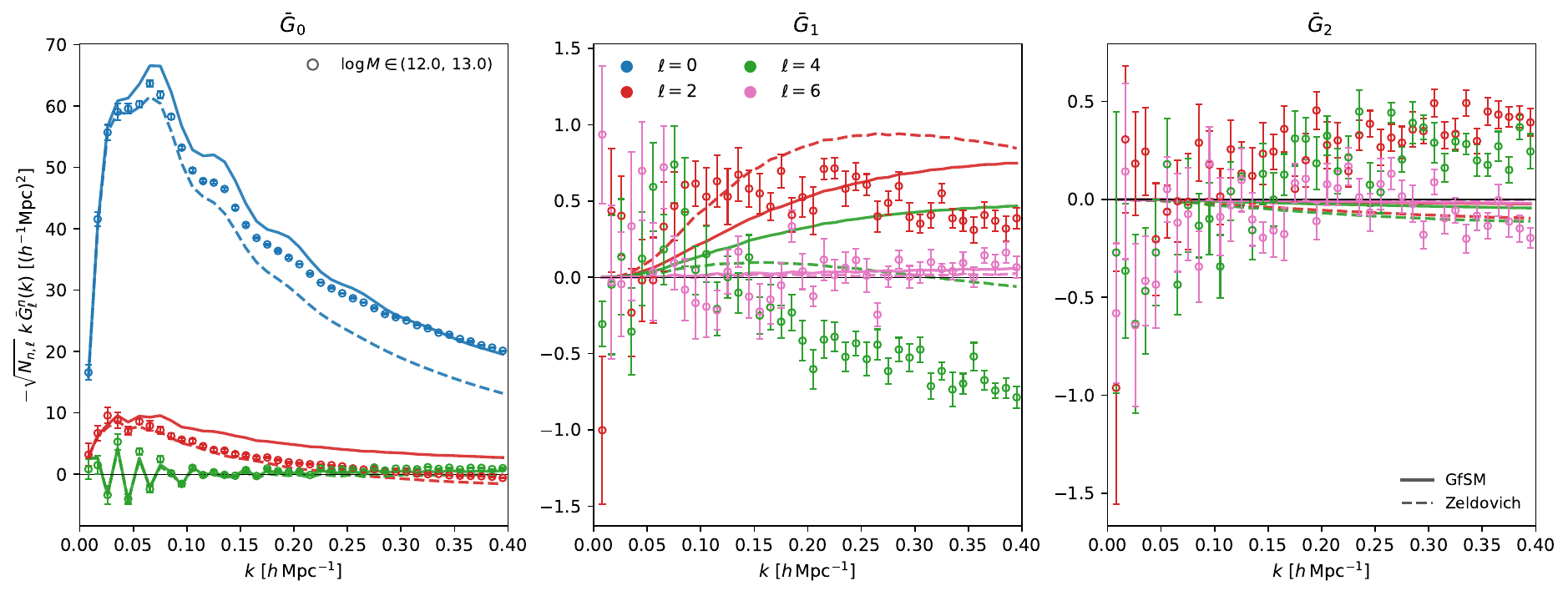}
    \caption{Halo density-shape form factors. For comparison, the solid and dashed lines show predictions from the GfSM and Zeldovich models. Thicker (solid) theory curves indicate fits of the two models with leading counterterms included.}
    \label{fig:cross-halos}
\end{figure}

\begin{figure}
    \centering
    \includegraphics[width=\linewidth]{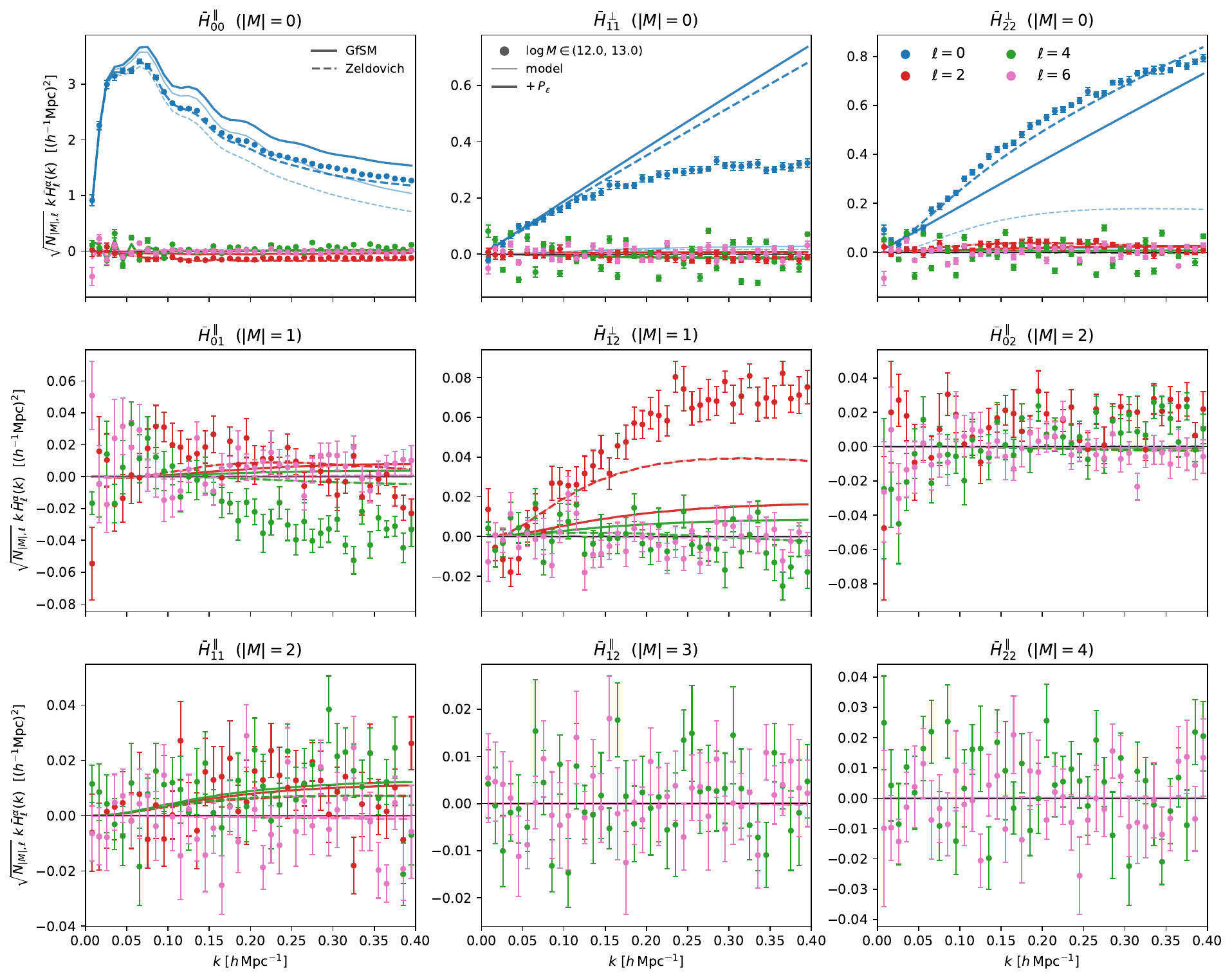}
    \caption{Halo shape-shape form factors. For comparison, the solid and dashed lines show predictions from the GfSM and Zeldovich models. Thicker (solid) theory curves indicate fits of the two models with leading counterterms and stochastic contributions included.}
    \label{fig:auto-halos}
\end{figure}

\begin{figure}
    \centering
    \includegraphics[width=\linewidth]{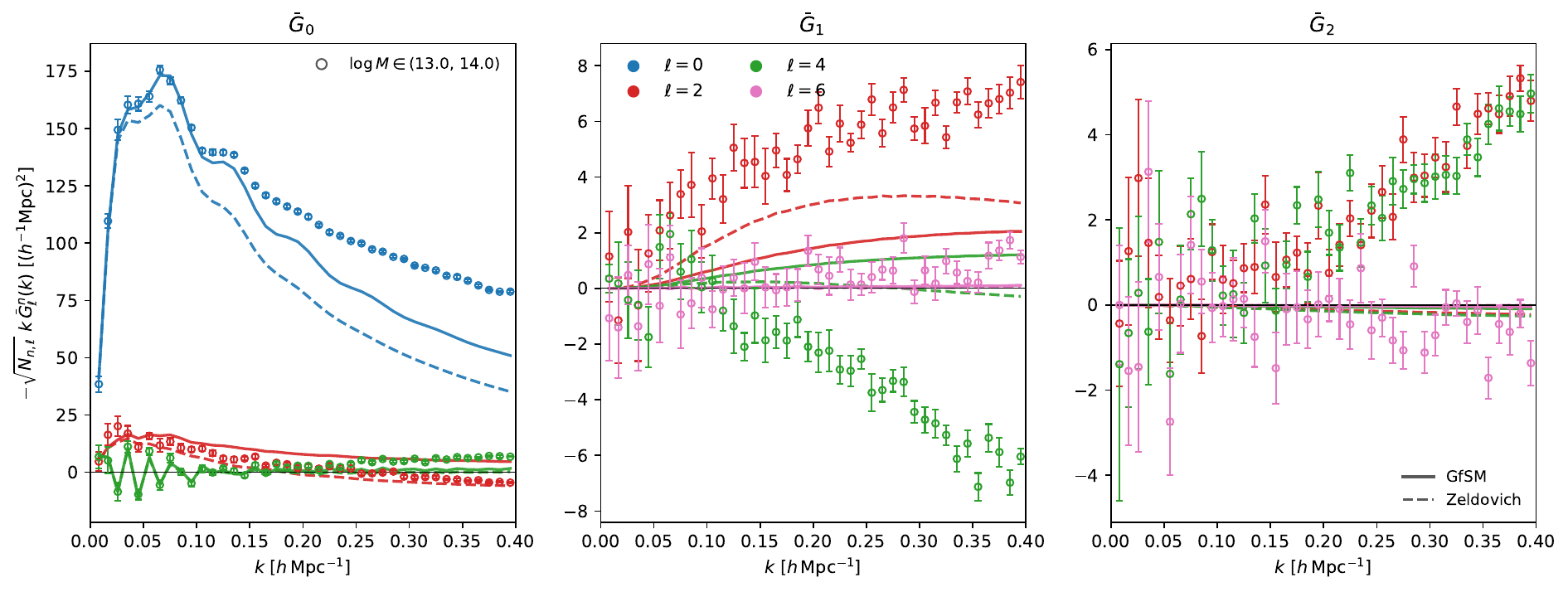}
    \caption{Same as Figure~\ref{fig:cross-halos} but for the higher mass halo sample.}
    \label{fig:cross-halos-highmass}
\end{figure}

\begin{figure}
    \centering
    \includegraphics[width=\linewidth]{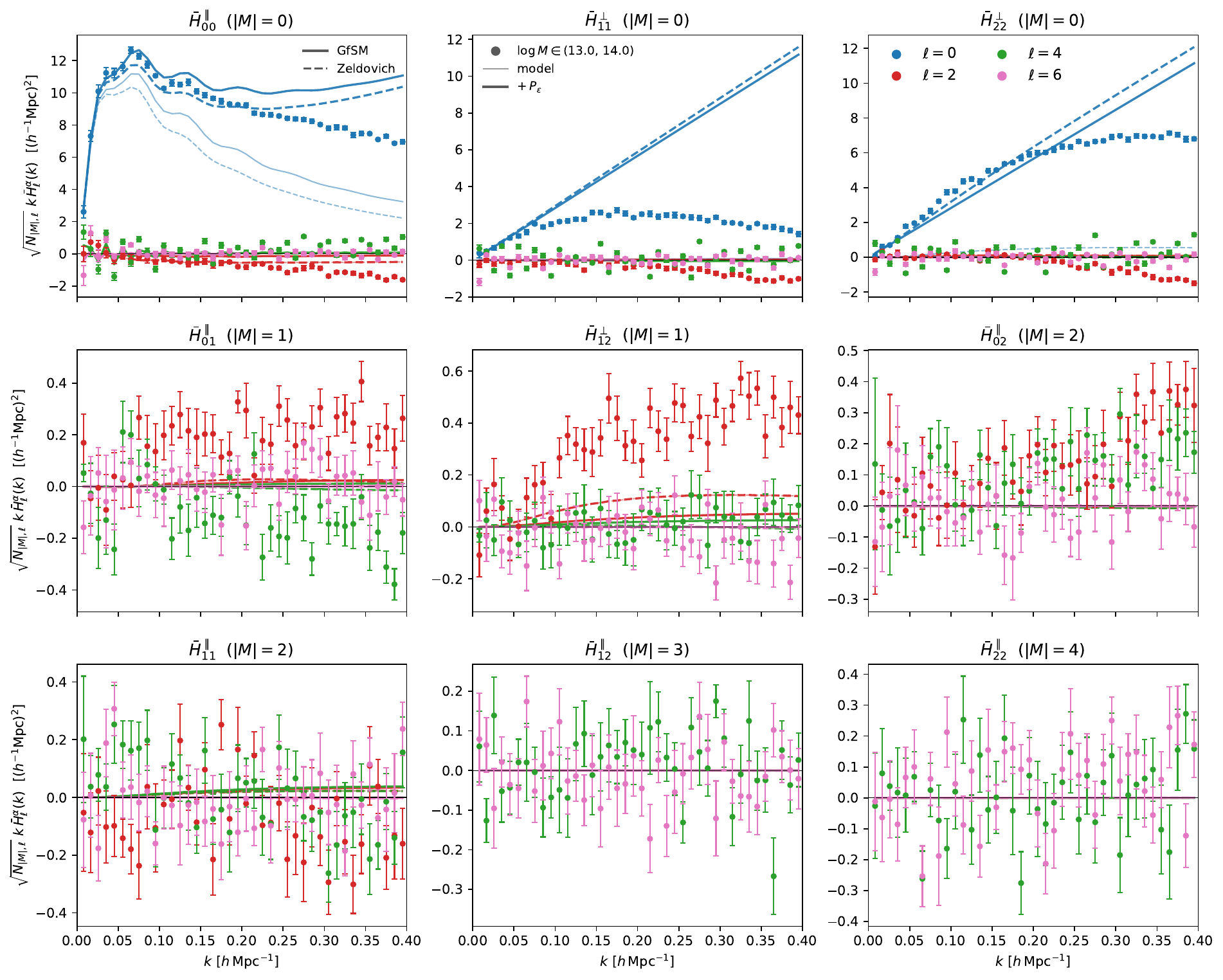}
    \caption{Same as Figure~\ref{fig:auto-halos} but for the higher mass halo sample.}
    \label{fig:auto-halos-highmass}
\end{figure}

Figures~\ref{fig:cross-halos} and \ref{fig:auto-halos} show the halo density and shape power spectra measured in the $\mathcal{Q}, \mathcal{Y}$ bases for halos in the low mass bin. We obtain significant detections of nonzero form factors for every channel in the auto and cross spectra except for the highest angular momentum channels $H^\parallel_{12},\, H^\parallel_{22}$. For comparison, in both figures we show the predictions of two simple models of halo density and shape cross correlations, with linear density and shape biases set by measuring the cross correlation between the density and shape fields with the initial conditions: the GfSM model, as described in the previous section and shown as solid lines, and a Zeldovich model, shown via dashed lines where both the galaxy density and shape fields have their perturbations sourced by the initial conditions but are then advected to their final positions by both the linear theory velocity \textit{and} displacement $\Psi = -\nabla^{-1} \delta_0$. In both cases, we subtract off the shape noise contribution to the $M = 0$ autospectrum form factors. In both cases, the Zeldovich model does a somewhat better job reproducing the observed form factors beyond $M=0$ on large scales, particularly the higher signal-to-noise measurements of $\bar{G}_1$ and $\bar{H}^\perp_{12}$, but neither is successful in reproducing the sign and amplitude of all form factor channels simultaneously. Figures~\ref{fig:cross-halos-highmass} and \ref{fig:auto-halos-highmass} show the equivalent plots for the high mass bin. While the results are quantitatively similar, the form factor amplitudes in the high mass bin, when detected, are notably higher than the predictions of the simple models, highlighting the importance of accounting for higher-order mode coupling and velocities.

\begin{figure}
    \centering
    \includegraphics[width=\linewidth]{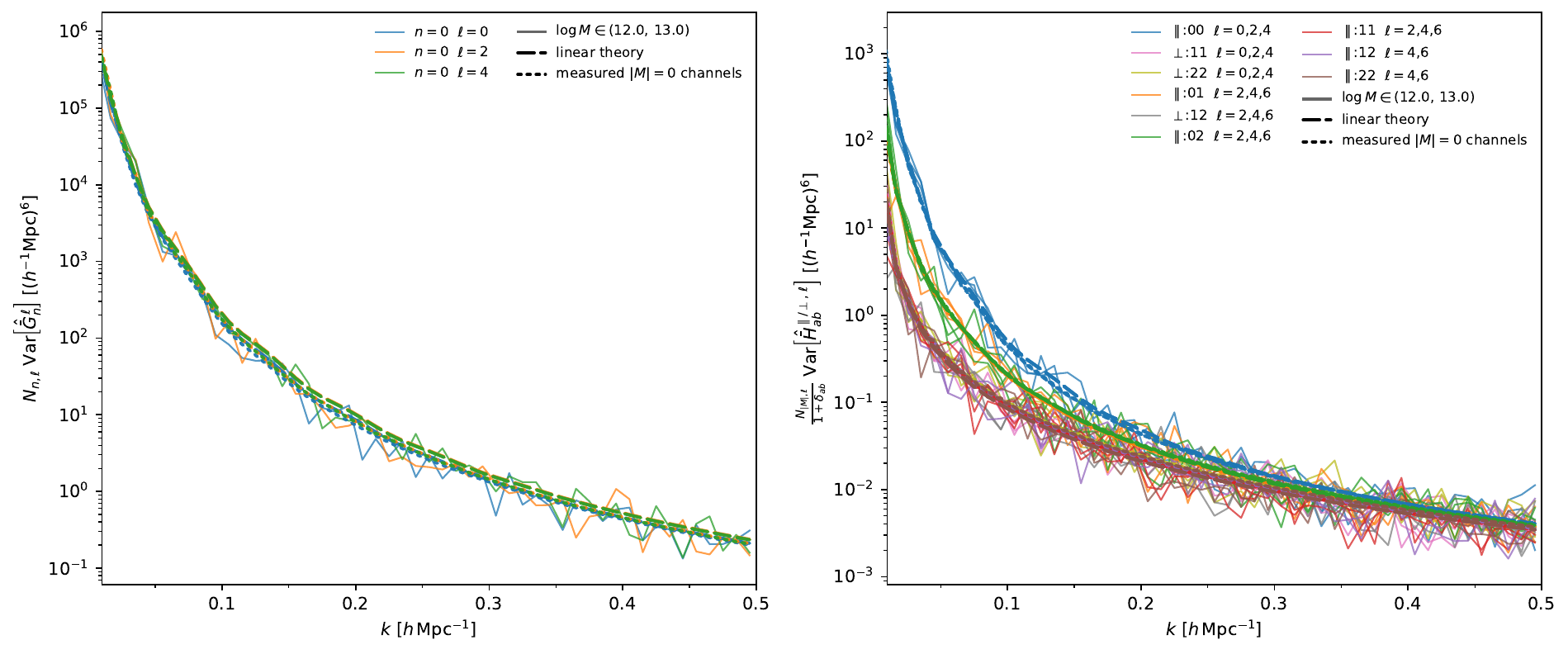}
    \caption{Same as Figure~\ref{fig:cov-mocks} but for simulated halos with $\log M \in (12.0, 13.0)$. The disconnected covariance remains an excellent approximation even for fully nonlinear halos, even when only linear theory predictions are used for the input power spectra. The tensor-tensor covariance at high $k$ (i.e. in the shot-noise dominated limit) converges across form factors as expected.}
    \label{fig:cov-halos}
\end{figure}

\begin{figure}
    \centering
    \includegraphics[width=\linewidth]{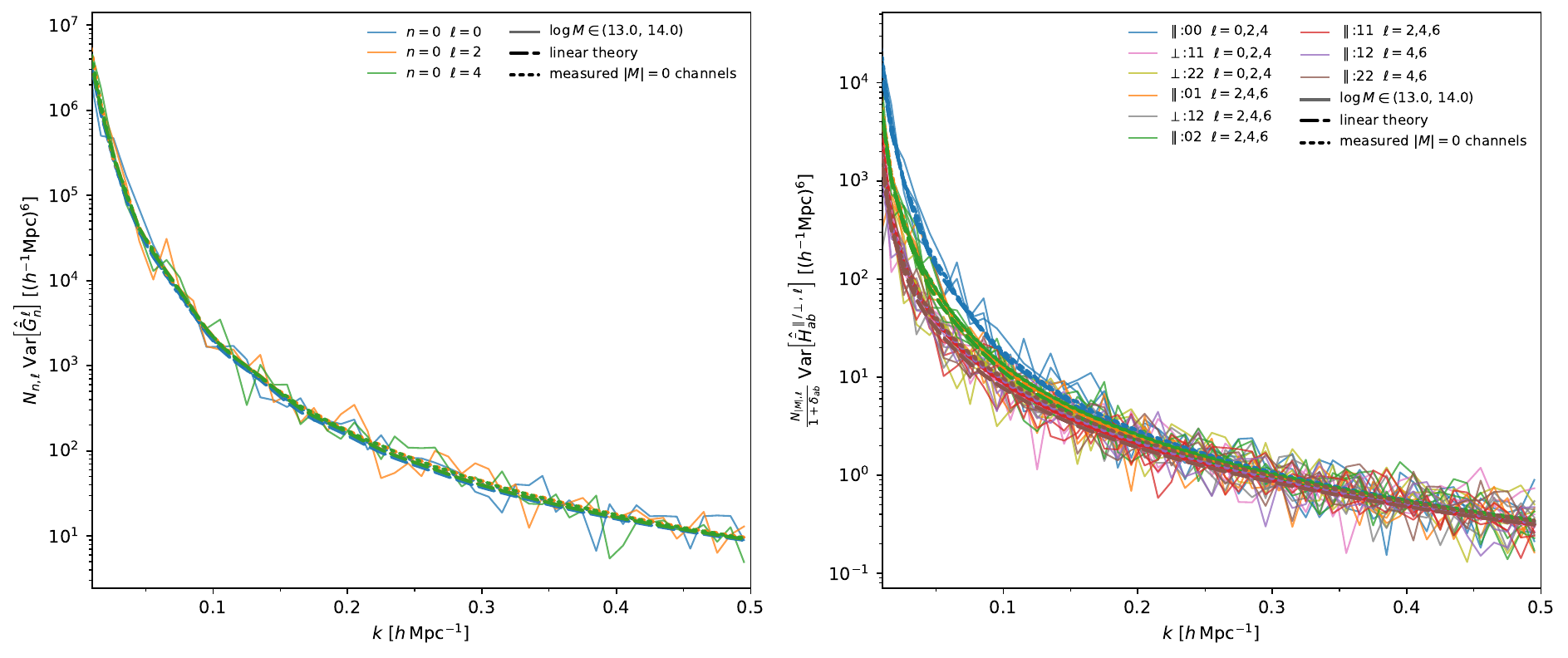}
    \caption{Same as Figure~\ref{fig:cov-halos} but for the higher mass halo sample.}
    \label{fig:cov-halos-highmass}
\end{figure}

Finally, Figures~\ref{fig:cov-halos} and \ref{fig:cov-halos-highmass} show the covariance of the scalar-tensor and tensor-tensor form factors of halos discussed above, compared to the disconnected predictions using linear theory (long-dashed) and measured form factors (dashed). The measured covariances are in excellent agreement with theoretical expectations, with the not-shown off-diagonal covariance elements consistent with zero in all cases. Unlike in Figure~\ref{fig:cov-mocks}, the various curves converge much faster due to the dominance of the shape noise in halo shape statistics.

%==========================================================================
\section{Summary and Conclusions} 
\label{sec:conclusion}

The large-scale statistics of galaxy shapes are emerging as a powerful cosmological observable, both for calibrating the contribution of intrinsic alignments in state-of-the-art weak galaxy lensing surveys and, since they are \textit{tensor} tracers of large scale structure, as unique probes of new physics beyond the scalar sector, including inflationary dynamics, gravitational waves, and parity violation. While often studied in weak lensing surveys as a projected observable, galaxy shapes are inherently three-dimensional, and our aim in this paper was to establish a general framework to analyze the observed statistics of galaxy shapes---and tensor observables in general---in 3D, allowing for the breaking of rotational symmetry by the observer's line of sight. This symmetry breaking exists in essentially all three-dimensional cosmological observables due to so-called \textit{redshift-space distortions}, perturbations to objects' observed positions due to Doppler shifts from their peculiar velocities. To this end, we have focused on the structural consequences of the nonlinear redshift-space mapping, particularly the role of symmetries, while separating out properties due to particular dynamical models (e.g. the effective field theory of large-scale structure) to an upcoming work. Indeed, we stress that the framework constructed in this paper is equally valid and relevant to analytic and simulations-based studies of galaxy intrinsic alignments, in the sense that we provide an optimal and model-agnostic framework to describe the statistics of galaxy shapes. Our findings can be summarized as follows

\begin{itemize}
    \item \textit{Kinematic structure, symmetries and form factors.} After reviewing the nonlinear redshift space mapping and selection rules due to homogeneity, isotropy, parity and exchange in Sections~\ref{sec:rsd} and \ref{sec:symm}, we show how the redshift-space mapping generates a kinematic basis $\mathcal{M}$ describing the tensor structure of scalar-tensor and tensor-tensor correlators via form factors multiplying products of $\hk, \hn, \delta_{ij}$. Importantly, the form factors in this basis are explicitly polynomial in the line-of-sight angle $\mu = \hk \cdot \hn$, terminating at finite order corresponding to the power of the velocity field sourcing redshift-space distortions, and indirectly to the perturbative order in effective-theory calculations. We show that this construction leads to $3$ and $13$ independent form factors, respectively, with the latter reducing to $9$ for correlators even under the exchange operator, like auto-correlations.

    \item \textit{Selection rules, spin weights, and associated Legendre polynomials.} Galaxy shape correlators can also be expressed in terms of a parity-adapted basis via rank-2 basis tensors $\mathcal Q$ constructed from triads based on the $\hk, \hn$ plane. These $\mathcal Q$ are closely related to spin-2 spherical harmonic tensors $Y^{(m)}_{ij}$ built from the same triads, except that all but one vanish in the collinear $\mu \rightarrow 1$ limit where $\hk$ and $\hn$ are degenerate. In this basis, the 3 and 13 independent form factors enumerated above are clearly seen to be due to the parity, and the reduction to 9 due to the exchange selection rules. 
    
    The parity-adapted basis preserves the polynomiality of the kinematic basis for the scalar-tensor form factors; however, it fails to do so in the tensor-tensor sector. We show that polynomiality is preserved in a \textit{total-helicity basis}, $\mathcal{Y}$, given in terms of the parity-adapted basis or, equivalently, parity-even combinations of spherical harmonics $Y^{m}_{ij} Y^{m'}_{kl}$, multiplied by spin weights $s^{|M|} = (1 - \mu^2)^{|M|/2}$, where $|M|$ is the total angular momentum along $\hk$ of the tensor basis. In Section~\ref{sec:spin_weights}, we show that this spin weight counts the symmetry-breaking powers of $\hn$ in redshift-space mapping, and similarly explains the scalar-tensor case, which was already a total-helicity basis by construction. Our results are equivalent to the statement that, expressed in terms of a \textit{normalized} basis (e.g. $\bar{\mathcal Y}$), form factors corresponding to basis elements with total angular momentum $|M|$ can be expressed as a finite series of associated Legendre polynomials $P^{|M|}_\ell$ up to a given order in velocity. The $1$ and $3$ spectra with $|M| = 0$, which can be produced with no insertions of $\hn$, correspond to the nonzero equal-helicity spectra in real space.

    \item \textit{Estimators and projections.} Finally, we construct estimators for tensor-correlator form factors derived in the previous sections based on projecting tensor fields into the \textit{un-normalized} helicity basis. We derive the covariance for these estimators and show that they are diagonal to leading order in the total-helicity basis. We further show that the \textit{normalized} total-helicity basis, and the associated Legendre multipole moments, correspond to the optimal weighting in $\mu$ for the form factors, and write down their covariance. As a final connection to observations of projected galaxy shapes, we write down the projections of each form factor into shape $E$- and $B$ modes, which map exclusively from the parity-even and -odd components of the $\mathcal Q$ basis, respectively. 
\end{itemize}

In order to validate our formalism, we construct mocks in a toy model, the ``Gaussian-field Streaming Model'' (GfSM), where galaxy densities, shapes and velocities are described by the Gaussian-field limit, but the redshift-space mapping is fully nonlinear. We analytically compute the form factors in the GfSM and show that they are indeed polynomial in the chosen bases, and moreover that all enumerated form factors are produced in this simple model. We further generate mock catalogs in the GfSM to confirm our calculations and show that the estimators and covariance behave as expected. As a final validation, we measure the shapes of halos in N-body simulations and find that we detect all scalar-tensor and tensor-tensor channels with $|M| \leq 2$, with only the two highest $|M|$ form factors, which are highly suppressed by their spin weights, undetected beneath the shape noise.

Our work suggests several immediate applications and extensions. First and foremost, our derivations establish the normalized, total-helicity basis as the basis-of-choice for summarizing three-dimensional galaxy shape information at any scale, carrying higher signal-to-noise and more complete tensorial information beyond what can be measured from projected $E$ and $B$ modes. This makes it an especially powerful tool to extract intrinsic-alignment information from e.g. hydrodynamical simulations with realistic galaxy formation physics, particularly in comparison to measurements from spectroscopic surveys. From the analytic side, our basis will be indispensable in organizing perturbative contributions to galaxy shape statistics in 3D in the effective field theory (EFT) of large-scale structure. Our comparisons of the GfSM and Zeldovich models of shape statistics to the measured shapes of halos, especially at the high-mass end, suggest that linear biasing and dynamics are insufficient to explain the measured density-shape and shape-shape spectra even when nonlinear redshift-space mapping itself is carried to arbitrarily high order, highlighting the need for nonlinear dynamics and galaxy bias. On the flip side, the rich tensor structure of our formalism should enable tighter constraints on EFT parameters than can be achieved using projected statistics or in real space. We will explore these dynamical issues in Paper II.

Finally, while we have focused on tensor contributions from the redshift space mapping in this paper, the bases we have constructed are fully general and could be used to isolate new physics. One particularly fruitful avenue may be to look for channels that are forbidden by selection rules---for example, while we have focused on parity-even channels in this paper, any parity-violating process could generate the additional parity-odd channels ($2$ and $12$ in the scalar-tensor and tensor-tensor cases), which would act as uncontaminated probes of new physics. Equally, if nonlinear corrections to the higher angular-momentum form factors are small, as is suggested by our toy models and halo measurements, they too could act as clean probes of new, tensor-mediated physics, assuming that we can use EFT predictions to form a theoretical error due to standard mode coupling. More pragmatically, we can also investigate effects beyond the plane-parallel approximation and non-relativistic limit, e.g. via odd multipoles, exchange-odd channels coming from cross correlations between distinct samples, or line-of-sight selection effects that involve further contractions with the line-of-sight vector. Finally, our formalism can be applied to other tensor observables beyond galaxy shapes, including vectors (peculiar velocities) and pseudovectors (galaxy angular momenta), and new, more complete probes of galaxy shapes such as kinematic lensing. We intend to return to all these topics in future work.

%==========================================================================
\begin{acknowledgments}
Support for this work was provided by NASA through
the NASA Hubble Fellowship grant HST-HF2-51572.001
awarded by the Space Telescope Science Institute, which
is operated by the Association of Universities for Research in Astronomy, Inc., for NASA, under contract
NAS5-26555. Z.V. acknowledges the support of the Croatian Science Foundation (grant number IP-2025-02-1338).
Numerical computation was carried out on the Cray XD2000 at the Center for Computational Astrophysics, National Astronomical Observatory of Japan. This work was performed in part at the Aspen Center for Physics, which is supported by National
Science Foundation grant PHY-2210452. 
\end{acknowledgments}  

\appendix
%====================================================%

%====================================================
\section{Direct tensor decomposition of tensor correlators}
\label{app:direct_tensor}

Here, we construct a general decomposition algorithm for scalar-tensor and tensor-tensor two-point correlators. The idea is to rely on underlying correlator symmetries (such as translation, rotational, and parity invariance) to construct the tensor basis in which any given tensor correlator can be represented. Before continuing to the general construction of our RSD-related tensors, we cover a few simple cases to motivate the general procedure. 

There is a general structure to how statistically isotropic (and parity-invariant) tensor correlators of higher spin $L$ (i.e., with $L$ indices) can be decomposed. These correlators are functions of a single vector $\vec{r}$, so their index structure can only be built from the unit vector $\hat{r}_i \equiv r_i / r$ and the Kronecker delta $\delta_{ij}$, and combinations of these that are symmetric, parity-even/odd, and transform correctly under rotations. For any rank-$L$ tensor correlator, statistical isotropy and parity invariance implies
\eeq{
\xi_{i_1 \ldots i_L}(\vec{r}) = \sum_{n=0}^{\lfloor L/2 \rfloor} A_n(r) \cdot \mathcal{S} \left[ \delta_{(i_1 i_2} \cdots \delta_{i_{2n-1} i_{2n}} \hat{r}_{i_{2n+1}} \cdots \hat{r}_{i_L)} \right],
\label{eq:xi_L_symmetric}
}
where $A_n(r)$ are scalar functions of $r = |\vec{r}|$ and $\mathcal{S}$ denotes symmetrization over all indices (and removal of traces if the correlator is traceless). The terms are built from $n$ Kronecker deltas (which eat up $2n$ indices) and $L - 2n$ unit vectors $\hat{r}_i$.  The expression is parity-even/odd and rotationally covariant. 

For spin 1 (vector) we thus have
\eeq{
\xi_i(\vec{r}) = A(r) \hat{r}_i \, ,
}
while for spin 2 (rank-2 tensor) we have
\eeq{
\label{eq:quadrupole}
\xi_{ij}(\vec{r}) = A(r)\delta_{ij} + B(r)\hat{r}_i \hat{r}_j \, .
}
For spin 3 we can construct symmetric combinations of three $\hat{r}_i$’s and 1 $\delta_{ij} \hat{r}_k$ to get
\eeq{
\xi_{ijk}(\vec{r}) = A(r)\hat{r}_i \hat{r}_j \hat{r}_k + B(r)\left( \delta_{ij} \hat{r}_k + \delta_{ik} \hat{r}_j + \delta_{jk} \hat{r}_i \right) \, .
}
Similarly, for spin 4 we have
\eeq{
\xi_{ijkl}(\vec{r}) = A(r)\hat{r}_i \hat{r}_j \hat{r}_k \hat{r}_l + B(r)\left(\delta_{ij} \hat{r}_k \hat{r}_l + \cdots \right) + C(r)\left(\delta_{ij} \delta_{kl} + \cdots \right) \, .
}
The general expansion uses all possible symmetric, parity-even combinations of $\hat{r}_i$ and $\delta_{ij}$, subject to rotational covariance.

This construction procedure allows us to count independent terms. For a rank-$L$ symmetric tensor correlator built from $\hat{r}_i$ and $\delta_{ij}$,  the number of independent scalar functions $A_n(r)$ that appear is $M(L) = \lfloor L/2 \rfloor + 1$. That is, all even partitions of $L$ indices into delta functions and remaining indices into $\hat{r}_i$'s. The geometric (in contrast to intrinsic) parity of these correlators is completely determined by the number of $\hat r$ terms.

Often in cosmology we also require $\xi_{i_1\cdots i_L}$ to be traceless and symmetric (i.e., an irreducible representation of SO(3)).  Then, the expansion needs to include only traceless symmetric combinations, which reduces the number of terms. For instance, for spin 2 we then have  $ {\rm TF} \left[ \xi_{ij} \right] = A(r) \left( \hat{r}_i \hat{r}_j - \frac{1}{3}\delta_{ij} \right)$ and similarly for higher spin, one can build traceless symmetric tensors from $\hat{r}_i$ and subtract traces appropriately.

As a further example, let us examine the cases of cross-correlating the scalar, vector and the tensor field of rank two.  We consider first the three-index correlators $\xi_{ijk}(\vec r)= \la \phi(\vec{0})\,T_{ij}(\vec r)\,V_k(\vec r)\ra$ and, as we have noted earlier, the expansion for the $\la V_k(\vec 0) \,T_{ij}(\vec r) \ra$ correlator is identical. Under homogeneity, isotropy, and no Levi-Civita contributions (forbidden by the $i\leftrightarrow j$ symmetry), both are rank-3 tensors symmetric only in $i\leftrightarrow j$ and built from $\delta_{ij}$ and $\hat r_i$. The most general (geometrically) parity-odd, $i\leftrightarrow j$ symmetric form is
\eeq{
\xi_{ijk}(r)
= A_1(r)\,\hat r_i\hat r_j\hat r_k + A_2(r)\,\delta_{ij}\,\hat r_k + A_3(r)\,\big(\delta_{ik}\hat r_j+\delta_{jk}\hat r_i\big)\, .
}
If $T_{ij}$ is traceless we can impose $\delta^{ij}\xi_{ijk}=0$ condition. Tracing the general form gives $\delta^{ij}\xi_{ijk} =\lb A_1+3A_2+2A_3\rb \,\hat r_k=0$ and thus $A_1+3 A_2+2 A_3=0$, so only two independent functions remain. It is then convenient to use an explicitly traceless basis
\eeq{
\xi_{ijk}(r)
= A(r)\,\big(\hat r_i\hat r_j-\tfrac13\delta_{ij}\big)\hat r_k
+ B(r)\,\Big(\delta_{ik}\hat r_j+\delta_{jk}\hat r_i-\tfrac{2}{3}\delta_{ij}\hat r_k\Big)\, .
}

Lastly, we consider the case with two vector fields and one tensor field. As an example we can consider the correlator 
\eeq{
\xi_{ijlm}(\vec r)\equiv \big\langle V_l(\vec x)\,V_m(\vec x)\,T_{ij}(\vec x+\vec r)\big\rangle \, ,
}
where there is algebraic symmetry $i\leftrightarrow j$ (from $T_{ij}$) and also $l\leftrightarrow m$ since $V$'s are identical. We thus allow rank-4 tensors made from $\delta$'s and $\hat r_i$ with even powers of $\hat r$ and symmetric in $i\leftrightarrow j$ and $l\leftrightarrow m$. There are six terms
\eq{
\xi_{ijlm}(r) =
A_1(r)\,\hat r_i\hat r_j\hat r_l\hat r_m
& + A_2(r)\,\delta_{ij}\,\hat r_l\hat r_m
+ A_3(r)\,\delta_{lm}\,\hat r_i\hat r_j \\
& + A_4(r)\,\Big(\delta_{il}\hat r_j\hat r_m+\delta_{im}\hat r_j\hat r_l+\delta_{jl}\hat r_i\hat r_m+\delta_{jm}\hat r_i\hat r_l\Big) \non\\
& + A_5(r)\,\delta_{ij}\,\delta_{lm}
+ A_6(r)\,\big(\delta_{il}\delta_{jm}+\delta_{im}\delta_{jl}\big) \, . \non
}
This basis is already symmetric under $l\leftrightarrow m$, given that our two vectors are identical fields, that symmetry is automatic. If the two vectors are different fields, the correlator need not be symmetric under $l\leftrightarrow m$, but parity evenness plus $i\leftrightarrow j$ symmetry leaves the same even-parity basis above (no additional parity-even, $i\leftrightarrow j$-symmetric structure exists that is odd under $l\leftrightarrow m$).

Moreover, if we impose $\delta^{ij}\xi_{ijlm}=0$. we get two independent constraints (because $\hat r_l\hat r_m$ and $\delta_{lm}$ are independent rank-2 tensors)
\eq{
&\text{Coeff of } \hat r_l\hat r_m:\quad A_1+3A_2+4A_4=0\, ,\\
&\text{Coeff of } \delta_{lm}:\quad A_3+3A_5+2A_6=0 \, . \non
}
Hence only four independent scalar functions remain.
It’s convenient to switch to an explicitly traceless basis in the $ij$ indices. One clean choice is:
\eq{
\xi_{ijlm}(r) =&
B_1(r)\,\Big(\hat r_i\hat r_j-\tfrac13\delta_{ij}\Big)\hat r_l\hat r_m
+ B_2(r)\,\Big(\hat r_i\hat r_j-\tfrac13\delta_{ij}\Big)\delta_{lm} \\
&+ B_3(r)\,\Big[\delta_{il}\hat r_j\hat r_m+\delta_{im}\hat r_j\hat r_l+\delta_{jl}\hat r_i\hat r_m+\delta_{jm}\hat r_i\hat r_l
-\tfrac{4}{3}\delta_{ij}\hat r_l\hat r_m\Big] \non\\
&+ B_4(r)\,\Big[\delta_{il}\delta_{jm}+\delta_{im}\delta_{jl}-\tfrac{2}{3}\delta_{ij}\delta_{lm}\Big] \, \non .
}
Each bracket is symmetric in $i\leftrightarrow j$ and traceless in $ij$. It is also easy to see how this procedure directly leads to the usual real space tensor tensor-tensor results used in \cite{Vlah2020} and obtained using the decomposition into the irreducible spherical tensors, as mentioned in Sec.~\ref{sec:irreducible_cartesian_basis}. 

%------------------------------------------------------%
\subsection{Scalar-Tensor Correlator}

Let us now give a general procedure for the decomposition of the correlator of one symmetric traceless rank-2 tensor field $T_{ij}$and $L$ identical vector fields $V_{i_1},\dots,V_{i_L}$ that is statistically homogeneous and isotropic (and no $\eps$). When cross-correlating the scalar and shape fields in redshift space (see Sec.~\ref{sec:rsd}), we are interested in correlators of type 
\eeq{
\big\langle \phi(\mathbf x) V_{i_1}\cdots V_{i_{L'}}(\mathbf x)\,  V_{i_{L'+1}} \cdots V_{i_L}(\mathbf x + \mathbf r) \,T_{ij}(\mathbf x+\mathbf r)\big\rangle\, ,
}
where $L'\leq L$, and $\phi(\mathbf x)$ is a scalar field (which is present here so that the $L'=0$ case is not trivial).  Since we eventually use fully symmetric contractions in the $L$ vector indices, e.g. $N_{i_1\cdots i_L}=N_{(i_1\cdots i_L)} = \hat n_{i_1}\ldots \hat n_{i_L}$, we may replace any correlator $\xi_{i_1\cdots i_L;ij}$ by its totally symmetrised projection in those indices, 
\eeq{
\xi_{(i_1\cdots i_L);ij}\equiv \frac{1}{L!}\sum_{\pi\in S_L}\xi_{i_{\pi(1)}\cdots i_{\pi(L)};ij}\, ,
}
since $N_{i_1\cdots i_L}\xi_{i_1\cdots i_L;ij}=N_{i_1\cdots i_L}\xi_{(i_1\cdots i_L);ij}$. Once symmetrized, all of these correlators share the underlying symmetries in $L$ indices and $i$ and $j$, and thus, according to the discussion presented in the Sec.~\ref{sec:symm} we expect them to share the same structure in terms of the decomposition into the basis tensors. Consequently, we can just have a look at the decomposition of the representative correlators of type
\eeq{
\label{eq:V(N)T}
\xi_{i_1\cdots i_L;\,ij}(\mathbf r)
\;\equiv\;
\big\langle \phi(\mathbf x) V_{i_1}\cdots V_{i_L}(\mathbf x)\,T_{ij}(\mathbf x+\mathbf r)\big\rangle\, ,
\qquad T_{ij}=T_{ji}\, ,\quad T_{ii}=0\, ,
}
into a linear combination of isotropic tensors made only from $\delta_{ij}$ and $\hat r_i=r_i/r$, with scalar functions of $r=|\mathbf r|$. In our representative correlator all $V_i$'s are at the same point, hence symmetric.  The index symmetries enforce symmetry of the $L$ vector indices $(i_1,\ldots,i_L)$ among themselves (identical vectors at the `same' point), symmetry $i\leftrightarrow j$ for $T_{ij}$ and tracelessness on $i,j$: contraction with $\delta_{ij}$ gives zero.

We start by enumerating isotropic structures before taking the trace.
We build all the tensor structures from:
\begin{itemize}
\item $a$ V–V $\delta$ pairings among the $L$ vector indices,
\item $b\in\{0,1\}$ T–T $\delta$ (i.e. $\delta_{ij}$),
\item $c\in\{0,1,2\}$ V–T $\delta$ pairings that connect distinct V indices to $i$ and/or $j$,
\end{itemize}
These are subject to the index-budget constraints $2 a + c \le L$, and $2 b + c \le 2$. All unpaired indices (among $i_1,\dots,i_L$ and $i,j$) are carried by factors of $\hat r$.  Then symmetrize over the $L$ V-indices and over $i\leftrightarrow j$. We denote the resulting set of basis tensors by $\{S^{(p)}_{i_1\cdots i_L;\,ij}(\hat r)\}$.  Before imposing tracelessness, the number of independent basis tensors is
\eeq{
N(L)=
\begin{cases}
2L+2,& L\ \text{even}\, ,\\[2pt]
2L+1,& L\ \text{odd}\, .
\end{cases}
\label{eq:number_scalar_tensor}
}
The second step projects out the trace to obtain the basis traceless in $i,j$. For each pre-trace structure $S^{(p)}$, we can apply the symmetric-traceless projector on the $ij$ pair:
\eeq{
\label{eq:tracefree}
\mathrm{TF} \!\left[S^{(p)}\right]_{i_1\cdots i_L;\,ij}
= \frac{1}{2}\!\left(S^{(p)}_{i_1\cdots i_L;\,ij} + S^{(p)}_{i_1\cdots i_L;\,ji}\right)
-\frac{1}{3}\,\delta_{ij}\,S^{(p)}_{i_1\cdots i_L;\,kk}\, ,
}
and we then discard linearly dependent combinations to obtain a basis. We also note that the $b=1$ sector is pure trace and can be omitted before taking the trace-free part, and one can therefore generate a spanning set using only the $b=0$ family from the outset (and then apply $\mathrm{TF}$ to remove traces arising from other contractions).

The number of independent scalar functions after tracelessness is the difference of full number of independent basis tensors  and the number of traces
\eeq{
N_{\text{TF}}(L) \;=\; N(L) - M(L),\quad
M(L)=\Big\lfloor\frac{L}{2}\Big\rfloor+1
}
i.e.
\eeq{
N_{\text{TF}}(L)=
\begin{cases}
\displaystyle \frac{3L}{2}+1,& L\ \text{even},\\[6pt]
\displaystyle \frac{3L+1}{2},& L\ \text{odd}.
\end{cases}
}
Here $M(L)=\lfloor L/2\rfloor+1$ is the dimension of the space of fully symmetric rank-$L$ isotropic tensors (built from $\delta$ and $\hat r$), which is precisely the number of independent trace structures $\xi_{i_1\cdots i_L;kk}$ in Eq.~\eqref{eq:xi_L_symmetric}. 
The final decomposition is then
\eeq{
\xi_{i_1\cdots i_L;\,ij}(\mathbf r)\;=\;\sum_{q=1}^{N_{\text{TF}}(L)}
F_q(r)\;\widetilde S^{(q)}_{i_1\cdots i_L;\,ij}(\hat r),
}
with $\widetilde S^{(q)}$ an independent subset of $\mathrm{TF}[S^{(p)}]$ and $F_q(r)$ arbitrary radial functions.

Let us note the parity properties of these correlators.  Since our isotropic basis tensors are built only from $\delta_{ij}$ and $\hat r_i$, their \emph{geometric} parity is completely determined by the total number of $\hat r$ factors. Under spatial inversion $\hat r\mapsto-\hat r$, each factor of $\hat r$ contributes a minus sign, while $\delta_{ij}$ is invariant. In the $(a,b,c)$ construction, the number of $\hat r$'s carried by the $L$ vector indices is $L-2a-c$, and the number of $\hat r$'s carried by the $T$-pair $(ij)$ is $2-2b-c$. Hence the total number of $\hat r$'s is $N_{\hat r}=(L-2a-c)+(2-2b-c)=L+2-2(a+b+c)$ so that its parity is independent of $(a,b,c)$ and results in $(-1)^{N_{\hat r}} = (-1)^L$.  Therefore \emph{every} isotropic basis tensor $S^{(q)}_{i_1\cdots i_L;ij}(\hat r)$ has the same geometric parity eigenvalue $\rho(S^{(q)}) = (-1)^L$.
If the ensemble is parity invariant, the correlator must satisfy
\eeq{
\xi_{i_1\cdots i_L;ij}(\mathbf r)
=\eta_\phi \eta_V^{\,L}\,\eta_T\;\xi_{i_1\cdots i_L;ij}(-\mathbf r),
}
so in a parity-diagonal isotropic basis one gets the global selection rule
\eeq{
\eta_\phi \eta_V^{\,L}\,\eta_T\,(-1)^L = +1
\qquad \Rightarrow \qquad
\text{correlator allowed},
}
whereas
\eeq{
\eta_\phi \eta_V^{\,L}\,\eta_T\,(-1)^L = -1
\qquad \Rightarrow \qquad
\xi_{i_1\cdots i_L;ij}(\mathbf r)= 0.
}
In particular, in our case of polar vectors and a true (non-pseudo) rank-2 tensor, $\eta_\phi= \eta_V=\eta_T=+1$, parity invariance implies that all odd-$L$ correlators vanish identically, unless they are further combined with another parity odd structure (like in the case of redshift-space distortions).

%------------------------------------------------------%
\subsection{Double Tensor Correlator}

We introduce a second symmetric rank-two tensor $W_{lm}$ into the correlator. For the same reason as in the case of single two-tensor, i.e. we will eventually use fully symmetric contractions in the $L$ vector indices, we focus on analysing the decomposition of the representative, totally symmetrised, correlator 
\eeq{
\xi_{i_1 \cdots i_L;\,ij;\,lm}(\mathbf r)
\;\equiv\;
\big\langle V_{i_1}\cdots V_{i_L}(\mathbf x)\,T_{ij}(\mathbf x)\,W_{lm}(\mathbf x+\mathbf r)\big\rangle\, ,
}
with $T_{ij} = T_{ji}$ and $W_{lm} = W_{ml}$. The vector block is still fully symmetric under permutations of the $L$ vector indices $(i_1,\dots,i_L)$. Tensor blocks are symmetric in $i\leftrightarrow j$ and in $l\leftrightarrow m$. If $T$ and $W$ are distinct, there is no exchange symmetry between $(ij)$ and $lm$ for any $L$. If they are identical and $L=0$, then we also have the exchange symmetry $X:\ (ij,lm,\hat r)\;\mapsto\;(lm,ij,-\hat r)$. 
It is useful to note that if the two tensor field would be at the same point $T_{ij}(\mathbf r)W_{lm}(\mathbf r)$ there would exist an additional internal symmetry $(ij) \leftrightarrow (lm)$ when the two fields would be equal, i.e.  $T=W$, however we are not considering such cases here. 

The building blocks are the various tensor-tensor contractions $T-T$ (via $\delta_{ij}$), $W-W$ (via $\delta_{lm}$), cross $T-W$ (via $\delta_{il},\;\delta_{im},\;\delta_{jl},\;\delta_{jm}$), vector-tensor contractions $V-T$ and $V-W$, and vector-vector $V-V$ contractions.  The leftover indices are carried by $\hat r$. Thus, the bookkeeping expands and one we choose the following parameterisation
\begin{itemize}
\item $a$: $V-V$ deltas,
\item $b_T\in\{0,1\}$: whether $i,j$ contracted together,
\item $b_W\in\{0,1\}$: whether $l,m$ contracted together,
\item $d$: number of $T-W$ deltas $(0,1,2)$,
\item $c_T$: how many of $i,j$ are contracted with $V$ indices,
\item $c_W$: how many of $l,m$ are contracted with $V$ indices.
\end{itemize}
For $d=2$, the only independent cross-contraction after $(ij)$ and $(lm)$ symmetrisation is $\delta_{il}\delta_{jm}+\delta_{jl}\delta_{im}$. Similarly for $d=1$ all four combinations of $\delta$ become equivalent once we symmetrise in $(ij)$ and $(lm)$. These are subject to constraints:
\eeq{
2a + c_T + c_W \;\le L\, , \qquad
2b_T + c_T + d \;\le 2\, , \qquad 2b_W + c_W + d \;\le 2\, .
}
Any unused index contributes with a factor of $\hat r$. After symmetrising over the indices, we can write the decomposition
\eeq{
\xi_{i_1\cdots i_L;\,ij;\,lm}(\mathbf r) = \sum_{q=1}^{N_L} F_q(r) \; S^{(q)}_{i_1\cdots i_L;\,ij;\,lm}(\hat r).
}
With two tensor blocks, the number of invariants $N_L$ grows faster, because we now have cross-pairings between the two tensor blocks. The formula for the number of terms is 
\eeq{
N(L)=
\begin{cases}
\displaystyle \frac{21}{2}L + 3 + 3 \delta_{L,0} ,& L\ \text{even}\, ,\\[6pt]
\displaystyle \frac{21}{2}L + \frac{1}{2} + \delta_{L,1} ,& L\ \text{odd}\, .
\end{cases}
}

The second step again is to project out the trace to obtain the basis traceless in $i$, $j$ and $l$, $m$.  This can be achieved by subtracting the trace twice, as done in Eq.~\eqref{eq:tracefree}, and correcting for double-counting, which gives 
\eeq{
\mathrm{TF} \!\left[S^{(q)}\right]_{i_1\cdots i_L;\,ij;\,lm}
=S^{(q)}_{i_1\cdots i_L;\,ij;\,lm}
- \frac{1}{3} \lb \,\delta_{ij}\, S^{(q)}_{i_1\cdots i_L;\,kk;\,lm} + \delta_{lm}\, S^{(q)}_{i_1\cdots i_L;\,ij;\,kk}  \rb
+ \frac{1}{9} \delta_{ij} \delta_{lm} \, S^{(q)}_{i_1\cdots i_L;\,kk;\,qq} \, ,
}
where TF means traceless in the $(ij)$ pair and traceless in the $(lm)$ pair (no condition on mixed traces like $\delta_{il}$).

The number of independent scalar functions after tracelessness is
\eeq{
N_{\text{TF}}(L) \;=\; N(L) - 2 N'(L) + M(L),\quad
M(L)=\Big\lfloor\frac{L}{2}\Big\rfloor+1
}
where $N'(L)$ is the number of terms in the case of a single symmetric rank-two tensor (not necessarily traceless), which we have obtained in Eq.~\eqref{eq:number_scalar_tensor}.
We obtain
\eeq{
N_{\text{TF}}(L)=
\begin{cases}
 7 L + 3 \delta_{L,0} ,& L\ \text{even}\, ,\\[2pt]
 7 L -1 + \delta_{L,1} ,& L\ \text{odd}\, .
\end{cases}
}
Lastly, we can discuss parity properties. As in the case of single tensor, also in the two tensor case  every basis tensor is made from $\delta$'s and $\hat r$'s. The total rank is $L+4$, so the number of has the same geometric parity as $L$. Therefore the geometric parity of every basis tensor is the same, i.e.  $\rho(S^{(p)}) = (-1)^{N_{\hat r}} = (-1)^L$. Thus, if we impose parity invariance, you again get a global selection rule:
\eeq{
\eta_V^{\,L}\,\eta_T\, \eta_W\,(-1)^L = +1\, ,
\qquad \Rightarrow \qquad
\text{correlator allowed}\, ,
}
and it vanishes otherwise.

%====================================================%
\section{Analytic Formulae for Form Factors in Section~\ref{sec:toy_example}}
\label{app:simple_model}

%------------------------------------------------------%
\subsection{Predictions at 1-loop}

The correlators in Eq.~\eqref{eqn:linear_correlators} are given by the Hankel transforms
\eq{
    S_2(r) &= -\frac23 \xi^2_0(r)\, , \quad B_1(r) = \frac{2f}{15} \xi^1_{-1}(r)\, , \quad B_3(r) = \frac{2f}{5} \xi^3_{-1}(r)\, , \non \\
    V_1(r) &= - f\, \xi^1_{-1}(r)\, , \quad  A_0(r) = \frac{2f^2}{3} \left( \xi^0_{-2}(0) - \xi^0_{-2}(r) \right), \quad A_2(r) = \frac{4f^2}{3} \xi^2_{-2}(r) \, ,
}
where we have defined $\xi^\ell_n(r) = (2\pi^2)^{-1} \int dk\ k^{2+n} j_\ell(k r) P_{\rm lin}(k).$ In order to compute the form-factor coefficients we can simply dot the tensor expression in Eq.~\eqref{eqn:toy_model_cross} with the form factors, properly normalized. It is convenient to write the radial direction relative to the plane defined by the $\hk$ and $\hn$, i.e. $\hat{r} = (\sqrt{1-\nu^2} \cos\phi, \sqrt{1-\nu^2} \sin\phi, \nu)$ where $\nu = \hk \cdot \hat{r}$ and $\phi$ is the azimuthal angle relative to this plane. The above dot products depend on $\phi$ only as polynomials of sines and cosines, which can be analytically integrated, leaving us with integrals that are polynomial in $\nu$, which we can express as Hankel transforms
\begin{equation}
    G_n(k,\mu) = 4\pi \sum_{b_i \in \{1, b_1\}} \sum_{c_j \in \{ c_s \}} b_i c_j \sum_{N,\ell} \mu^N \int dr \ r^2 j_\ell(kr) f^{b_i c_j}_{n,N,\ell}(r)\, .
\end{equation}
We summarize these results as polynomials $N_n f^{b_i c_j}_{n,\ell}(r; k,\mu) = N_n \sum_N \mu^N f^{b_i c_j}_{n,N,\ell}(r)$ in Table~\ref{tab:simple_model_Gn}, where we have also defined $X, Y = A_0 - \frac12 A_2, \frac23 A_2$ for convenience. We similarly give expressions for the 1-loop shape autospectrum, in the form 
\begin{equation}
    H_{nm}^{\parallel / \perp}(k,\mu) = 4\pi c_s^2 \sum_{N,\ell} \mu^N i^\ell \int dr \ r^2 j_\ell(kr) f^{c_s^2,\parallel / \perp}_{n,N,\ell}(r)\, ,
\end{equation}
in Tables~\ref{tab:auto-kernels-1} and \ref{tab:auto-kernels-2}.

\begin{table}[h]
\label{tab:simple_model_Gn}
\centering
\label{tab:Gn-all}
\footnotesize
\resizebox{0.95\textwidth}{!}{%
\begin{tabular}{c | l l | l l | l l}
\toprule
 & \multicolumn{2}{c|}{$n=0$} & \multicolumn{2}{c|}{$n=1$} & \multicolumn{2}{c}{$n=2$} \\
$\ell$ & $c_s$ & $c_s b_1$ & $c_s$ & $c_s b_1$ & $c_s$ & $c_s b_1$ \\
\midrule
$0$ & 0 &
$\begin{aligned}[t]
&-\tfrac{1}{30}\,f^2 k^2\mu^2\,Y\,\xi^{2}_{0} \\
&+\,\tfrac{1}{15}\,f^2 k^2\mu^2\,[\xi^{1}_{-1}]^2 \\
&+\,\tfrac{1}{10}\,f^2 k^2\mu^4\,Y\,\xi^{2}_{0} \\
&-\,\tfrac{1}{5}\,f^2 k^2\mu^4\,[\xi^{1}_{-1}]^2
\end{aligned}$
 & 0 &
$\begin{aligned}[t]
&\tfrac{2}{15}\,f^2 k^2\mu^3\,Y\,\xi^{2}_{0} \\
&-\,\tfrac{4}{15}\,f^2 k^2\mu^3\,[\xi^{1}_{-1}]^2
\end{aligned}$
 & 0 &
$\begin{aligned}[t]
&\tfrac{1}{15}\,f^2 k^2\mu^2\,Y\,\xi^{2}_{0} \\
&-\,\tfrac{2}{15}\,f^2 k^2\mu^2\,[\xi^{1}_{-1}]^2
\end{aligned}$ \\
\midrule
$1$ &
$\begin{aligned}[t]
&\tfrac{2}{5}\,f k\mu^2\,\xi^{1}_{-1} \\
&-\,\tfrac{1}{5}\,f^3 k^3\mu^4\,X\,\xi^{1}_{-1} \\
&+\,\tfrac{3}{70}\,f^3 k^3\mu^4\,Y\,\xi^{3}_{-1} \\
&-\,\tfrac{3}{25}\,f^3 k^3\mu^6\,Y\,\xi^{1}_{-1} \\
&+\,\tfrac{3}{350}\,f^3 k^3\mu^6\,Y\,\xi^{3}_{-1}
\end{aligned}$
 & 0 &
$\begin{aligned}[t]
&\tfrac{2}{5}\,f k\mu\,\xi^{1}_{-1} \\
&-\,\tfrac{1}{5}\,f^3 k^3\mu^3\,X\,\xi^{1}_{-1} \\
&-\,\tfrac{1}{25}\,f^3 k^3\mu^3\,Y\,\xi^{1}_{-1} \\
&+\,\tfrac{8}{175}\,f^3 k^3\mu^3\,Y\,\xi^{3}_{-1} \\
&-\,\tfrac{4}{25}\,f^3 k^3\mu^5\,Y\,\xi^{1}_{-1} \\
&+\,\tfrac{2}{175}\,f^3 k^3\mu^5\,Y\,\xi^{3}_{-1}
\end{aligned}$
 & 0 &
$\begin{aligned}[t]
&-\tfrac{2}{25}\,f^3 k^3\mu^4\,Y\,\xi^{1}_{-1} \\
&+\,\tfrac{1}{175}\,f^3 k^3\mu^4\,Y\,\xi^{3}_{-1}
\end{aligned}$
 & 0 \\
\midrule
$2$ & 0 &
$\begin{aligned}[t]
&\xi^{2}_{0} \\
&-\,\tfrac{1}{2}\,f^2 k^2\mu^2\,X\,\xi^{2}_{0} \\
&-\,\tfrac{5}{42}\,f^2 k^2\mu^2\,Y\,\xi^{2}_{0} \\
&+\,\tfrac{1}{15}\,f^2 k^2\mu^2\,[\xi^{1}_{-1}]^2 \\
&-\,\tfrac{6}{35}\,f^2 k^2\mu^2\,\xi^{1}_{-1}\xi^{3}_{-1} \\
&-\,\tfrac{1}{7}\,f^2 k^2\mu^4\,Y\,\xi^{2}_{0} \\
&+\,\tfrac{1}{5}\,f^2 k^2\mu^4\,[\xi^{1}_{-1}]^2 \\
&-\,\tfrac{3}{35}\,f^2 k^2\mu^4\,\xi^{1}_{-1}\xi^{3}_{-1}
\end{aligned}$
 & 0 &
$\begin{aligned}[t]
&-\tfrac{2}{21}\,f^2 k^2\mu^3\,Y\,\xi^{2}_{0} \\
&+\,\tfrac{2}{15}\,f^2 k^2\mu^3\,[\xi^{1}_{-1}]^2 \\
&-\,\tfrac{2}{35}\,f^2 k^2\mu^3\,\xi^{1}_{-1}\xi^{3}_{-1}
\end{aligned}$
 & 0 &
$\begin{aligned}[t]
&\tfrac{2}{21}\,f^2 k^2\mu^2\,Y\,\xi^{2}_{0} \\
&-\,\tfrac{2}{15}\,f^2 k^2\mu^2\,[\xi^{1}_{-1}]^2 \\
&+\,\tfrac{2}{35}\,f^2 k^2\mu^2\,\xi^{1}_{-1}\xi^{3}_{-1}
\end{aligned}$ \\
\midrule
$3$ &
$\begin{aligned}[t]
&\tfrac{3}{5}\,f k\mu^2\,\xi^{3}_{-1} \\
&-\,\tfrac{3}{10}\,f^3 k^3\mu^4\,X\,\xi^{3}_{-1} \\
&-\,\tfrac{1}{10}\,f^3 k^3\mu^4\,Y\,\xi^{3}_{-1} \\
&+\,\tfrac{2}{25}\,f^3 k^3\mu^6\,Y\,\xi^{1}_{-1} \\
&-\,\tfrac{4}{75}\,f^3 k^3\mu^6\,Y\,\xi^{3}_{-1}
\end{aligned}$
 & 0 &
$\begin{aligned}[t]
&-\tfrac{2}{5}\,f k\mu\,\xi^{3}_{-1} \\
&-\,\tfrac{1}{25}\,f^3 k^3\mu^3\,Y\,\xi^{1}_{-1} \\
&+\,\tfrac{1}{5}\,f^3 k^3\mu^3\,X\,\xi^{3}_{-1} \\
&+\,\tfrac{7}{75}\,f^3 k^3\mu^3\,Y\,\xi^{3}_{-1} \\
&+\,\tfrac{1}{25}\,f^3 k^3\mu^5\,Y\,\xi^{1}_{-1} \\
&-\,\tfrac{2}{75}\,f^3 k^3\mu^5\,Y\,\xi^{3}_{-1}
\end{aligned}$
 & 0 &
$\begin{aligned}[t]
&-\tfrac{2}{25}\,f^3 k^3\mu^4\,Y\,\xi^{1}_{-1} \\
&+\,\tfrac{4}{75}\,f^3 k^3\mu^4\,Y\,\xi^{3}_{-1}
\end{aligned}$
 & 0 \\
\midrule
$4$ & 0 &
$\begin{aligned}[t]
&-\tfrac{3}{35}\,f^2 k^2\mu^2\,Y\,\xi^{2}_{0} \\
&-\,\tfrac{6}{35}\,f^2 k^2\mu^2\,\xi^{1}_{-1}\xi^{3}_{-1} \\
&+\,\tfrac{9}{35}\,f^2 k^2\mu^4\,Y\,\xi^{2}_{0} \\
&+\,\tfrac{18}{35}\,f^2 k^2\mu^4\,\xi^{1}_{-1}\xi^{3}_{-1}
\end{aligned}$
 & 0 &
$\begin{aligned}[t]
&-\tfrac{8}{35}\,f^2 k^2\mu^3\,Y\,\xi^{2}_{0} \\
&-\,\tfrac{16}{35}\,f^2 k^2\mu^3\,\xi^{1}_{-1}\xi^{3}_{-1}
\end{aligned}$
 & 0 &
$\begin{aligned}[t]
&\tfrac{1}{35}\,f^2 k^2\mu^2\,Y\,\xi^{2}_{0} \\
&+\,\tfrac{2}{35}\,f^2 k^2\mu^2\,\xi^{1}_{-1}\xi^{3}_{-1}
\end{aligned}$ \\
\midrule
$5$ &
$\begin{aligned}[t]
&-\tfrac{1}{7}\,f^3 k^3\mu^4\,Y\,\xi^{3}_{-1} \\
&+\,\tfrac{5}{21}\,f^3 k^3\mu^6\,Y\,\xi^{3}_{-1}
\end{aligned}$
 & 0 &
$\begin{aligned}[t]
&\tfrac{1}{21}\,f^3 k^3\mu^3\,Y\,\xi^{3}_{-1} \\
&-\,\tfrac{5}{21}\,f^3 k^3\mu^5\,Y\,\xi^{3}_{-1}
\end{aligned}$
 & 0 &
$\tfrac{1}{21}\,f^3 k^3\mu^4\,Y\,\xi^{3}_{-1}$
 & 0 \\
\bottomrule
\end{tabular}%
}
\caption{Radial kernels $N_n f^{(b_i c_j)}_{n,\ell}(r;k,\mu)$ for $n\in\{0,1,2\}$ and $b_i c_j \in\{c_s,\,c_s b_1\}$, by Hankel order $\ell$.}
\end{table}
\renewcommand{\arraystretch}{1.2}

\begin{table}[h]
\centering
\label{tab:auto-kernels-1}
\footnotesize
{
\begin{tabular}{c | l | l | l}
\toprule
$\ell$ & $H^{\parallel}_{00}$ & $H^{\perp}_{11}$ & $H^{\perp}_{22}$ \\
\midrule
$0$
 & $\begin{aligned}[t]
&\tfrac{2}{15}\,\xi^{0}_{0} \\
&-\,\tfrac{1}{15}f^{2}k^{2}\mu^{2}\,X\,\xi^{0}_{0} \\
&-\,\tfrac{1}{45}f^{2}k^{2}\mu^{2}\,Y\,\xi^{0}_{0} \\
&-\,\tfrac{2}{315}f^{2}k^{2}\mu^{2}\,Y\,\xi^{2}_{0} \\
&-\,\tfrac{2}{225}f^{2}k^{2}\mu^{2}\,[\xi^{1}_{-1}]^2 \\
&-\,\tfrac{4}{175}f^{2}k^{2}\mu^{2}\,[\xi^{3}_{-1}]^2 \\
&+\,\tfrac{2}{105}f^{2}k^{2}\mu^{4}\,Y\,\xi^{2}_{0} \\
&-\,\tfrac{2}{75}f^{2}k^{2}\mu^{4}\,[\xi^{1}_{-1}]^2 \\
&-\,\tfrac{2}{175}f^{2}k^{2}\mu^{4}\,[\xi^{3}_{-1}]^2
\end{aligned}$
 & $\begin{aligned}[t]
&\tfrac{2 \sqrt{2}}{15}\,\xi^{0}_{0} \\
&-\,\tfrac{\sqrt{2}}{15}f^{2}k^{2}\mu^{2}\,X\,\xi^{0}_{0} \\
&-\,\tfrac{\sqrt{2}}{45}f^{2}k^{2}\mu^{2}\,Y\,\xi^{0}_{0} \\
&-\,\tfrac{\sqrt{2}}{315}f^{2}k^{2}\mu^{2}\,Y\,\xi^{2}_{0} \\
&-\,\tfrac{\sqrt{2}}{75}f^{2}k^{2}\mu^{2}\,[\xi^{1}_{-1}]^2 \\
&-\,\tfrac{13 \sqrt{2}}{525}f^{2}k^{2}\mu^{2}\,[\xi^{3}_{-1}]^2 \\
&+\,\tfrac{\sqrt{2}}{105}f^{2}k^{2}\mu^{4}\,Y\,\xi^{2}_{0} \\
&-\,\tfrac{\sqrt{2}}{75}f^{2}k^{2}\mu^{4}\,[\xi^{1}_{-1}]^2 \\
&-\,\tfrac{\sqrt{2}}{175}f^{2}k^{2}\mu^{4}\,[\xi^{3}_{-1}]^2
\end{aligned}$
 & $\begin{aligned}[t]
&\tfrac{2 \sqrt{2}}{15}\,\xi^{0}_{0} \\
&-\,\tfrac{\sqrt{2}}{15}f^{2}k^{2}\mu^{2}\,X\,\xi^{0}_{0} \\
&-\,\tfrac{\sqrt{2}}{45}f^{2}k^{2}\mu^{2}\,Y\,\xi^{0}_{0} \\
&+\,\tfrac{2 \sqrt{2}}{315}f^{2}k^{2}\mu^{2}\,Y\,\xi^{2}_{0} \\
&-\,\tfrac{2 \sqrt{2}}{75}f^{2}k^{2}\mu^{2}\,[\xi^{1}_{-1}]^2 \\
&-\,\tfrac{16 \sqrt{2}}{525}f^{2}k^{2}\mu^{2}\,[\xi^{3}_{-1}]^2 \\
&-\,\tfrac{2 \sqrt{2}}{105}f^{2}k^{2}\mu^{4}\,Y\,\xi^{2}_{0} \\
&+\,\tfrac{2 \sqrt{2}}{75}f^{2}k^{2}\mu^{4}\,[\xi^{1}_{-1}]^2 \\
&+\,\tfrac{2 \sqrt{2}}{175}f^{2}k^{2}\mu^{4}\,[\xi^{3}_{-1}]^2
\end{aligned}$ \\
\midrule
$2$
 & $\begin{aligned}[t]
&-\,\tfrac{4}{21}\,\xi^{2}_{0} \\
&+\,\tfrac{2}{21}f^{2}k^{2}\mu^{2}\,X\,\xi^{2}_{0} \\
&+\,\tfrac{1}{45}f^{2}k^{2}\mu^{2}\,Y\,\xi^{0}_{0} \\
&+\,\tfrac{10}{441}f^{2}k^{2}\mu^{2}\,Y\,\xi^{2}_{0} \\
&+\,\tfrac{4}{245}f^{2}k^{2}\mu^{2}\,Y\,\xi^{4}_{0} \\
&+\,\tfrac{2}{225}f^{2}k^{2}\mu^{2}\,[\xi^{1}_{-1}]^2 \\
&-\,\tfrac{8}{175}f^{2}k^{2}\mu^{2}\,\xi^{1}_{-1}\,\xi^{3}_{-1} \\
&-\,\tfrac{4}{175}f^{2}k^{2}\mu^{2}\,[\xi^{3}_{-1}]^2 \\
&-\,\tfrac{1}{15}f^{2}k^{2}\mu^{4}\,Y\,\xi^{0}_{0} \\
&+\,\tfrac{4}{147}f^{2}k^{2}\mu^{4}\,Y\,\xi^{2}_{0} \\
&-\,\tfrac{12}{245}f^{2}k^{2}\mu^{4}\,Y\,\xi^{4}_{0} \\
&-\,\tfrac{2}{25}f^{2}k^{2}\mu^{4}\,[\xi^{1}_{-1}]^2 \\
&+\,\tfrac{32}{175}f^{2}k^{2}\mu^{4}\,\xi^{1}_{-1}\,\xi^{3}_{-1} \\
&-\,\tfrac{4}{175}f^{2}k^{2}\mu^{4}\,[\xi^{3}_{-1}]^2
\end{aligned}$
 & $\begin{aligned}[t]
&-\,\tfrac{2 \sqrt{2}}{21}\,\xi^{2}_{0} \\
&+\,\tfrac{\sqrt{2}}{21}f^{2}k^{2}\mu^{2}\,X\,\xi^{2}_{0} \\
&+\,\tfrac{\sqrt{2}}{45}f^{2}k^{2}\mu^{2}\,Y\,\xi^{0}_{0} \\
&+\,\tfrac{5 \sqrt{2}}{441}f^{2}k^{2}\mu^{2}\,Y\,\xi^{2}_{0} \\
&-\,\tfrac{8 \sqrt{2}}{735}f^{2}k^{2}\mu^{2}\,Y\,\xi^{4}_{0} \\
&-\,\tfrac{2 \sqrt{2}}{75}f^{2}k^{2}\mu^{2}\,[\xi^{1}_{-1}]^2 \\
&-\,\tfrac{6 \sqrt{2}}{175}f^{2}k^{2}\mu^{2}\,\xi^{1}_{-1}\,\xi^{3}_{-1} \\
&-\,\tfrac{4 \sqrt{2}}{525}f^{2}k^{2}\mu^{2}\,[\xi^{3}_{-1}]^2 \\
&-\,\tfrac{\sqrt{2}}{15}f^{2}k^{2}\mu^{4}\,Y\,\xi^{0}_{0} \\
&+\,\tfrac{2 \sqrt{2}}{147}f^{2}k^{2}\mu^{4}\,Y\,\xi^{2}_{0} \\
&+\,\tfrac{8 \sqrt{2}}{245}f^{2}k^{2}\mu^{4}\,Y\,\xi^{4}_{0} \\
&+\,\tfrac{4 \sqrt{2}}{75}f^{2}k^{2}\mu^{4}\,[\xi^{1}_{-1}]^2 \\
&+\,\tfrac{22 \sqrt{2}}{175}f^{2}k^{2}\mu^{4}\,\xi^{1}_{-1}\,\xi^{3}_{-1} \\
&-\,\tfrac{4 \sqrt{2}}{175}f^{2}k^{2}\mu^{4}\,[\xi^{3}_{-1}]^2
\end{aligned}$
 & $\begin{aligned}[t]
&\tfrac{4 \sqrt{2}}{21}\,\xi^{2}_{0} \\
&-\,\tfrac{2 \sqrt{2}}{21}f^{2}k^{2}\mu^{2}\,X\,\xi^{2}_{0} \\
&+\,\tfrac{\sqrt{2}}{45}f^{2}k^{2}\mu^{2}\,Y\,\xi^{0}_{0} \\
&-\,\tfrac{10 \sqrt{2}}{441}f^{2}k^{2}\mu^{2}\,Y\,\xi^{2}_{0} \\
&+\,\tfrac{2 \sqrt{2}}{735}f^{2}k^{2}\mu^{2}\,Y\,\xi^{4}_{0} \\
&+\,\tfrac{2 \sqrt{2}}{75}f^{2}k^{2}\mu^{2}\,[\xi^{1}_{-1}]^2 \\
&-\,\tfrac{4 \sqrt{2}}{175}f^{2}k^{2}\mu^{2}\,\xi^{1}_{-1}\,\xi^{3}_{-1} \\
&+\,\tfrac{8 \sqrt{2}}{175}f^{2}k^{2}\mu^{2}\,[\xi^{3}_{-1}]^2 \\
&-\,\tfrac{\sqrt{2}}{15}f^{2}k^{2}\mu^{4}\,Y\,\xi^{0}_{0} \\
&-\,\tfrac{4 \sqrt{2}}{147}f^{2}k^{2}\mu^{4}\,Y\,\xi^{2}_{0} \\
&-\,\tfrac{2 \sqrt{2}}{245}f^{2}k^{2}\mu^{4}\,Y\,\xi^{4}_{0} \\
&-\,\tfrac{2 \sqrt{2}}{75}f^{2}k^{2}\mu^{4}\,[\xi^{1}_{-1}]^2 \\
&+\,\tfrac{4 \sqrt{2}}{175}f^{2}k^{2}\mu^{4}\,\xi^{1}_{-1}\,\xi^{3}_{-1} \\
&-\,\tfrac{8 \sqrt{2}}{175}f^{2}k^{2}\mu^{4}\,[\xi^{3}_{-1}]^2
\end{aligned}$ \\
\midrule
$4$
 & $\begin{aligned}[t]
&\tfrac{12}{35}\,\xi^{4}_{0} \\
&-\,\tfrac{6}{35}f^{2}k^{2}\mu^{2}\,X\,\xi^{4}_{0} \\
&-\,\tfrac{4}{245}f^{2}k^{2}\mu^{2}\,Y\,\xi^{2}_{0} \\
&-\,\tfrac{114}{2695}f^{2}k^{2}\mu^{2}\,Y\,\xi^{4}_{0} \\
&+\,\tfrac{8}{175}f^{2}k^{2}\mu^{2}\,\xi^{1}_{-1}\,\xi^{3}_{-1} \\
&-\,\tfrac{12}{1925}f^{2}k^{2}\mu^{2}\,[\xi^{3}_{-1}]^2 \\
&+\,\tfrac{12}{245}f^{2}k^{2}\mu^{4}\,Y\,\xi^{2}_{0} \\
&-\,\tfrac{24}{539}f^{2}k^{2}\mu^{4}\,Y\,\xi^{4}_{0} \\
&+\,\tfrac{24}{175}f^{2}k^{2}\mu^{4}\,\xi^{1}_{-1}\,\xi^{3}_{-1} \\
&-\,\tfrac{96}{1925}f^{2}k^{2}\mu^{4}\,[\xi^{3}_{-1}]^2
\end{aligned}$
 & $\begin{aligned}[t]
&-\,\tfrac{8 \sqrt{2}}{35}\,\xi^{4}_{0} \\
&+\,\tfrac{4 \sqrt{2}}{35}f^{2}k^{2}\mu^{2}\,X\,\xi^{4}_{0} \\
&-\,\tfrac{2 \sqrt{2}}{245}f^{2}k^{2}\mu^{2}\,Y\,\xi^{2}_{0} \\
&+\,\tfrac{76 \sqrt{2}}{2695}f^{2}k^{2}\mu^{2}\,Y\,\xi^{4}_{0} \\
&-\,\tfrac{8 \sqrt{2}}{175}f^{2}k^{2}\mu^{2}\,\xi^{1}_{-1}\,\xi^{3}_{-1} \\
&+\,\tfrac{52 \sqrt{2}}{1925}f^{2}k^{2}\mu^{2}\,[\xi^{3}_{-1}]^2 \\
&+\,\tfrac{6 \sqrt{2}}{245}f^{2}k^{2}\mu^{4}\,Y\,\xi^{2}_{0} \\
&+\,\tfrac{16 \sqrt{2}}{539}f^{2}k^{2}\mu^{4}\,Y\,\xi^{4}_{0} \\
&-\,\tfrac{8 \sqrt{2}}{175}f^{2}k^{2}\mu^{4}\,\xi^{1}_{-1}\,\xi^{3}_{-1} \\
&-\,\tfrac{68 \sqrt{2}}{1925}f^{2}k^{2}\mu^{4}\,[\xi^{3}_{-1}]^2
\end{aligned}$
 & $\begin{aligned}[t]
&\tfrac{2 \sqrt{2}}{35}\,\xi^{4}_{0} \\
&-\,\tfrac{\sqrt{2}}{35}f^{2}k^{2}\mu^{2}\,X\,\xi^{4}_{0} \\
&+\,\tfrac{4 \sqrt{2}}{245}f^{2}k^{2}\mu^{2}\,Y\,\xi^{2}_{0} \\
&-\,\tfrac{19 \sqrt{2}}{2695}f^{2}k^{2}\mu^{2}\,Y\,\xi^{4}_{0} \\
&+\,\tfrac{4 \sqrt{2}}{175}f^{2}k^{2}\mu^{2}\,\xi^{1}_{-1}\,\xi^{3}_{-1} \\
&-\,\tfrac{46 \sqrt{2}}{1925}f^{2}k^{2}\mu^{2}\,[\xi^{3}_{-1}]^2 \\
&-\,\tfrac{12 \sqrt{2}}{245}f^{2}k^{2}\mu^{4}\,Y\,\xi^{2}_{0} \\
&-\,\tfrac{4 \sqrt{2}}{539}f^{2}k^{2}\mu^{4}\,Y\,\xi^{4}_{0} \\
&-\,\tfrac{4 \sqrt{2}}{175}f^{2}k^{2}\mu^{4}\,\xi^{1}_{-1}\,\xi^{3}_{-1} \\
&+\,\tfrac{116 \sqrt{2}}{1925}f^{2}k^{2}\mu^{4}\,[\xi^{3}_{-1}]^2
\end{aligned}$ \\
\midrule
$6$
 & $\begin{aligned}[t]
&\tfrac{2}{77}f^{2}k^{2}\mu^{2}\,Y\,\xi^{4}_{0} \\
&+\,\tfrac{4}{77}f^{2}k^{2}\mu^{2}\,[\xi^{3}_{-1}]^2 \\
&-\,\tfrac{6}{77}f^{2}k^{2}\mu^{4}\,Y\,\xi^{4}_{0} \\
&-\,\tfrac{12}{77}f^{2}k^{2}\mu^{4}\,[\xi^{3}_{-1}]^2
\end{aligned}$
 & $\begin{aligned}[t]
&-\,\tfrac{4 \sqrt{2}}{231}f^{2}k^{2}\mu^{2}\,Y\,\xi^{4}_{0} \\
&-\,\tfrac{8 \sqrt{2}}{231}f^{2}k^{2}\mu^{2}\,[\xi^{3}_{-1}]^2 \\
&+\,\tfrac{4 \sqrt{2}}{77}f^{2}k^{2}\mu^{4}\,Y\,\xi^{4}_{0} \\
&+\,\tfrac{8 \sqrt{2}}{77}f^{2}k^{2}\mu^{4}\,[\xi^{3}_{-1}]^2
\end{aligned}$
 & $\begin{aligned}[t]
&\tfrac{\sqrt{2}}{231}f^{2}k^{2}\mu^{2}\,Y\,\xi^{4}_{0} \\
&+\,\tfrac{2 \sqrt{2}}{231}f^{2}k^{2}\mu^{2}\,[\xi^{3}_{-1}]^2 \\
&-\,\tfrac{\sqrt{2}}{77}f^{2}k^{2}\mu^{4}\,Y\,\xi^{4}_{0} \\
&-\,\tfrac{2 \sqrt{2}}{77}f^{2}k^{2}\mu^{4}\,[\xi^{3}_{-1}]^2
\end{aligned}$ \\
\bottomrule
\end{tabular}%
}
\caption{Radial kernels $f^{c_s^2,\parallel / \perp}_{n,\ell}(r;k,\mu)$ of the shape auto-spectrum form factors for the basis elements $\mathcal Y^{\parallel}_{00}$, $\mathcal Y^{\perp}_{11}$, $\mathcal Y^{\perp}_{22}$, by Hankel order $\ell$.}
\end{table}
\begin{table}[h]
\centering
\label{tab:auto-kernels-2}
\footnotesize
{
\begin{tabular}{c | l | l | l | l}
\toprule
$\ell$ & $H^{\parallel}_{01}\ (=H^{\parallel}_{10})$ & $H^{\parallel}_{02}\ (=H^{\parallel}_{20})$ & $H^{\parallel}_{11}$ & $H^{\perp}_{12}\ (=H^{\perp}_{21})$ \\
\midrule
$0$
 & $\begin{aligned}[t]
&\tfrac{2 \sqrt{3}}{315}f^{2}k^{2}\mu^{3}\,Y\,\xi^{2}_{0} \\
&-\,\tfrac{2 \sqrt{3}}{225}f^{2}k^{2}\mu^{3}\,[\xi^{1}_{-1}]^2 \\
&-\,\tfrac{2 \sqrt{3}}{525}f^{2}k^{2}\mu^{3}\,[\xi^{3}_{-1}]^2
\end{aligned}$
 & $\begin{aligned}[t]
&-\,\tfrac{2 \sqrt{3}}{315}f^{2}k^{2}\mu^{2}\,Y\,\xi^{2}_{0} \\
&+\,\tfrac{2 \sqrt{3}}{225}f^{2}k^{2}\mu^{2}\,[\xi^{1}_{-1}]^2 \\
&+\,\tfrac{2 \sqrt{3}}{525}f^{2}k^{2}\mu^{2}\,[\xi^{3}_{-1}]^2
\end{aligned}$
 & $\begin{aligned}[t]
&\tfrac{\sqrt{2}}{105}f^{2}k^{2}\mu^{2}\,Y\,\xi^{2}_{0} \\
&-\,\tfrac{\sqrt{2}}{75}f^{2}k^{2}\mu^{2}\,[\xi^{1}_{-1}]^2 \\
&-\,\tfrac{\sqrt{2}}{175}f^{2}k^{2}\mu^{2}\,[\xi^{3}_{-1}]^2
\end{aligned}$
 & $\begin{aligned}[t]
&\tfrac{2 \sqrt{2}}{105}f^{2}k^{2}\mu^{3}\,Y\,\xi^{2}_{0} \\
&-\,\tfrac{2 \sqrt{2}}{75}f^{2}k^{2}\mu^{3}\,[\xi^{1}_{-1}]^2 \\
&-\,\tfrac{2 \sqrt{2}}{175}f^{2}k^{2}\mu^{3}\,[\xi^{3}_{-1}]^2
\end{aligned}$ \\
\midrule
$2$
 & $\begin{aligned}[t]
&\tfrac{2 \sqrt{3}}{441}f^{2}k^{2}\mu^{3}\,Y\,\xi^{2}_{0} \\
&-\,\tfrac{4 \sqrt{3}}{147}f^{2}k^{2}\mu^{3}\,Y\,\xi^{4}_{0} \\
&-\,\tfrac{2 \sqrt{3}}{45}f^{2}k^{2}\mu^{3}\,[\xi^{1}_{-1}]^2 \\
&+\,\tfrac{2 \sqrt{3}}{105}f^{2}k^{2}\mu^{3}\,\xi^{1}_{-1}\,\xi^{3}_{-1}
\end{aligned}$
 & $\begin{aligned}[t]
&\tfrac{4 \sqrt{3}}{441}f^{2}k^{2}\mu^{2}\,Y\,\xi^{2}_{0} \\
&-\,\tfrac{\sqrt{3}}{147}f^{2}k^{2}\mu^{2}\,Y\,\xi^{4}_{0} \\
&-\,\tfrac{2 \sqrt{3}}{225}f^{2}k^{2}\mu^{2}\,[\xi^{1}_{-1}]^2 \\
&+\,\tfrac{2 \sqrt{3}}{75}f^{2}k^{2}\mu^{2}\,\xi^{1}_{-1}\,\xi^{3}_{-1} \\
&+\,\tfrac{2 \sqrt{3}}{525}f^{2}k^{2}\mu^{2}\,[\xi^{3}_{-1}]^2
\end{aligned}$
 & $\begin{aligned}[t]
&-\,\tfrac{2 \sqrt{2}}{147}f^{2}k^{2}\mu^{2}\,Y\,\xi^{2}_{0} \\
&-\,\tfrac{2 \sqrt{2}}{147}f^{2}k^{2}\mu^{2}\,Y\,\xi^{4}_{0} \\
&-\,\tfrac{2 \sqrt{2}}{75}f^{2}k^{2}\mu^{2}\,[\xi^{1}_{-1}]^2 \\
&-\,\tfrac{6 \sqrt{2}}{175}f^{2}k^{2}\mu^{2}\,\xi^{1}_{-1}\,\xi^{3}_{-1} \\
&-\,\tfrac{4 \sqrt{2}}{525}f^{2}k^{2}\mu^{2}\,[\xi^{3}_{-1}]^2
\end{aligned}$
 & $\begin{aligned}[t]
&\tfrac{2 \sqrt{2}}{147}f^{2}k^{2}\mu^{3}\,Y\,\xi^{2}_{0} \\
&+\,\tfrac{2 \sqrt{2}}{147}f^{2}k^{2}\mu^{3}\,Y\,\xi^{4}_{0} \\
&+\,\tfrac{2 \sqrt{2}}{75}f^{2}k^{2}\mu^{3}\,[\xi^{1}_{-1}]^2 \\
&+\,\tfrac{6 \sqrt{2}}{175}f^{2}k^{2}\mu^{3}\,\xi^{1}_{-1}\,\xi^{3}_{-1} \\
&+\,\tfrac{4 \sqrt{2}}{525}f^{2}k^{2}\mu^{3}\,[\xi^{3}_{-1}]^2
\end{aligned}$ \\
\midrule
$4$
 & $\begin{aligned}[t]
&-\,\tfrac{8 \sqrt{3}}{735}f^{2}k^{2}\mu^{3}\,Y\,\xi^{2}_{0} \\
&-\,\tfrac{4 \sqrt{3}}{539}f^{2}k^{2}\mu^{3}\,Y\,\xi^{4}_{0} \\
&+\,\tfrac{4 \sqrt{3}}{525}f^{2}k^{2}\mu^{3}\,\xi^{1}_{-1}\,\xi^{3}_{-1} \\
&+\,\tfrac{4 \sqrt{3}}{275}f^{2}k^{2}\mu^{3}\,[\xi^{3}_{-1}]^2
\end{aligned}$
 & $\begin{aligned}[t]
&-\,\tfrac{2 \sqrt{3}}{735}f^{2}k^{2}\mu^{2}\,Y\,\xi^{2}_{0} \\
&+\,\tfrac{6 \sqrt{3}}{539}f^{2}k^{2}\mu^{2}\,Y\,\xi^{4}_{0} \\
&-\,\tfrac{2 \sqrt{3}}{75}f^{2}k^{2}\mu^{2}\,\xi^{1}_{-1}\,\xi^{3}_{-1} \\
&+\,\tfrac{2 \sqrt{3}}{1925}f^{2}k^{2}\mu^{2}\,[\xi^{3}_{-1}]^2
\end{aligned}$
 & $\begin{aligned}[t]
&\tfrac{\sqrt{2}}{245}f^{2}k^{2}\mu^{2}\,Y\,\xi^{2}_{0} \\
&+\,\tfrac{12 \sqrt{2}}{539}f^{2}k^{2}\mu^{2}\,Y\,\xi^{4}_{0} \\
&-\,\tfrac{8 \sqrt{2}}{175}f^{2}k^{2}\mu^{2}\,\xi^{1}_{-1}\,\xi^{3}_{-1} \\
&-\,\tfrac{18 \sqrt{2}}{1925}f^{2}k^{2}\mu^{2}\,[\xi^{3}_{-1}]^2
\end{aligned}$
 & $\begin{aligned}[t]
&-\,\tfrac{8 \sqrt{2}}{245}f^{2}k^{2}\mu^{3}\,Y\,\xi^{2}_{0} \\
&+\,\tfrac{2 \sqrt{2}}{539}f^{2}k^{2}\mu^{3}\,Y\,\xi^{4}_{0} \\
&-\,\tfrac{6 \sqrt{2}}{175}f^{2}k^{2}\mu^{3}\,\xi^{1}_{-1}\,\xi^{3}_{-1} \\
&+\,\tfrac{74 \sqrt{2}}{1925}f^{2}k^{2}\mu^{3}\,[\xi^{3}_{-1}]^2
\end{aligned}$ \\
\midrule
$6$
 & $\begin{aligned}[t]
&\tfrac{8 \sqrt{3}}{231}f^{2}k^{2}\mu^{3}\,Y\,\xi^{4}_{0} \\
&+\,\tfrac{16 \sqrt{3}}{231}f^{2}k^{2}\mu^{3}\,[\xi^{3}_{-1}]^2
\end{aligned}$
 & $\begin{aligned}[t]
&-\,\tfrac{\sqrt{3}}{231}f^{2}k^{2}\mu^{2}\,Y\,\xi^{4}_{0} \\
&-\,\tfrac{2 \sqrt{3}}{231}f^{2}k^{2}\mu^{2}\,[\xi^{3}_{-1}]^2
\end{aligned}$
 & $\begin{aligned}[t]
&-\,\tfrac{2 \sqrt{2}}{231}f^{2}k^{2}\mu^{2}\,Y\,\xi^{4}_{0} \\
&-\,\tfrac{4 \sqrt{2}}{231}f^{2}k^{2}\mu^{2}\,[\xi^{3}_{-1}]^2
\end{aligned}$
 & $\begin{aligned}[t]
&-\,\tfrac{4 \sqrt{2}}{231}f^{2}k^{2}\mu^{3}\,Y\,\xi^{4}_{0} \\
&-\,\tfrac{8 \sqrt{2}}{231}f^{2}k^{2}\mu^{3}\,[\xi^{3}_{-1}]^2
\end{aligned}$ \\
\bottomrule
\end{tabular}%
}
\caption{As in Table \ref{tab:auto-kernels-1}, for the remaining basis elements.}
\end{table}

%------------------------------------------------------%
\subsection{Predictions at Arbitrary Order}

We collect here the angular machinery entering Eqs.~\eqref{eqn:gfsm_full_cross}--\eqref{eqn:gfsm_shifted_angular}. Expanding the exponential in Eq.~\eqref{eqn:gfsm_E_coefficients} order by order in $a$ gives us the closed form for the angular coefficients, 
\eeq{
\mathcal E^{(n)}_\ell(a) = \sum_{p=0}^{\infty} \frac{(-a)^p}{p!}\, \frac{2\ell+1}{2} \int_{-1}^{1} dy\; y^n \ls \mathcal L_2(y) \rs^p \mathcal L_\ell(y) \, ,
}
where each term is an elementary integral of a product of Legendre polynomials. The parity of the integrand gives the selection rule $\mathcal E^{(n)}_\ell = 0$ for $\ell+n$ odd quoted in Sec.~\ref{subsec:gfsm_full}.

Consequently, the angular integrals required are
\eeq{
\int d\Omega_{\hat r}\; e^{i k r \nu}\, M^{(p)}_\lambda(\hat r,\hn)\, \mathcal L_\ell(\hn\cdot\hat r) \equiv 4\pi\, i^{\ell+p}\, \mathcal T^{(p)}_{\lambda \ell}(kr,\mu) \, , \qquad p = 0,1 \, ,
}
where $M^{(0)}_\lambda$ and $M^{(1)}_\lambda$ are the projections onto $\mathcal Q_\lambda$ basis of the two tensor structures carried by $\la s_{ab}\delta \ra$ and $\hn_i \la \Delta \vec u_i s_{ab}\ra$ respectively. These can be derived from the scalar master integral $\int d\Omega_{\hat r} e^{iz\nu}\mathcal L_\ell(\hn\cdot\hat r) = 4\pi i^\ell j_\ell(z)\mathcal L_\ell(\mu)$ by differentiation with respect to $k_i$. Writing $\Phi_\ell(z,\mu) \equiv j_\ell(z)\mathcal L_\ell(\mu)$ and $\Lambda_\ell \equiv \ell(\ell+1)$, we have
\eq{
\mathcal T^{(0)}_{0\ell} &= \frac{1}{z^2}\ls 2z\partial_z + \frac23 z^2 - \Lambda_\ell \rs \Phi_\ell \, , \qquad
\mathcal T^{(0)}_{1\ell} = -\frac{s^2}{z^2}\lb z\partial_z - 1 \rb \partial_\mu \Phi_\ell \, , \qquad
\mathcal T^{(0)}_{2\ell} = -\frac{s^4}{2z^2}\partial_\mu^2 \Phi_\ell \, , \non \\
\mathcal T^{(1)}_{0\ell} &= -\frac{2}{15 z}\lb 2\mu z \partial_z - s^2 \partial_\mu \rb \Phi_\ell \, , \qquad
\mathcal T^{(1)}_{1\ell} = -\frac{s^2}{5z}\lb z\partial_z + \mu \partial_\mu \rb \Phi_\ell \, , \qquad
\mathcal T^{(1)}_{2\ell} = -\frac{s^4}{5z}\partial_\mu \Phi_\ell \, .
}
Although our usual norms $N_1$ and $N_2$ carry inverse powers of $s$, the kernels supply compensating factors $s^2$ and $s^4$, so that the products entering the form factors are regular as $\mu \rightarrow \pm 1$.

Furthermore, our result above obtain the derivatives of the spherical Bessel function that we would like to remove. Using $z j'_\ell = \ell j_\ell - z j_{\ell+1}$ and $\mathcal L'_\ell = (\ell+1) F_\ell$, we can introduce
\eeq{
F_\ell(\mu) = \frac{1}{\ell+1}\sum_{n=0}^{\lfloor (\ell-1)/2 \rfloor} \lb 2\ell - 4n - 1 \rb \mathcal L_{\ell-1-2n}(\mu) \, ,
}
a polynomial of degree $\ell-1$ ($\mathcal L'_\ell$ that is manifestly regular at $\mu = \pm1$, despite $\mathcal L'_\ell = (\ell+1)\lb \mu \mathcal L_\ell - \mathcal L_{\ell+1}\rb/s^2$), and where $F'_\ell$ can be obtained by term-by-term differentiation. Thus, every kernel reduces to a combination of $j_\ell$ and $j_{\ell+1}$ with polynomial coefficients in $\mu$ and no derivatives.

Moreover, we can shift the Bessel function terms around to obtain a single Bessel order. The remaining inverse powers of $z$ are removed with the spherical-Bessel recurrence, each trading one power of $1/z$ for a unit shift in the Bessel order. The $M^{(0)}$ kernels then involve only $j_{\ell-2},j_\ell,j_{\ell+2}$ and the $M^{(1)}$ kernels only $j_{\ell-1},j_{\ell+1}$, with
\eeq{
N_\lambda \mathcal T^{(0)}_{\lambda\ell} = \sum_{\Delta = -2,0,2} C^{(\lambda,\Delta)}_\ell(\mu)\, j_{\ell+\Delta}(z) \, , \qquad
N_\lambda \mathcal T^{(1)}_{\lambda\ell} = \sum_{\Delta = -1,1} D^{(\lambda,\Delta)}_\ell(\mu)\, j_{\ell+\Delta}(z) \, ,
}
where each $C^{(\lambda,\Delta)}_\ell$ factorises into a purely $\ell$-dependent rational prefactor times an angular function, i.e. $C^{(\lambda,\Delta)}_\ell = c^{(\lambda,\Delta)}_\ell X^{(\lambda)}_\ell(\mu)$ with $X^{(0)}_\ell = \mathcal L_\ell$, $X^{(1)}_\ell = F_\ell$, $X^{(2)}_\ell = F'_\ell$, and
\eeq{
c^{(\lambda,\Delta)}_\ell =
\begin{array}{c|ccc}
 & \Delta=-2 & \Delta=0 & \Delta=2 \\
\hline
\lambda=0 & \displaystyle-\frac{3\ell(\ell-1)}{2(2\ell-1)(2\ell+1)} & \displaystyle\frac{\ell(\ell+1)}{(2\ell-1)(2\ell+3)} & \displaystyle-\frac{3(\ell+1)(\ell+2)}{2(2\ell+1)(2\ell+3)} \\[10pt]
\lambda=1 & \displaystyle-\frac{2(\ell-1)(\ell+1)}{(2\ell-1)(2\ell+1)} & \displaystyle\frac{2(\ell+1)}{(2\ell-1)(2\ell+3)} & \displaystyle\frac{2(\ell+1)(\ell+2)}{(2\ell+1)(2\ell+3)} \\[10pt]
\lambda=2 & \displaystyle-\frac{\ell+1}{(2\ell-1)(2\ell+1)} & \displaystyle-\frac{2(\ell+1)}{(2\ell-1)(2\ell+3)} & \displaystyle-\frac{\ell+1}{(2\ell+1)(2\ell+3)}
\end{array}
}
while the $D$ coefficients all share the common prefactor and we can write
\eq{
D^{(0,-1)}_\ell &= \frac{1}{5(2\ell+1)}\Big[ -2\ell\, \mu\, \mathcal L_\ell(\mu) + (\ell+1)\, s^2 F_\ell(\mu) \Big] \, , \non \\
D^{(0,+1)}_\ell &= \frac{1}{5(2\ell+1)}\Big[ \phantom{-} 2(\ell+1)\, \mu\, \mathcal L_\ell(\mu) + (\ell+1)\, s^2 F_\ell(\mu) \Big] \, , \non \\
D^{(1,-1)}_\ell &= \frac{1}{5(2\ell+1)}\Big[ -2\ell\, \mathcal L_\ell(\mu) - 2(\ell+1)\, \mu\, F_\ell(\mu) \Big] \, , \non \\
D^{(1,+1)}_\ell &= \frac{1}{5(2\ell+1)}\Big[ \phantom{-} 2(\ell+1)\, \mathcal L_\ell(\mu) - 2(\ell+1)\, \mu\, F_\ell(\mu) \Big] \, , \non \\
D^{(2,-1)}_\ell &= D^{(2,+1)}_\ell = -\frac{2(\ell+1)}{5(2\ell+1)}\, F_\ell(\mu) \, .
}
Reindexing the sum by the final Bessel order and separating even from odd orders using the $\mathcal E^{(n)}_\ell$ selection rule then produces the shifted combinations $\widetilde{C}^{(n)}_{\lambda L}$, $\widetilde{D}^{(n)}_{\lambda L}$ of Eq.~\eqref{eqn:gfsm_shifted_angular} and hence Eq.~\eqref{eqn:gfsm_full_cross}.

The tensor-tensor case proceeds identically, with the rank-two projections $M^{(p)}_\lambda$ replaced by their rank-four counterparts and the master integral differentiated up to fourth order; the resulting kernels involve $j_{\ell+\Delta}$ with $\Delta = -4,\ldots,4$, and the selection rule leaves only even final orders, as quoted in Eq.~\eqref{eqn:gfsm_full_auto}.

%==========================================================================
\section*{References}
\bibliography{main} 

\end{document}